\pdfoutput=1
\documentclass[sigconf,nonacm]{acmart}
\setcopyright{none}
\renewcommand\footnotetextcopyrightpermission[1]{}
\usepackage{mathtools}
\usepackage{algorithm}
\usepackage[noend]{algpseudocode}
\usepackage{multirow}
\usepackage{array}
\usepackage{enumitem}
\usepackage{float}
\usepackage{stfloats}
\usepackage{listings}
\usepackage{placeins}
\usepackage{needspace}
\lstnewenvironment{promptbox}{\endgraf\nopagebreak[4]\smallskip\noindent\minipage{\linewidth}\lstset{basicstyle=\ttfamily\scriptsize,breaklines=true,breakatwhitespace=false,columns=fullflexible,keepspaces=true,frame=single,framesep=3pt,upquote=true,aboveskip=0pt,belowskip=0pt}}{\endminipage\endgraf\smallskip}
\theoremstyle{acmplain}
\newtheorem{theorem}{Theorem}

\newtheorem{proposition}{Proposition}

\newcommand{\ncccN}{9{,}903}
\newcommand{\AuthorOneName}{Liang Yue}
\newcommand{\AuthorOneEmail}{liyu2195@colorado.edu}
\newcommand{\AuthorTwoEmail}{Eric.Vance@colorado.edu}
\newcommand{\RepoURL}{https://github.com/liyu2195/you-is-not-toxic}

\newcommand{\E}{\mathbb{E}}
\newcommand{\R}{\mathbb{R}}
\newcommand{\ind}{\mathbf{1}}
\newcommand{\ip}[2]{\langle #1, #2\rangle_p}
\DeclareMathOperator{\Var}{Var}
\DeclareMathOperator{\Cov}{Cov}
\newcommand{\ts}{\textsc{Threadsafe}}

\algrenewcommand\alglinenumber[1]{\footnotesize #1}

\begin{document}
\emergencystretch=1em
\raggedbottom
\title{\texorpdfstring{``You''}{\textquotedblleft You\textquotedblright} Is Not Toxic: How LLM Labelers Rewrite the Web's Conversation Laws}

\author{\AuthorOneName}
\affiliation{\institution{University of Colorado Boulder}\city{Boulder}\state{Colorado}\country{USA}}
\email{\AuthorOneEmail}
\author{Eric A. Vance}
\affiliation{\institution{University of Colorado Boulder}\city{Boulder}\state{Colorado}\country{USA}}
\email{\AuthorTwoEmail}

\begin{abstract}
What the Web community believes about toxic conversations, such as that toxicity rises in longer and deeper threads, was measured mostly with Google's Perspective API, which is shutting down; large language models (LLMs) will make the next measurements. We find that LLM labelers mistake second-person address for toxicity, and that this one bias manufactures laws that human labels do not show. On 1.25 million news comments with crowd labels, Claude Haiku flags 12\% of the non-toxic comments that contain no ``you'' and 51\% of those with four or more; a single crowd rater flags 6\% and 10\%. Deeper replies address someone more often, so under Haiku toxicity rises with reply depth (+0.93 percentage points per level), while under the crowd's labels it slightly declines ($-$0.12). Rewriting ``you'' in the third person removes two thirds of the false rise; flagging only as many comments as the crowd does not remove it, and showing Haiku the parent comment enlarges it. All seven LLMs we test over-flag ``you'', and at the crowd's flag rate, without the parent, any false rise of the open ones vanishes at a fixed number of second-person words; Haiku also over-flags ``you'' on Reddit, Gab and Wikipedia talk pages. Agreement with human labels does not reveal the bias: we prove that a fixed Cohen's $\kappa$ leaves a law nearly unconstrained. We therefore certify the law instead of the labeler. \ts{} uses the labeler's scores to choose which comments human raters label and to predict the others, and corrects the predictions with the human labels; linear laws are estimated without bias whatever the labeler does. In simulated audits its intervals for the laws keep nominal coverage from 250 labels on, where uniform-sample methods such as PPI++ fall short, and at 1{,}000 to 8{,}000 labels are 0.75 times as wide as those of design-based supervised learning (DSL); a thousand human labels rule out Haiku's rising law.
\end{abstract}


\maketitle
\hypersetup{pdftitle={\textquotedblleft You\textquotedblright\ Is Not Toxic: How LLM Labelers Rewrite the Web's Conversation Laws}}

\section{Introduction}

Three regularities organize what the Web community believes about toxic conversations. Discussions escalate: toxic conversations have deeper reply trees~\cite{saveski2021structure}, and civil ones derail into personal attacks~\cite{zhang2018cga}. Longer conversations are more toxic, on eight platforms and over three decades~\cite{avalle2024persistent}. Toxicity is contagious: a toxic comment draws toxic replies~\cite{saveski2021structure,cheng2017troll,kim2021distorting}. Most of these measurements used an automatic labeler, usually Google's Perspective API. Perspective stopped admitting new users in February 2026 and ends service at the end of 2026~\cite{perspective2026,hartmann2026perspective}, and the next measurements will use large language models (LLMs), which already label text at scale~\cite{gilardi2023chatgpt,ziems2024css}. We measure each regularity as a statistic of comment labels and conversation structure, a \emph{law}: the slope of toxicity on reply depth, the slope of a conversation's toxic share on its size (the law Avalle et al.\ report, which we measure over reply threads and over articles' comment sections), and the risk ratio along reply edges (Section~\ref{sec:laws}). Will LLM labelers measure the same laws?

\begin{figure}[t]
\centering
\includegraphics[width=\columnwidth]{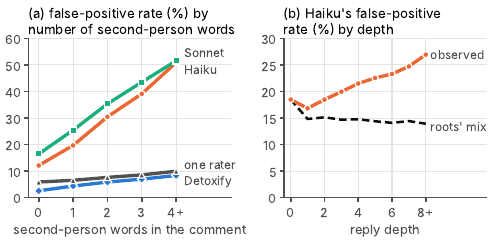}
\vspace{-6mm}
\caption{Second-person address drives the LLMs' false positives, among the Civil Comments posts that crowd raters call non-toxic. (a)~False-positive rate by the number of second-person words, for Haiku and Sonnet flagging what they call toxic (22\% and 27\% of comments; Sonnet on a 5\% sample), Detoxify flagging as many comments as the crowd calls toxic (7.6\%), and a single rater judged against the other raters' majority. (b)~Haiku's false-positive rate by reply depth, observed and with every depth given the roots' mix of second-person words (0.5 per root comment, 2.0 at depth 8 or more): at the roots' mix the rise disappears, and replies are flagged about four points less often than roots (13.9\% to 15.1\%, against 18.5\%).}
\Description{Two line charts. Left: against the number of second-person words from 0 to 4 or more, Haiku's false-positive rate rises from about 12 to 51 percent and Sonnet's from about 16 to 52 percent, while a single human rater's rises from about 6 to 10 percent and Detoxify's from about 3 to 8 percent. Right: against reply depth 0 to 8 or more, Haiku's observed false-positive rate rises from about 17 percent at depth 1 to 27 percent at depth 8, while the rate standardized to the roots' mix of second-person words stays near 14 to 15 percent after depth 0.}
\label{fig:you}
\end{figure}

They do not, and one bias is enough to change them: LLM labelers read second-person address as toxicity. Claude Haiku calls the reply ``I recommend you also read this awesome article'' toxic and ``I recommend they also read this awesome article'' not toxic, and its false positives climb with every ``you'' a comment contains (Figure~\ref{fig:you}a). Address is part of a conversation's structure. Replies address the comment they answer, and the deeper they sit, the more often they do: of the comments that raters call non-toxic in our news corpus, 23\% contain a second-person word at the root of a thread and 61\% six or more replies down. A labeler that over-flags address therefore over-flags deep comments, and it reports that conversations escalate (Figure~\ref{fig:you}b).

A law compares comments in different places of a conversation, so a labeler's errors can create or reverse it only where they differ between those places (errors that are the same everywhere only shrink it), and overall agreement with human labels, which pools all places, cannot show whether they do. We make this exact (Proposition~\ref{prop:decomp}). Under a labeler, every linear law (a weighted sum of the comments' labels, such as the depth law) equals the human-labeled law scaled by the labeler's true-positive rate minus its false-positive rate, plus two terms that measure how its false positives and its true positives follow the structure. Errors that ignore structure only shrink a law; to create or reverse one, errors must follow structure, as address does. A labeler that agrees better with human labels can therefore report the wrong law. On a 5\% sample of our corpus, when both flag as many comments as the crowd, Haiku asked for a 0--100 rating (Haiku-100) agrees with the crowd better than Gemma does ($\kappa$ = 0.31 against 0.26), yet under Haiku-100 toxicity rises with depth (+0.23 points per level) and under Gemma, as under the crowd, it falls ($-$0.22, against $-$0.16). \emph{Placebo statistics}, the law computed from the labeler's flags on the comments that humans call non-toxic, measure the false-positive term directly; like a placebo outcome, they show no effect when the false positives ignore structure.

We compare seven LLM labelers and 14 classifier settings with \emph{gold labels}, the majority vote of each comment's crowd raters, on 1.25 million news comments and five other corpora. An LLM flags either what it calls toxic (its \emph{own cut}) or as many comments as gold calls toxic (\emph{gold's flag rate}). Section~\ref{sec:setup} names the main settings, Table~\ref{tab:terms} defines the terms we use throughout, and Appendix~\ref{app:labelers} lists the settings with their agreement with gold and gives every prompt. We find:
\begin{itemize}[leftmargin=*,topsep=2pt,itemsep=1pt]
\item \textbf{LLMs manufacture escalation.} Under gold labels, toxicity falls slightly with reply depth ($-$0.12 percentage points per level); under Claude Haiku~5.5 at its own cut it rises (+0.93), and under Sonnet~5.5 too. Structured false positives make up 90\% of Haiku's \emph{false rise}, its distortion of the law, which persists at gold's flag rate and at the cut a study would choose by maximizing $\kappa$ on 500 gold labels. Haiku's thread size law, the rise of toxicity with thread length, is 12.7 times gold's.
\item \textbf{``You'' drives the false rise.} At a fixed number of second-person words Haiku's false rise disappears, and on the 5\% sample editing only the pronouns, rewriting ``you'' in the third person, removes two thirds of it, and all of it within threads. Context does not cure it: shown the parent comment, Haiku's false rise among replies grows from +1.29 to +1.59, and among the replies of the 5\% sample whose text or parent addresses someone, rewriting both comments removes only 25\% of it.
\item \textbf{Contagion depends on how gold is made.} Gold's reply risk ratio is 2.39 with all of a comment's raters and 1.63 with one. The Claude models report less at their own cut (Haiku 1.80), but only because they flag so much: the raters' votes, flagging as many comments as Haiku, give 1.60, less than Haiku's. Most classifier settings (12 of 14) report more contagion than gold, and those at gold's flag rate about as much as gold defined at a vote threshold of 0.6. Shown the parent, every LLM reports less contagion, while raters who see it report more (2.48 to 3.36).
\item \textbf{The bias is general and changes conclusions.} On the 5\% sample, at gold's flag rate, every LLM flags gold negatives with four or more second-person words 3.3 to 5.1 times as often as those with none, and the Claude models' 0--100 ratings distort escalation most (+0.39 and +0.49 points per level, against +0.18 to +0.26 for four of the five open models); without the parent, at a fixed number of second-person words the open models' false rises vanish. Haiku over-flags ``you'' on Reddit, Gab and Wikipedia talk pages even when asked each forum's own question, though not in chatbot dialogues. In samples of 10{,}000 to 100{,}000 comments, Haiku's verdict on escalation (a significant rise, fall or neither) differs from gold's in 70\%--100\% of samples.
\item \textbf{Agreement does not identify the law.} A labeler with $\kappa = 0.9$ at gold's flag rate can report any depth law from $-$1.30 to +1.65 points per level (Proposition~\ref{prop:bounds}). Higher agreement helps on average, but of the seven LLMs at gold's flag rate the two that agree best with gold, the Claude models, distort escalation most, and the textbook correction for validated error rates more than doubles Haiku's rising law, to +2.02 points per level.
\end{itemize}

We therefore certify the law instead of the labeler, in the spirit of risk-limiting audits, which certify an election outcome rather than the machines that counted it~\cite{stark2020shangrla}. \ts{} buys gold labels from human raters for a sample of comments drawn by a randomized design. The labeler's scores predict every comment's label and steer the sample to the comments that move the law most and are hardest to predict; the gold labels then correct the predictions. The output is the \emph{certificate}: an estimate of the human-labeled law with an approximate 95\% confidence interval. A labeler whose law falls outside the interval is ruled out, and the estimate, not the labeler's law, is what a study reports. Because the sampling probabilities are known, the estimate of every linear law is exactly unbiased for every labeler, even one that knows the gold labels and tries to mislead the audit (Theorem~\ref{thm:valid}).

Prediction-powered inference with power tuning (PPI++) and design-based supervised learning (DSL) are unbiased in the same way, but they sample uniformly, and the size laws put their largest weights on the few comments in small threads, which a small uniform sample rarely draws. In simulated audits on fully labeled corpora, \ts's intervals keep nominal coverage from 250 labels on, where PPI++'s cover in as few as 55\% of the audits for some labeler, corpus and statistic, and DSL's in 79\%. \ts{} learns where to sample in waves, from the labeler, cheap classifiers and the conversation's structure, and samples reply edges for contagion (Section~\ref{sec:method}). At 1{,}000 to 8{,}000 labels its intervals for the laws are 0.75 times as wide as DSL's, which uses the same predictions, and 0.81 times as wide as those of active inference, which samples by the labeler's uncertainty. A thousand gold labels rule out Haiku's rising law. We release the labels, prompts and code, with scripts that rebuild every gold-labeled population (\RepoURL).

\paragraph{Outline.} Section~\ref{sec:setup} describes the corpora, the labelers and the operating points. Section~\ref{sec:laws} defines the laws and proves what a labeler's errors do to them. Sections~\ref{sec:escalation}--\ref{sec:corpora} give the evidence: the false rise in escalation and its cause, the size laws and contagion, and the other corpora. Section~\ref{sec:consequences} shows what the distortions do to the conclusions a study would draw, and Section~\ref{sec:agreement} proves what agreement can and cannot certify and why error-rate corrections do not prevent the distortions. Sections~\ref{sec:method} and~\ref{sec:eval} present \ts{} and evaluate it on every labeler. Section~\ref{sec:related} reviews related work and Section~\ref{sec:limits} the limitations. The appendices give every prompt, sample and proof, and the results that the main text summarizes.
\section{Data, Labelers and Operating Points}
\label{sec:setup}

\paragraph{News comments.} Civil Comments~\cite{borkan2019nuanced} archives the comments posted under the articles of news sites that used the Civil Comments platform, with the votes of crowd raters, who saw each comment alone, without the comment it replies to~\cite{xenos2022toxicity}. A comment is toxic under \emph{gold} when at least half of its raters call it toxic or very toxic; the other comments are the \emph{gold negatives}. Most comments (66\%) have four raters, and some have many more. We rebuild the reply threads of the corpus's training file and keep the threads of at least two comments in which every comment has Haiku's and Qwen's labels, each without and with the parent shown: 1{,}251{,}487 comments in 270{,}967 reply threads under 57{,}769 articles, linked by 980{,}520 reply edges (Table~\ref{tab:pops}). The rule leaves out 7{,}717 of the 278{,}684 threads of at least two comments (2.8\%), in 7{,}716 of them because the root replies to a comment outside the training file, which the LLMs could not be shown as its parent. Gold calls 7.6\% of the comments toxic. CCC~\cite{xenos2022toxicity}, the subset of the corpus that Xenos et al.\ relabeled with raters who saw the parent, covers \ncccN{} of the replies.

\paragraph{Five more corpora.} Each has gold labels for its own construct (Table~\ref{tab:pops}; Appendix~\ref{app:data}). CAD~\cite{vidgen2021cad}, the Contextual Abuse Dataset, labels Reddit threads for abuse in the context of the thread. Qian et al.'s Reddit and Gab conversations~\cite{qian2019benchmark}, retrieved by hate keywords, are labeled for hate speech. CGA-Wiki~\cite{zhang2018cga}, Conversations Gone Awry, holds Wikipedia talk-page conversations labeled for personal attacks; half of them end at an attack. BAD~\cite{xu2021bad}, the Bot-Adversarial Dialogue, labels the turns of human--chatbot dialogues as safe or unsafe.

\begin{table*}[t]
\caption{The seven populations: source, the construct of their gold labels and their unit, with the number of comments, units and reply edges, and gold's share of positives. In CAD a submission's title and body are both roots and a few comments have no parent in the release, and in CGA-Wiki comments that reply to a section header or to themselves have none (1{,}401 and 1{,}220 comments beyond one root per unit), so their reply edges fall short of the comments minus the units.}
\label{tab:pops}
\footnotesize
\begin{tabular}{@{}llllrrrr@{}}
\toprule
Population & Source & Gold construct & Unit & Comments & Units & Reply edges & Gold positives \\
\midrule
Civil Comments reply threads & news comments & toxicity & reply thread & 1{,}251{,}487 & 270{,}967 & 980{,}520 & 7.6\% \\
\quad 5\% sample &  &  & reply thread & 60{,}668 & 13{,}467 & 47{,}201 & 7.5\% \\
CAD & Reddit & abuse & thread & 26{,}550 & 1{,}395 & 23{,}754 & 16.2\% \\
Qian et al.'s Reddit & Reddit & hate speech & conversation & 22{,}309 & 5{,}020 & 17{,}289 & 23.6\% \\
Qian et al.'s Gab & Gab & hate speech & conversation & 33{,}776 & 11{,}825 & 21{,}951 & 43.3\% \\
CGA-Wiki & Wikipedia talk pages & personal attack & conversation & 5{,}340 & 1{,}168 & 2{,}952 & 10.9\% \\
BAD & human--chatbot dialogue & unsafe turn & dialogue & 78{,}874 & 5{,}784 & 73{,}090 & 39.2\% \\
\bottomrule
\end{tabular}
\end{table*}

\paragraph{Samples.} The labelers that cost most label a \emph{5\% sample} of the articles: 60{,}668 comments in 13{,}467 threads under 2{,}896 articles. It leaves out the 198 threads of the sampled articles in which Sonnet answered some comment without an option letter (9\% of their comments are toxic, against 7.5\% in the sample). Three more samples serve the experiments that edit comments: the 5\% sample with every second-person comment rewritten; the \emph{edit sample} of 4{,}000 second-person comments from the whole corpus (1{,}500 clearly non-toxic ones at depth 2 or more, 1{,}500 clearly non-toxic roots and 1{,}000 gold positives, where a comment is \emph{clearly non-toxic} when fewer than 20\% of its raters call it toxic); and a \emph{reverse sample} of comments that refer to someone as \emph{they} and contain no ``you''. Appendix~\ref{app:data} defines every sample.

\paragraph{LLM labelers.} Claude Haiku~5.5 and Sonnet~5.5 answer the Civil Comments raters' four-level question, very toxic, toxic, hard to say or not toxic, with the raters' definitions (the \emph{main prompt}), and flag the comments they call toxic or very toxic: their \emph{own cut}. To compare them at gold's flag rate, both also rate each comment of the 5\% sample from 0 to 100 under the definition of toxicity of Perspective, the instrument the LLMs replace, and we flag the highest-rated comments, as many as gold calls toxic (\emph{Haiku-100} and \emph{Sonnet-100}). Five open LLMs, Qwen2.5-7B-Instruct~\cite{qwen2024}, Gemma~3~27B~\cite{gemma2025}, OLMo-2-1124-7B-Instruct~\cite{olmo2024}, Granite-3.1-8B-Instruct~\cite{granite2024} and Llama-3.1-8B-Instruct~\cite{llama2024}, answer the main prompt; we read the probability of the two toxic answers and flag at gold's flag rate, and Qwen also at its own cut, where that probability exceeds one half (57.0\% of comments flagged). Qwen labels all reply threads and the other open LLMs the 5\% sample. Every LLM also labels each reply with the replied-to comment shown (\emph{+ parent}). Roots keep their label without one, so a setting ``+ parent'' is evaluated on replies only: its depth law among replies and its contagion over edges between two replies. Appendix~\ref{app:labelers} gives every prompt and setting.

\paragraph{Classifiers.} Seven classifiers, each at its usual threshold and at gold's flag rate, give 14 settings. Detoxify~\cite{hanu2020detoxify}, trained on data from Jigsaw, Perspective's maker, stands in for Perspective, which admits no new users; Avalle et al.\ checked Perspective with Detoxify and with IMSYPP~\cite{kralj2022imsypp}. The other five are Detoxify-unbiased, the s-nlp RoBERTa toxicity classifier, the martin-ha DistilBERT toxicity classifier, the Dynabench hate-speech model and CardiffNLP's offensive-language model. Detoxify-unbiased, s-nlp RoBERTa and martin-ha DistilBERT are trained at least in part on Civil Comments labels ($^{\S}$ in tables).

\paragraph{Human benchmarks.} Raters disagree, so we read every labeler against human labelers that make errors too. The \emph{single rater} is one of a comment's raters drawn at random, judged against the majority of the others (\emph{gold B}). The \emph{vote-fraction instrument} flags the comments with the largest shares of toxic votes, as many as a given labeler flags: it is how the raters would label the corpus at that labeler's flag rate. CCC's raters saw the parent. Appendix~\ref{app:humans} adds the \emph{noise-free majority}, the majority of infinitely many raters, estimated from each comment's votes with a model of the raters.

\begin{table}[t]
\caption{Terms used throughout.}
\label{tab:terms}
\footnotesize
\begin{tabular}{@{}>{\raggedright\arraybackslash}p{2.4cm}p{5.6cm}@{}}
\toprule
Term & Meaning \\
\midrule
Gold; gold negative & A comment is toxic when at least half of its raters call it toxic or very toxic; the gold negatives are the others. \\
Law & A statistic of the gold labels and the conversation's structure: the depth law, the size laws, contagion (Section~\ref{sec:laws}). \\
Distortion & A labeler's law minus gold's on the same comments. \\
Placebo statistic & A law computed from the labeler's flags on the gold negatives alone. \\
Own cut & An LLM flags the comments it calls toxic or very toxic (Qwen: those whose toxic answers have probability above one half). \\
Gold's flag rate & The labeler flags as many comments as gold calls toxic (7.6\% on news comments). \\
Haiku-100, Sonnet-100 & The model's 0--100 rating, flagging at gold's flag rate unless another rate is given. \\
+ parent & Each reply labeled with the comment it replies to shown; measured among replies. \\
5\% sample, $^{\dagger}$ & The reply threads of a 5\% sample of Civil Comments articles. On Civil Comments, Sonnet, Haiku-100, Sonnet-100 and the open LLMs other than Qwen label only this sample and are always reported on it; $^{\dagger}$ marks the other numbers on it. \\
False rise & A positive distortion of the depth law where gold's law falls or is flat. \\
Single rater (One rater in tables); gold B & One of a comment's raters drawn at random; gold B, the majority of its other raters, is its reference. \\
Noise-free majority & The majority of infinitely many raters, estimated from each comment's votes (Appendix~\ref{app:humans}). \\
Vote-fraction instrument & Flags the comments with the largest shares of toxic votes, as many as a given labeler flags. \\
Toxicity question & The main prompt, asked on the other corpora, whose gold labels measure other constructs (Section~\ref{sec:corpora}). \\
Certificate & \ts's estimate of the human-labeled law with its approximate 95\% interval (Section~\ref{sec:method}). \\
Cell & One labeler on one population, the unit of the simulated audits (Section~\ref{sec:eval}). \\
\bottomrule
\end{tabular}
\end{table}

\paragraph{Names and what we report.} Table~\ref{tab:settings} in Appendix~\ref{app:labelers} lists the main settings with their flag rates, $\kappa$ and error rates, and Table~\ref{tab:terms} defines the terms. A bare name always means the same one: \emph{Haiku} is Haiku at its own cut on all reply threads and \emph{Sonnet} is Sonnet at its own cut on the 5\% sample. Relabeling the 5\% sample in a later batch changes 2.4\% of Haiku's flags ($\kappa$ between the runs 0.93) and moves its depth law by +0.02 points per level; we quote Haiku's distortion of the depth law as +1.05 on all reply threads and +1.19 on the 5\% sample, and an experiment that relabels comments in its own batch is compared with its own relabeled originals. For each law of Section~\ref{sec:laws}, a labeler's \emph{distortion} is its law minus gold's on the same comments, in percentage points per reply level for the depth law, per doubling for the size laws and in log units for the reply risk ratio. Intervals are 95\% bootstrap intervals over articles on Civil Comments and over threads, conversations or dialogues elsewhere, and a distortion is significant when its interval excludes 0.
\section{Laws and What a Labeler's Errors Do to Them}
\label{sec:laws}

A corpus is a set of comments $i = 1, \dots, N$, grouped into units (reply threads, conversations or the comment sections of articles) and linked by reply edges $(p, c)$ from a parent to its reply. A comment's reply depth is 0 at the root of its thread and grows by one with each reply. Each comment has a gold label $Z_i \in \{0,1\}$ from its human raters, which makes it a \emph{gold positive} or a \emph{gold negative}, and a \emph{labeler} (an LLM, a classifier or a human) gives it a flag $Y_i$ and usually a score. A \emph{law} is a statistic of the gold labels and the structure; the labeler's version is the same statistic of its flags, and its \emph{distortion} is the difference, $L(Y) - L(Z)$. We measure the three laws as follows:
\begin{itemize}[leftmargin=*,topsep=2pt,itemsep=1pt]
\item \emph{escalation}, measured by the \emph{depth law}: the least-squares slope of $Z$ on reply depth (capped at 10), in percentage points per level, over all comments, among replies, or \emph{within threads} (with thread fixed effects, so that only the comparison of comments of one thread counts);
\item the \emph{size laws}: the least-squares slope of a unit's toxic share on $\log_2$ of its size, in points per doubling, over units with at least two comments: reply threads (the \emph{thread size law}) and, on news comments, also the comment sections of articles (the \emph{article size law}, the level at which Avalle et al.\ report it~\cite{avalle2024persistent});
\item \emph{contagion}: the reply risk ratio $\mathrm{RR}(Z) = P(Z_c = 1 \mid Z_p = 1)/P(Z_c = 1 \mid Z_p = 0)$ over reply edges.
\end{itemize}
The depth and size laws are \emph{linear}: $L(v) = \sum_i w_i v_i$, with known weights that sum to zero because each law is a contrast between comments. For the depth law, $w_i = (x_i - \bar x)/\mathrm{SS}$, with $x_i$ the capped depth and $\mathrm{SS} = \sum_i (x_i - \bar x)^2$: comments far from the mean depth weigh most. For a size law, a comment's weight is its unit's weight, the unit's centered $\log_2$ size over the units' sum of squares, divided by the unit's size, so comments of small units weigh most.

\subsection{Which part of a law does a labeler produce?}

Let TPR and FPR be the shares of the corpus's gold positives and gold negatives that the labeler flags.

\begin{proposition}[Exact decomposition]\label{prop:decomp}
For every linear law,
\begin{align}
L(Y) - L(Z) = {}& (\mathrm{TPR} - \mathrm{FPR} - 1)\, L(Z) && \text{(unstructured errors)}\nonumber\\
&+ \textstyle\sum_{Z_i = 0} w_i (Y_i - \mathrm{FPR}) && (S_{\mathrm{FP}}\text{: false positives})\nonumber\\
&+ \textstyle\sum_{Z_i = 1} w_i (Y_i - \mathrm{TPR}) && (S_{\mathrm{TP}}\text{: true positives}).\label{eq:decomp}
\end{align}
For escalation, $S_{\mathrm{FP}} = (\mathrm{SS}_0/\mathrm{SS})\,\beta_0$, with $\beta_0$ the slope of $Y$ on depth among gold negatives and $\mathrm{SS}_0/\mathrm{SS}$ their share of the variation in depth.
\end{proposition}

The three terms answer three different questions. Errors \emph{ignore structure} when, within each gold class, the flags are exchangeable: given its gold label, a comment's chance of being flagged does not depend on its place in the conversation. Each flag then has expectation FPR or TPR, the corpus-wide rates against which the split is taken, so $S_{\mathrm{FP}}$ and $S_{\mathrm{TP}}$ have expectation zero. The first term is the classical misclassification bias~\cite{bross1954misclassification}: errors that ignore structure shrink a law by the factor $\mathrm{TPR} - \mathrm{FPR}$ and, when $\mathrm{TPR} > \mathrm{FPR}$, never change its sign. Shrinking a law toward zero is a distortion of the opposite sign: gold's depth law on news comments falls ($-$0.12 points per level), so unstructured errors alone give a small \emph{positive} distortion, the gray bars of Figure~\ref{fig:decomp}. $S_{\mathrm{FP}}$ and $S_{\mathrm{TP}}$ measure how the false and the true positives follow the structure: they are differential misclassification~\cite{gustafson2003measurement} with respect to structure, and only they can manufacture or reverse a law. A small example shows how little it takes. If gold toxicity is 10\% at every depth (depths 0--6 equally common) and a labeler finds 80\% of toxic comments but flags 5\% of non-toxic ones at depth 0 and one point more per level, its $\kappa$ is 0.58 and it reports a rise of 0.9 points per level, all of it $S_{\mathrm{FP}}$.

Contagion is not linear, so we split its distortion by permutation. Shuffling the flags within gold positives and within gold negatives keeps TPR and FPR but removes all structure, which gives the unstructured term; restoring the actual false positives and the actual true positives, averaged over both orders (Shapley values, Appendix~\ref{app:proofs}), gives the other two. For a linear law the same construction returns the three terms of Proposition~\ref{prop:decomp}.

\paragraph{Placebo statistics.} The false-positive term involves only the gold negatives. A \emph{placebo statistic} is a law computed from the labeler's flags among them: the \emph{placebo depth slope} $\beta_0$, in points per level; the \emph{placebo size slope}, the slope of a thread's share of flagged gold negatives on $\log_2$ of its size; and the \emph{placebo risk ratio}, over reply edges whose two ends are gold negatives. A labeler whose false positives ignore structure has placebo slopes of 0 and a placebo risk ratio of 1. Comments that most raters call non-toxic but some call toxic can have structure of their own, so we read each placebo statistic against a human benchmark, the single rater of Section~\ref{sec:setup}: how much structure do one human's false positives have?
\section{LLMs Manufacture Escalation}
\label{sec:escalation}

Under gold labels, toxicity falls slightly with reply depth on news comments: $-$0.12 points per level (CI $-$0.16 to $-$0.09; Table~\ref{tab:placebo}). Under Haiku it rises (+0.93), a distortion of +1.05, and under Sonnet too (+1.24, a distortion of +1.40 against gold's $-$0.16 on the 5\% sample). From the root to the fifth reply level, Haiku's law adds 4.7 percentage points of toxicity, in a corpus where 7.6\% of comments are toxic; the raters' law takes away 0.6. The false rise holds within threads, where only comments of the same thread are compared (Haiku's distortion is +0.68), and among replies (+1.29).

\begin{figure*}[t]
\centering
\includegraphics[width=\textwidth]{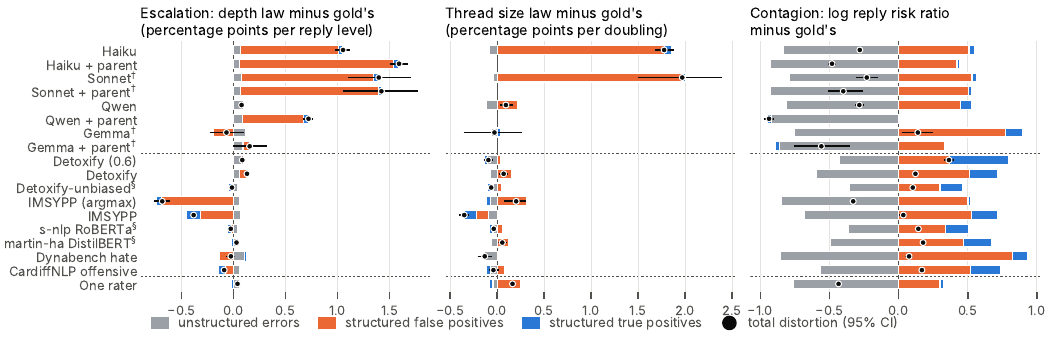}
\vspace{-6mm}
\caption{Which part of each distortion the labeler produces, on Civil Comments reply threads: each bar splits a labeler's law minus gold's into the three terms of Proposition~\ref{prop:decomp} (for contagion, unstructured errors are the permutation baseline); dots mark the total with its 95\% interval. Gold's depth law falls, so unstructured errors, which shrink it toward zero, give a small positive distortion. Structured false positives make up 81\%--94\% of the escalation distortions of the Claude models, with and without the parent, and of Qwen shown the parent. Rows ``+ parent'' label replies with their parent, so their escalation is measured among replies (Table~\ref{tab:esc} in Appendix~\ref{app:news}), their contagion over edges between replies, and their size law is not shown. Classifiers flag at gold's flag rate unless a threshold is given. $^{\dagger}$5\% sample. $^{\S}$Trained at least in part on Civil Comments labels. One rater: against the other raters' majority.}
\Description{Three panels of horizontal stacked bars, one row per labeler. Escalation and thread size: the Claude rows and Qwen shown the parent have long orange bars (structured false positives) to the right; most classifier rows are near zero, and IMSYPP's escalation bars point left. Contagion: most rows have a long gray bar to the left (unstructured errors), and every classifier row except IMSYPP at argmax has orange and blue bars to the right that are as long or longer.}
\label{fig:decomp}
\end{figure*}

Proposition~\ref{prop:decomp} says which errors produce it (Figure~\ref{fig:decomp}). Structured false positives make up 90\% of Haiku's distortion and 91\% of Sonnet's: the comments that the raters call non-toxic are flagged more often the deeper they sit. Haiku's placebo depth slope is +1.02 points per level, +5.4\% of its false-positive rate per level; a single rater's is +0.01 (+0.1\% of its rate). The text classifiers show no comparable slope (at most +0.07, Detoxify at gold's flag rate), and under them escalation stays flat or falls. IMSYPP at its argmax, one of the two checks Avalle et al.\ ran, reports an exaggerated fall ($-$0.81 points per level, against gold's $-$0.12): its false positives, a third of the gold negatives, fall with depth (placebo slope $-$0.74). Structured errors can push a law either way.

\begin{table*}[t]
\caption{Placebo statistics and laws on Civil Comments reply threads, with 95\% intervals. Placebo statistics use the labeler's flags among gold negatives: the placebo depth slope, in points per level and as a share of the labeler's false-positive rate per level, and the placebo risk ratio over edges with two gold-negative ends. Depth laws are in points per level, thread size laws in points per doubling. Rows flag at gold's flag rate except at an own cut (Qwen's flags the comments to which it gives the two toxic answers a probability above one half) and at the thresholds given. One rater: a single rater against the other raters' majority, the median of 20 draws with the range over draws in brackets where computed (under that majority the laws are $-$0.12, +0.10 and 2.52). $^{\dagger}$5\% sample (gold: $-$0.16 and 2.28; Sonnet's size laws in Table~\ref{tab:size}, Gemma's size-law distortions in Appendix~\ref{app:news}, Tables~\ref{tab:decompa} and~\ref{tab:decompb}). The five other classifiers, at both cuts, are in Appendix~\ref{app:news} (Table~\ref{tab:classifiers}).}
\label{tab:placebo}
\footnotesize
\setlength{\tabcolsep}{4pt}
\begin{tabular}{@{}llcccccc@{}}
\toprule
 & & \multicolumn{3}{c}{Placebo statistics (gold negatives)} & \multicolumn{3}{c}{The law under the labeler} \\
\cmidrule(lr){3-5}\cmidrule(l){6-8}
Labeler & Cut & Depth slope & \% of FPR per level & Risk ratio & Depth law & Thread size law & Risk ratio \\
\midrule
Gold &  & -- & -- & -- & $-$0.12 [$-$0.16, $-$0.09] & +0.15 [+0.09, +0.21] & 2.39 [2.34, 2.44] \\
One rater &  & +0.01 & +0.1 [$-$0.3, +0.4] & 1.43 [1.39, 1.49] & $-$0.08 [$-$0.10, $-$0.06] & +0.26 [+0.22, +0.30] & 1.63 [1.60, 1.66] \\
\midrule
Haiku & own cut & +1.02 [+0.93, +1.09] & +5.4 [+5.0, +5.8] & 1.83 [1.81, 1.85] & +0.93 [+0.84, +1.01] & +1.93 [+1.82, +2.03] & 1.80 [1.79, 1.82] \\
Sonnet$^{\dagger}$ & own cut & +1.37 [+1.02, +1.71] & +5.9 [+4.5, +7.3] & 1.83 [1.75, 1.92] & +1.24 [+0.84, +1.63] & -- & 1.81 [1.74, 1.90] \\
Qwen & gold's rate & $-$0.01 [$-$0.04, +0.02] & $-$0.1 [$-$0.6, +0.4] & 1.75 [1.70, 1.80] & $-$0.04 [$-$0.08, $-$0.01] & +0.24 [+0.19, +0.30] & 1.80 [1.76, 1.84] \\
 & own cut ($p > 0.5$) & $-$0.60 [$-$0.69, $-$0.52] & $-$1.1 [$-$1.3, $-$1.0] & 1.34 [1.33, 1.35] & $-$0.63 [$-$0.72, $-$0.55] & +1.28 [+1.16, +1.40] & 1.35 [1.34, 1.36] \\
Gemma$^{\dagger}$ & gold's rate & $-$0.21 [$-$0.37, $-$0.04] & $-$3.8 [$-$6.8, $-$0.7] & 2.80 [2.45, 3.19] & $-$0.22 [$-$0.43, 0.00] & -- & 2.62 [2.40, 2.86] \\
\midrule
Detoxify & 0.6 & +0.02 [+0.01, +0.03] & +1.4 [+0.5, +2.4] & 4.01 [3.67, 4.49] & $-$0.04 [$-$0.06, $-$0.01] & +0.06 [+0.02, +0.10] & 3.44 [3.32, 3.57] \\
 & gold's rate & +0.07 [+0.05, +0.09] & +1.9 [+1.3, +2.5] & 2.91 [2.80, 3.03] & +0.01 [$-$0.03, +0.05] & +0.22 [+0.16, +0.28] & 2.70 [2.64, 2.75] \\
\addlinespace[1pt]
IMSYPP & argmax & $-$0.74 [$-$0.83, $-$0.67] & $-$2.2 [$-$2.5, $-$2.0] & 1.75 [1.73, 1.76] & $-$0.81 [$-$0.90, $-$0.72] & +0.35 [+0.23, +0.48] & 1.72 [1.71, 1.73] \\
 & gold's rate & $-$0.34 [$-$0.37, $-$0.32] & $-$7.6 [$-$8.2, $-$7.2] & 2.46 [2.38, 2.55] & $-$0.50 [$-$0.54, $-$0.47] & $-$0.20 [$-$0.26, $-$0.15] & 2.47 [2.42, 2.53] \\
\bottomrule
\end{tabular}
\end{table*}

\subsection{The word \texorpdfstring{``you''}{\textquotedblleft you\textquotedblright}}
\label{sec:you}

The LLMs' false positives climb with every second-person word a comment contains (\emph{you}, \emph{your}, \emph{yours}, \emph{yourself}, \emph{yourselves} and their contractions, generic uses of ``you'' included; Figure~\ref{fig:you}a). Haiku flags 12.1\% of the gold negatives without one and 50.9\% of those with four or more; Sonnet flags 16.5\% and 51.5\%. A single rater's false-positive rate rises far less, from 5.9\% to 10.0\%, and so does Detoxify's (2.6\% to 8.4\%). The climb holds within each of five bands of comment length, so it is not that longer comments both say ``you'' more often and offend more often. It also holds among the gold negatives on which no rater voted toxic or marked a toxicity subtype such as an insult (Section~\ref{sec:construct}).

Address grows with depth. Of the gold negatives, 23\% contain a second-person word at the root of a thread and 61\% at depth 6 or more, and the mean number of such words grows from 0.49 at the root to 1.96 at depth 8 or more. Holding that number fixed removes the rise in Haiku's false positives. Given the roots' mix of second-person words, its false-positive rate no longer rises with depth (Figure~\ref{fig:you}b). Within strata of the number of second-person words, its placebo depth slope is $-$0.16 points per level (CI $-$0.22 to $-$0.10), against +1.02 overall: at a fixed number, deeper comments are flagged slightly less often. Strata that record only whether a comment has a second-person word leave +0.20. Most of Sonnet's slope goes too (+1.37 overall, +0.21 at a fixed number). At a fixed number, Haiku's distortion of the depth law is $-$0.05 (CI $-$0.11 to +0.01).

Holding the number fixed also holds fixed whatever comes with addressing someone, such as a question or a disagreement. To learn whether the words themselves cause the flags, we change them.

\subsection{Rewriting \texorpdfstring{``you''}{\textquotedblleft you\textquotedblright} turns the false positives off and on}
\label{sec:rewrite}

We edit comments by a fixed rule that replaces second-person pronouns with third-person ones (\emph{you} $\to$ \emph{they} or \emph{them}, \emph{your} $\to$ \emph{their}, \dots; Appendix~\ref{app:data:rules}) and label each edit together with its original in one batch, in shuffled order, so the originals are relabeled and their distortions differ by a few hundredths from the main labels'. Table~\ref{tab:rewrite} gives every result, and Table~\ref{tab:pairsthree} shows the three shortest comments that no rater called toxic and that the rewrite turns from toxic to not toxic for Haiku (Appendix~\ref{app:news}, Table~\ref{tab:pairs}, lists twelve).

\paragraph{The whole sample.} On the 5\% sample, rewriting every second-person comment (24{,}116 comments) removes 67\% of Haiku's distortion of the depth law, from +1.23 to +0.40; within threads it removes all of it, from +0.75 to 0.00. It also removes 54\% of Haiku's inflation of the thread size law. For Sonnet the three shares are 73\%, 83\% and 70\%. The effect grows with the dose, up to three words: rewriting one, two, three, or four or more second-person words changes Haiku's flag rate on gold negatives by $-$4.8, $-$8.7, $-$13.0 and $-$12.0 points, and Sonnet's by $-$7.1 to $-$15.2, the causal counterpart of Figure~\ref{fig:you}a.

\paragraph{Single comments.} On the edit sample, the rule lowers Haiku's flag rate on clearly non-toxic replies at depth 2 or more from 28.4\% to 19.5\% and on roots only from 22.7\% to 19.9\%, so it closes the gap between deep replies and roots, from +5.7 points to $-$0.4. Sonnet's paraphrases, which keep a comment's meaning, tone and politeness and either keep addressing the person as ``you'' or refer to them in the third person, give the same result: Haiku flags 26.7\% of the paraphrases that keep ``you'' and 18.1\% of those in the third person, and the depth gap falls from +6.4 to +1.2. Sonnet as the labeler responds more strongly still: the rule lowers its flag rate on deep replies from 33.6\% to 21.9\%, and the paraphrases from 34.4\% to 22.2\%.

\begin{table}[t]
\caption{The three shortest comments of the edit sample that no rater called toxic and that Haiku calls toxic as written and not toxic after the fixed rewrite (twelve in Appendix~\ref{app:news}, Table~\ref{tab:pairs}).}
\label{tab:pairsthree}
\scriptsize
\begin{tabular}{@{}p{0.47\columnwidth}p{0.47\columnwidth}@{}}
\toprule
As written (flagged) & Rewritten (``not toxic'') \\
\midrule
You're lecturing a fence post. & They're lecturing a fence post. \\
Dude, are you stalking him? Kinda creepy. & Dude, are they stalking him? Kinda creepy. \\
Clearly, you ate lead paint chips as a child. & Clearly, they ate lead paint chips as a child. \\
\bottomrule
\end{tabular}
\end{table}

\paragraph{The other direction.} Rewriting \emph{they} and its forms as \emph{you} in clearly non-toxic comments that contain no ``you'' raises Haiku's flag rate at depth 2 or more from 9.8\% to 14.6\%, and at the root by as much (+5.1 points): a new ``you'' counts the same at every depth, and the gap between depths does not change ($-$0.3 points). The ``you''s that comments already contain count more deep in a thread: removing them lowers Haiku's flags by 8.9 points at depth 2 or more but by 2.9 at the root. That is not because deep comments contain more of them. The edit sample's deep replies and roots have about as many (2.25 and 2.10 on average), and given the roots' mix of counts, removal still lowers the deep replies' flags by 8.7 points (CI 7.1 to 10.1; Sonnet: 11.3 against 3.8). Deep in a thread, ``you'' usually addresses the person replied to, and that is what the LLMs flag; depth itself adds nothing.

\paragraph{The response makes the distortion.} How strongly a labeler responds to the pronoun predicts how much its false positives follow depth: across Haiku, Sonnet and the 14 classifier settings, the rise in the flag rate when ``you'' is added has a rank correlation of +0.74 with the placebo depth slope relative to the false-positive rate, and the relative rise one of +0.90 (Appendix~\ref{app:news}); the two Claude models top the ranking by the rise in points and by the relative placebo slope. Adding ``you'' raises Haiku's flag rate by +49\% and Sonnet's by +45\%; the classifiers' flag rates move by $-$0.6 to +2.7 points, which for classifiers that flag almost none of these comments can be a large relative change.

The rewrite also lowers the flags on toxic comments: on the edit sample's second-person gold positives, the rule lowers Haiku's flag rate from 74.0\% to 65.1\%, so ``you'' is part of what it reads in toxic comments too. Second-person address itself thus causes most of the false rise: read alone, a comment is toxic to the LLMs partly for whom it addresses, not only for what it says.

\begin{table*}[t]
\caption{Rewriting ``you'' with a fixed rule, and with Sonnet's paraphrases, changes what the LLMs flag; each edit is labeled with its original in one batch, in shuffled order, so the originals' values (``as written'') differ slightly from the main labels' (Table~\ref{tab:esc} in Appendix~\ref{app:news}). Distortions are the labeler's law minus gold's (points per level; per doubling for the thread size law); flag rates in percent; changes in points with paired 95\% intervals. Top: all 24{,}116 second-person comments of the 5\% sample rewritten, the other comments keeping their labels; the flag rates are over all gold positives and all gold negatives of the sample. Edit sample: clearly non-toxic second-person replies at depth 2 or more (fewer than 20\% of raters voting toxic), roots and gold positives. Reverse rewrite: comments with \emph{they} and no ``you''. Dose: gold negatives of the 5\% sample with that many second-person words. Shown the parent: the replies of the 5\% sample whose text or parent addresses someone, relabeled with the reply rewritten and with both rewritten; the distortion is measured among these replies.}
\label{tab:rewrite}
\footnotesize
\begin{tabular}{@{}p{6.0cm}ll@{}}
\toprule
 & Haiku & Sonnet \\
\midrule
\multicolumn{3}{@{}l}{\emph{Every second-person comment of the 5\% sample rewritten; distortion as written $\to$ rewritten (change [CI]; share removed)}} \\
Depth law & +1.23 $\to$ +0.40 ($-$0.82 [$-$0.95, $-$0.70]; 67\%) & +1.43 $\to$ +0.38 ($-$1.05 [$-$1.16, $-$0.92]; 73\%) \\
Within threads & +0.75 $\to$ 0.00 ($-$0.75 [$-$0.90, $-$0.61]; 101\%) & +1.07 $\to$ +0.19 ($-$0.88 [$-$1.01, $-$0.74]; 83\%) \\
Thread size law & +1.98 $\to$ +0.92 ($-$1.07 [$-$1.24, $-$0.89]; 54\%) & +2.05 $\to$ +0.62 ($-$1.43 [$-$1.61, $-$1.28]; 70\%) \\
Gold positives flagged & 65.0\% $\to$ 60.7\% & 74.6\% $\to$ 71.0\% \\
Gold negatives flagged & 18.8\% $\to$ 15.8\% & 23.2\% $\to$ 19.2\% \\
\midrule
\multicolumn{3}{@{}l}{\emph{Edit sample; flag rate as written $\to$ edited (change, points [CI])}} \\
Rule rewrite, replies at depth $\ge 2$ & 28.4\% $\to$ 19.5\% ($-$8.9 [$-$10.5, $-$7.5]) & 33.6\% $\to$ 21.9\% ($-$11.8 [$-$13.6, $-$10.0]) \\
\quad roots & 22.7\% $\to$ 19.9\% ($-$2.9 [$-$4.4, $-$1.3]) & 21.9\% $\to$ 18.1\% ($-$3.8 [$-$5.4, $-$2.1]) \\
\quad gold positives & 74.0\% $\to$ 65.1\% ($-$8.9 [$-$11.2, $-$6.7]) & 78.8\% $\to$ 70.2\% ($-$8.6 [$-$10.8, $-$6.5]) \\
Paraphrase with ``you'' $\to$ in the third person & 26.7\% $\to$ 18.1\% ($-$8.7 [$-$10.4, $-$7.0]) & 34.4\% $\to$ 22.2\% ($-$12.2 [$-$14.3, $-$10.3]) \\
\quad gold positives & 70.2\% $\to$ 60.2\% ($-$10.0 [$-$12.6, $-$7.5]) & 76.4\% $\to$ 67.8\% ($-$8.6 [$-$11.0, $-$6.1]) \\
Depth gap (depth $\ge 2$ minus roots), rule rewrite & +5.7 [+2.4, +9.1] $\to$ $-$0.4 [$-$3.5, +2.6] & +11.7 [+8.4, +15.1] $\to$ +3.8 [+0.7, +6.8] \\
\quad change of the gap & $-$6.1 [$-$8.4, $-$3.9] & $-$8.0 [$-$10.3, $-$5.6] \\
Depth gap, paraphrase with ``you'' $\to$ third person & +6.4 [+3.3, +9.7] $\to$ +1.2 [$-$1.9, +4.2] & +12.3 [+8.9, +15.7] $\to$ +4.3 [+1.2, +7.3] \\
\quad change of the gap & $-$5.2 [$-$7.7, $-$2.6] & $-$8.0 [$-$10.7, $-$5.2] \\
\midrule
\multicolumn{3}{@{}l}{\emph{Reverse rewrite of clearly non-toxic comments without ``you'': third person $\to$ ``you''}} \\
Replies at depth $\ge 2$ & 9.8\% $\to$ 14.6\% (+4.8 [+3.7, +6.1]) & 13.8\% $\to$ 20.1\% (+6.3 [+4.7, +7.8]) \\
\quad roots & 17.0\% $\to$ 22.1\% (+5.1 [+3.7, +6.6]) & 17.6\% $\to$ 23.6\% (+6.0 [+4.5, +7.4]) \\
\quad gold positives & 63.1\% $\to$ 68.6\% (+5.5 [+3.7, +7.5]) & 70.8\% $\to$ 75.1\% (+4.3 [+2.6, +6.1]) \\
Depth gap (depth $\ge 2$ minus roots) & $-$7.2 $\to$ $-$7.5 ($-$0.3 [$-$2.2, +1.6]) & $-$3.8 $\to$ $-$3.5 (+0.3 [$-$1.7, +2.4]) \\
\midrule
\multicolumn{3}{@{}l}{\emph{Rewrite dose: change in the flag rate of gold negatives (points [CI]), by second-person words rewritten}} \\
1 (n = 11{,}160) & $-$4.8 [$-$5.3, $-$4.3] & $-$7.1 [$-$7.7, $-$6.4] \\
2 (n = 4{,}996) & $-$8.7 [$-$9.6, $-$7.9] & $-$12.0 [$-$13.1, $-$11.0] \\
3 (n = 2{,}500) & $-$13.0 [$-$14.4, $-$11.5] & $-$14.2 [$-$15.7, $-$12.7] \\
4 or more (n = 3{,}296) & $-$12.0 [$-$13.3, $-$10.8] & $-$15.2 [$-$16.6, $-$13.8] \\
\midrule
\multicolumn{3}{@{}l}{\emph{Shown the parent, among replies: distortion as written $\to$ reply rewritten $\to$ reply and parent rewritten}} \\
Depth law among replies & +1.66 $\to$ +1.41 $\to$ +1.25 & +1.38 $\to$ +0.82 $\to$ +0.72 \\
\quad change, both rewritten (share removed) & $-$0.41 [$-$0.58, $-$0.27] (25\%) & $-$0.66 [$-$0.80, $-$0.51] (48\%) \\
Reply and parent address ``you'': FPR at depth 1 $\to$ 5+, as written & 36\% $\to$ 50\% & 32\% $\to$ 46\% \\
\quad both rewritten & 30\% $\to$ 41\% & 24\% $\to$ 33\% \\
\bottomrule
\end{tabular}
\end{table*}

\subsection{Shown the parent}
\label{sec:parent}

The raters labeled each comment without the comment it replies to, so an LLM shown the parent might read address as what it is, a reply. It does not. Among replies, Haiku's distortion is +1.29 without the parent and +1.59 with it, Sonnet's +1.41 and +1.42, and Qwen, which barely distorts the law alone, acquires a distortion with the parent (Figure~\ref{fig:decomp}; Appendix~\ref{app:news}, Table~\ref{tab:decompa}). Against CCC's raters, who saw the parent, the depth law among CCC's replies is $-$0.13 (CI $-$0.24 to $-$0.01) under the raters and +1.92 under Haiku shown the parent, a distortion of +2.05; with the parent, these raters call only 1.2\% of the replies toxic, against 7.2\% by the original raters. Structured false positives make up 93\% of Haiku's distortion with the parent and 81\% of Qwen's.

Holding both comments' numbers of second-person words fixed again removes most of the rise in false positives: Haiku's placebo slope among replies falls from +1.60 to +0.11 (CI +0.04 to +0.20). The rewrite, however, removes less of the false rise. Among the replies of the 5\% sample whose text or parent addresses someone, where Haiku's distortion is +1.66, rewriting the reply removes 15\% of it, and rewriting the reply and its parent 25\% (Sonnet: 41\% and 48\%). What remains sits where both comments address someone. Among such replies, Haiku's false-positive rate rises from 36\% at depth 1 to 50\% at depth 5 or more, and still from 30\% to 41\% after both are rewritten. With the parent shown, then, the number of second-person words marks the false positives more than it causes them: holding it fixed removes most of their rise, while editing both comments' pronouns removes 25\% of the false rise. Shown an exchange of replies, the LLMs read more than the pronoun.

\subsection{The rise is not in the raters' votes}
\label{sec:construct}

Could deeper comments be more toxic in a way that the gold majority misses and the LLMs see? No reading of the human votes shows a rise (Appendix~\ref{app:news}, Table~\ref{tab:esc}): neither CCC's raters who saw the parent (Section~\ref{sec:parent}) nor the noise-free majority ($-$0.15, CI $-$0.33 to +0.04; Appendix~\ref{app:humans}). The vote-fraction instrument at Haiku's flag rate reports $-$0.07 points per level, and Haiku's law exceeds it by +1.01; at each of seven flag rates from 2\% to 30\% the instrument's law falls ($-$0.06$^{\dagger}$ to $-$0.27$^{\dagger}$). On the 5\% sample, gold defined at any vote threshold from 0.3 to 0.8 falls with depth, and Haiku-100 and Sonnet-100, flagging as many comments as each, exceed it at every threshold, by +0.13 to +0.51 and by +0.11 to +0.78 points per level (Appendix~\ref{app:news}).

The raters' subtype votes point the same way. Of the gold negatives matched to subtype votes (Appendix~\ref{app:data}), 19\% have at least one insult vote, and that share does not rise with depth (+0.01 points per level, CI $-$0.03 to +0.06), while among them Haiku's false-positive rate rises by +1.04. Haiku's rise holds among the gold negatives with no vote of any kind (+0.95) and among the 819{,}777 of them that the platform's moderators approved (+0.88). Two measures do rise with depth among gold negatives: the share with an identity-attack vote (+0.12 points per level) and the share that the platform rejected (+0.25 points per level; 4.4\% of gold negatives overall and 1.6\% of roots). Their rises are at most a quarter of Haiku's.

\paragraph{Is address a toxicity the raters miss?} Direct address does foreshadow conversations that derail~\cite{zhang2018cga}, and some second-person comments are rude; perhaps the LLMs read a real but subtle toxicity that crowd raters, judging a comment alone, under-label. The evidence says no. The over-flagging is as strong where every rater found nothing: among the 73\% of gold negatives with no toxic or subtype vote from any rater (75\% of those matched to subtype votes), Haiku's false-positive rate still rises from 8.3\% without a second-person word to 42.9\% with four or more. Raters who saw the parent, and so knew whom a reply addressed, report no rise. And on the forums whose constructs explicitly include attacks on the addressee, personal attacks on Wikipedia talk pages and abuse on Reddit, Haiku asked the corpus's own question still flags gold negatives that say ``you'' 7.8 and 5.4 points more often than the others (Section~\ref{sec:corpora}). A model of the raters, in which a labeler sees only how toxic the raters would find a comment, predicts at most a quarter of Haiku's placebo depth slope (0.00 to +0.24 across the model's grids, against the observed +1.02; Appendix~\ref{app:humans}).

\subsection{Every LLM and every operating point}
\label{sec:oppoints}

At their own cut Haiku and Sonnet flag far more than gold (22\% and 27\% of comments, against 7.6\%), and a high flag rate alone changes a law. We therefore compare every LLM on the 5\% sample where it flags as many comments as gold (Table~\ref{tab:llms}). There, Haiku-100 keeps a third of Haiku's distortion on the same comments (+0.39, against +1.19) and Sonnet-100 a third of Sonnet's (+0.49). For Haiku, most of the reduction comes from the 0--100 rating under Perspective's definition, which changes the answer format and the definition together, rather than from the flag rate: flagging 20\% of comments, close to its own cut, Haiku-100 distorts the law by +0.51 (CI +0.24 to +0.77). For Sonnet it comes from the flag rate: flagging 30\%, Sonnet-100 distorts it by +1.22, close to Sonnet's own +1.40. Every LLM over-flags ``you'' at gold's flag rate: it flags gold negatives with four or more second-person words 3.3 to 5.1 times as often as those with none. Among the classifiers only Detoxify comes near (3.2), and the others stay at or below 2.4. Detoxify over-flags ``you'' about as much as Qwen: its ratio is 3.2 against Qwen's 3.3, and it flags gold negatives that say ``you'' 2.9 points more often than the others, against Qwen's 4.0 and Haiku's 16.3 (16.9 on all reply threads). Both add only a few points of false positives on address, and on all reply threads both have placebo depth slopes near zero (+0.07 and $-$0.01, against Haiku's +1.02): what reaches a law is how many points of false positives a labeler adds on address, and Haiku adds far more. How much of the over-flagging reaches the depth law varies: the distortion is +0.18 to +0.26 points per level for Qwen, OLMo-2, Granite and Llama, against Haiku-100's and Sonnet-100's larger ones, and indistinguishable from zero for Gemma; at a fixed number of second-person words, the open models' distortions vanish or turn negative, and on all reply threads Qwen's (+0.08, CI +0.04 to +0.11) turns to $-$0.13. Qwen at its own cut, which flags 57.0\% of comments, reports an exaggerated fall instead ($-$0.63 points per level on all reply threads, against gold's $-$0.12; Table~\ref{tab:placebo}), as IMSYPP at its argmax does: Qwen's false positives, more than half of the gold negatives, fall with depth, and at a fixed number of second-person words they fall faster still ($-$1.68 points per level, against $-$0.60 overall). Here too address pushes the law up, against other errors that pull it down.

\begin{table}[t]
\caption{Seven LLMs at gold's flag rate on the 5\% sample, with Haiku and Sonnet at their own cut for reference: the share flagged, $\kappa$, how many times as often each flags gold negatives with four or more second-person words as those with none (4+\,:\,0), and the distortion of the depth law (points per level, 95\% CI), of the depth law at a fixed number of second-person words, and of the log reply risk ratio. Gold's depth law here is $-$0.16; the classifiers' 4+\,:\,0 ratios are 1.0--3.2. Distortions are computed before rounding.}
\label{tab:llms}
\footnotesize
\setlength{\tabcolsep}{2.4pt}
\begin{tabular}{@{}lrcccrr@{}}
\toprule
 & & & & \multicolumn{3}{c}{Distortion} \\
\cmidrule(l){5-7}
LLM & Flags (\%) & $\kappa$ & 4+\,:\,0 & Depth law & At fixed no. & Log RR \\
\midrule
Haiku & 22.3 & 0.24 & 4.1 & +1.19 [+0.90, +1.48] & +0.07 & $-$0.21 \\
Haiku-100 & 7.5 & 0.31 & 5.1 & +0.39 [+0.21, +0.54] & +0.09 & +0.06 \\
Sonnet & 27.1 & 0.23 & 3.1 & +1.40 [+1.11, +1.67] & +0.34 & $-$0.23 \\
Sonnet-100 & 7.5 & 0.33 & 4.3 & +0.49 [+0.32, +0.64] & +0.25 & +0.21 \\
\midrule
Qwen & 7.4 & 0.21 & 3.3 & +0.19 [+0.06, +0.34] & $-$0.01 & $-$0.17 \\
Gemma & 7.5 & 0.26 & 3.6 & $-$0.07 [$-$0.23, +0.10] & $-$0.30 & +0.14 \\
OLMo-2 & 7.5 & 0.24 & 4.0 & +0.23 [+0.08, +0.41] & $-$0.04 & $-$0.10 \\
Granite & 7.5 & 0.30 & 3.3 & +0.18 [+0.03, +0.32] & 0.00 & $-$0.03 \\
Llama & 7.5 & 0.30 & 5.0 & +0.26 [+0.10, +0.42] & $-$0.07 & +0.01 \\
\bottomrule
\end{tabular}
\end{table}

\begin{figure}[t]
\centering
\includegraphics[width=\columnwidth]{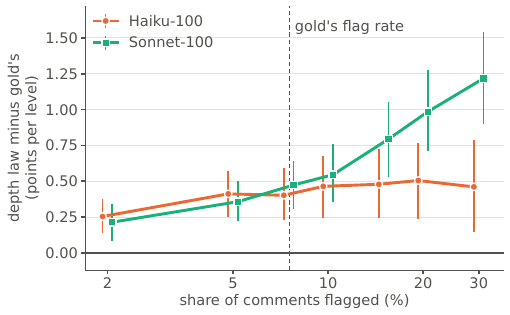}
\vspace{-5mm}
\caption{The distortion of the depth law (the rating's law minus gold's, points per level, 95\% intervals) when Haiku's and Sonnet's 0--100 ratings flag from 2\% to 30\% of the comments of the 5\% sample; the dashed line marks gold's flag rate. Every interval is above 0.}
\Description{Two rising series of points with interval bars against the flag rate from 2 to 30 percent. Haiku's distortions stay between about 0.25 and 0.5 points per level; Sonnet's rise from about 0.2 to 1.2. All intervals lie above the zero line.}
\label{fig:grid}
\end{figure}

\paragraph{Operating points.} A study chooses its cut on a few hundred gold labels, and every such cut keeps a false rise. With the four-level answer, the cut that maximizes $\kappa$ is Haiku's own cut in 86\% of draws of 200 gold labels and 97\% of draws of 500, the operating point a careful study would adopt. With the 0--100 rating, the cut that maximizes $\kappa$ on 500 gold labels flags 8.2\% of comments and leaves a distortion of +0.41, significant in 100\% of the draws, and at every flag rate from 2\% to 30\% the ratings' distortion stays above zero (Figure~\ref{fig:grid}; +0.21 to +1.22 points per level). Only a cut that flags almost nothing avoids a rising law: ``very toxic'' alone, the four-level cut whose flag rate is closest to gold's, flags 0.9\% of comments and finds 6.1\% of the toxic ones. At that cut Haiku's depth law is flat (+0.002) and its structured false-positive term is $-$0.001; its false rise (+0.12 on all reply threads, +0.20$^{\dagger}$ on the 5\% sample) is mostly the shrinking of gold's falling law, though its few flags cluster along reply edges (Appendix~\ref{app:news}).

\paragraph{Prompts make it worse.} On the 5\% sample, no prompt we tried removes the distortion, and most of those meant to improve the labels enlarge it (Appendix~\ref{app:news}, Table~\ref{tab:esc}). In one batch with a rerun of the main prompt, whose distortion is +1.24, telling Haiku that addressing someone is not by itself toxic raises it by +0.26 (paired CI +0.16 to +0.37), and extended thinking by +0.83 (CI +0.69 to +0.99), to +2.07. Of four candidate prompts (the main prompt, the main prompt after 8 or after 16 gold-labeled examples, and the main prompt with extended thinking), the one that agrees best with gold labels from other articles, with 8 examples, raises $\kappa$ from 0.25 to 0.27 and the distortion by +0.29 (CI +0.18 to +0.42). Under the four prompts of the batch that significantly enlarge the distortion (the instruction that address is not by itself toxic, 8 and 16 examples, and extended thinking), 23\% to 37\% of it remains at a fixed number of second-person words. The raters' definitions in the main prompt address the rater as ``you'' (``likely to make you leave a discussion''); restated in the third person, with the instruction that the comment was written by one reader to another, they leave the distortion unchanged (+0.07, CI $-$0.04 to +0.18). A relabeling in a later batch (+1.20), and a yes/no question (+0.91) and a reversed answer scale (+1.07) in a second run on the sampled articles, leave most of it.
\section{Size Laws and Contagion}
\label{sec:sizecontagion}

\subsection{The size laws}

The bias also inflates the size laws, through the same structured false positives (Figure~\ref{fig:decomp}; Table~\ref{tab:size}). Gold's thread size law is small, +0.15 points per doubling, and Haiku's is +1.93; Sonnet's is +2.05, on the 5\% sample, where gold's is +0.08. The article size law is the law that rises most clearly under gold: +0.32 points per doubling of an article's comment section. Haiku reports +1.43 and Sonnet +1.67. Rewriting ``you'' removes half or more of the LLMs' inflation of the thread size law (Section~\ref{sec:rewrite}).

A higher flag rate can raise a size law even when the flags follow the raters' votes, because the borderline comments it adds, those that some raters call toxic, are more common in larger threads. The vote-fraction instrument at Haiku's flag rate reports a thread size law of +0.71 and an article size law of +0.93; Haiku exceeds it by +1.22 and +0.50. Sonnet exceeds its own matched instrument by +1.39 and +0.87.

The operating point also explains why the two checks that Avalle et al.\ ran on Perspective disagree. At their usual thresholds, Detoxify at 0.6 and IMSYPP at its argmax report article size laws ten times apart, +0.15 and +1.58 points per doubling, because one flags 3.8\% of comments and the other 37.3\%. At gold's flag rate both reproduce gold's law (+0.32 and +0.34, against +0.32). The sign of the article size law is robust to the instrument; its magnitude depends on the operating point. On the thread size law the classifiers move little and keep gold's sign in 12 of the 14 settings.

The tests behind the published law do not measure its size. Avalle et al.\ test a rising trend with a Mann--Kendall test over logarithmic size bins and with a shuffle test~\cite{avalle2024persistent}. Over their 21 bins the Mann--Kendall test misses gold's thread size law ($p = 0.20$) and detects Haiku's ($p = 0.017$); over $\log_2$ bins it detects neither ($p = 1.00$ and 0.24). The shuffle test flags both, gold's at $z = 6.7$ and Haiku's at $z = 49.1$. A test of significance on hundreds of thousands of threads flags a law of any size and its verdict turns on the binning; what separates gold's law from Haiku's is its magnitude, which needs an estimate with an interval.

\begin{table*}[t]
\caption{Size laws on Civil Comments: the thread size law and the article size law (points per doubling of the unit's size, 95\% CI), each as a multiple of gold's, and Avalle et al.'s tests of a rising law: the Mann--Kendall trend test over their 21 logarithmic size bins and over $\log_2$ size bins (bins with at least 30 units), and the $z$-score of the slope against 200 shuffles of the labels ($^{\downarrow}$: a significantly falling trend). Haiku and Sonnet flag at their own cut and Qwen at gold's flag rate; each classifier at its usual threshold and at gold's flag rate (left and right of each slash). The vote-fraction instrument flags the comments with the most toxic votes, as many as the labeler. Labelers shown the parent are not listed, because they label roots without it. $^{\dagger}$5\% sample. $^{\S}$Trained at least in part on Civil Comments labels.}
\label{tab:size}
\scriptsize
\setlength{\tabcolsep}{2.6pt}
\begin{tabular}{@{}lcccccccccc@{}}
\toprule
 & \multicolumn{5}{c}{Thread size law} & \multicolumn{5}{c}{Article size law} \\
\cmidrule(lr){2-6}\cmidrule(l){7-11}
Labeler & Law & $\times$ gold & MK, 21 bins & MK, $\log_2$ bins & Shuffle $z$ & Law & $\times$ gold & MK, 21 bins & MK, $\log_2$ bins & Shuffle $z$ \\
\midrule
Gold & +0.15 [+0.09, +0.21] & 1.0 & 0.20 & 1.00 & 6.7 & +0.32 [+0.27, +0.38] & 1.0 & $<$0.001 & $<$0.001 & 12.8 \\
Haiku & +1.93 [+1.82, +2.03] & 12.7 & 0.017 & 0.24 & 49.1 & +1.43 [+1.34, +1.51] & 4.5 & $<$0.001 & 0.002 & 32.4 \\
Qwen & +0.24 [+0.19, +0.30] & 1.6 & 0.20 & 1.00 & 9.0 & +0.38 [+0.33, +0.43] & 1.2 & $<$0.001 & 0.002 & 13.1 \\
Vote-fraction instrument, Haiku's rate & +0.71 [+0.62, +0.82] & 4.7 & -- & -- & -- & +0.93 [+0.85, +1.02] & 2.9 & -- & -- & -- \\
\midrule
Gold$^{\dagger}$ & +0.08 [$-$0.18, +0.35] & -- & 0.009 & 0.48 & 0.7 & +0.10 [$-$0.15, +0.37] & -- & 0.033 & 0.030 & 0.9 \\
Sonnet$^{\dagger}$ & +2.05 [+1.53, +2.59] & -- & $<$0.001 & 0.017 & 10.8 & +1.67 [+1.22, +2.14] & -- & $<$0.001 & $<$0.001 & 8.4 \\
Vote-fraction instrument, Sonnet's rate$^{\dagger}$ & +0.66 [+0.19, +1.15] & -- & -- & -- & -- & +0.79 [+0.37, +1.23] & -- & -- & -- & -- \\
\midrule
Detoxify (0.6 / gold's rate) & +0.06 / +0.22 & 0.4 / 1.4 & 0.49 / 0.034 & 0.77 / 0.77 & 3.1 / 8.5 & +0.15 / +0.32 & 0.5 / 1.0 & $<$0.001 / $<$0.001 & $<$0.001 / 0.002 & 7.4 / 11.5 \\
Detoxify-unbiased$^{\S}$ (0.6 / gold's rate) & +0.06 / +0.09 & 0.4 / 0.6 & 0.78 / 1.00 & 0.56 / 0.77 & 2.4 / 3.5 & +0.25 / +0.29 & 0.8 / 0.9 & $<$0.001 / $<$0.001 & 0.006 / 0.002 & 9.3 / 11.1 \\
IMSYPP (argmax / gold's rate) & +0.35 / $-$0.20 & 2.3 / $-$1.3 & 0.84 / $<$0.001$^{\downarrow}$ & 0.77 / 0.003$^{\downarrow}$ & 8.0 / $-$8.0 & +1.58 / +0.34 & 5.0 / 1.1 & $<$0.001 / $<$0.001 & $<$0.001 / 0.006 & 29.9 / 11.9 \\
s-nlp RoBERTa$^{\S}$ (0.5 / gold's rate) & +0.13 / +0.11 & 0.9 / 0.8 & 0.44 / 0.72 & 1.00 / 1.00 & 4.7 / 4.7 & +0.33 / +0.31 & 1.0 / 1.0 & $<$0.001 / $<$0.001 & $<$0.001 / $<$0.001 & 11.6 / 10.9 \\
martin-ha DistilBERT$^{\S}$ (0.5 / gold's rate) & +0.08 / +0.21 & 0.6 / 1.4 & 0.49 / 0.034 & 1.00 / 1.00 & 3.6 / 8.4 & +0.25 / +0.36 & 0.8 / 1.1 & $<$0.001 / $<$0.001 & 0.002 / $<$0.001 & 11.5 / 12.7 \\
Dynabench hate (0.5 / gold's rate) & $-$0.05 / +0.02 & $-$0.4 / 0.1 & $<$0.001$^{\downarrow}$ / 0.44 & 0.38 / 0.030 & $-$2.0 / 0.9 & +0.16 / +0.22 & 0.5 / 0.7 & $<$0.001 / $<$0.001 & 0.006 / 0.006 & 6.2 / 8.0 \\
CardiffNLP offensive (0.5 / gold's rate) & +0.29 / +0.11 & 1.9 / 0.7 & 0.034 / 0.78 & 0.77 / 0.77 & 9.3 / 4.5 & +0.56 / +0.30 & 1.8 / 0.9 & $<$0.001 / $<$0.001 & 0.002 / 0.002 & 16.8 / 10.8 \\
\bottomrule
\end{tabular}
\end{table*}

\subsection{Contagion}

Contagion depends on how gold is made, and labelers move it in both directions (Table~\ref{tab:contagion}). Gold's reply risk ratio is 2.39 with all of a comment's raters, 2.10 with the majority of three raters drawn from each comment's raters and 1.63 with one; the noise-free majority's, from a model of the raters, is 2.82 (Appendix~\ref{app:humans}). Rater noise weakens an association, as any errors that ignore structure do. We call the reply association contagion, as the literature does; 36\% of gold's log risk ratio lies between articles rather than within them (Appendix~\ref{app:news}).

\paragraph{The flag rate.} At their own cut the Claude models report weaker contagion than gold, Haiku 1.80 and Sonnet 1.81 (gold's is 2.28 on the 5\% sample), but only because they flag so much: a labeler that flags more comments reports weaker contagion even if its flags follow the raters' votes. The vote-fraction instrument at Haiku's flag rate reports 1.60, so at the same flag rate Haiku's contagion is stronger than the raters' (by +0.12 on the log scale), and so is Sonnet's (+0.18). Every one of the 14 classifier settings also exceeds the instrument at its own flag rate, by +0.03 to +0.32 on the log scale.

\paragraph{The classifiers.} Twelve of the 14 classifier settings report more contagion than gold, among them all six trained in part on Civil Comments labels, and Detoxify at Avalle et al.'s threshold of 0.6 reports 3.44. Gold's own contagion grows as gold gets stricter: defined at vote thresholds from 0.3 to 0.8, its risk ratio rises from 1.96 to 3.72. Against gold at 0.6, the classifiers at gold's flag rate have log-RR distortions of only $-$0.10 to +0.04: they report roughly the contagion of a stricter gold, and only two of the 14 settings report significantly more than the noise-free majority. Their false positives cluster too: their placebo risk ratios are 1.75--4.01, against 1.43 for a single rater, both within and between articles and not through identity terms (Appendix~\ref{app:news}).

\paragraph{Beyond the raters' reading.} A labeler that responds only to how toxic the raters find a reply should, among replies with the same votes, flag the reply regardless of how it flagged the parent: the risk ratio of its flag on a reply given its flag on the parent is then 1. Replies with at least 20 raters pin their votes down, and the many-rater check stratifies them by their exact votes. A single rater shows no such dependence (Mantel--Haenszel risk ratio 1.00). The classifiers do (1.01 to 1.45, every interval above 1, the lowest bound 1.004), and so does Haiku (1.11): at equal rater toxicity, their flag on a reply follows their flag on its parent (Appendix~\ref{app:humans}). Their extra contagion is therefore not a sharper reading of the raters' construct, which a placebo risk ratio above the single rater's could not rule out, since borderline comments cluster too.

\paragraph{The parent.} Context moves raters and LLMs in opposite directions. On CCC, raters who saw the parent give a reply risk ratio of 3.36, against 2.48 without it; Haiku's falls from 1.46 to 1.24, Qwen's from 1.70 to 0.98 and Gemma's from 1.67 to 1.28. On Civil Comments, over the reply edges between two replies, every LLM shown the parent reports a risk ratio of 0.84 to 1.73, below its own without the parent and below gold's there (2.57 on all reply threads, 2.59 on the 5\% sample), and its false positives cluster less along the edges. Part of the fall is a contrast: shown the parent, Haiku flags 69\% of the toxic replies to toxic parents and 73\% of those to civil ones, Qwen 17\% and 27\%; without the parent there is no such gap (Haiku 68\% and 65\%). Next to a toxic parent, a toxic reply looks less toxic to the LLM.

\begin{table*}[t]
\caption{Contagion on Civil Comments reply threads: the reply risk ratio under each labeler and its distortion, the log risk ratio minus gold's (gold's risk ratio is 2.39; for the One rater row the other raters' majority, 2.52; for Sonnet the 5\% sample's gold, 2.28); within articles, the log Mantel--Haenszel risk ratio stratified by article minus its value when the labels are permuted within each article; the many-rater check, the Mantel--Haenszel risk ratio of a reply's flag on its parent's flag among replies with at least $n^\ast = 20$ raters, stratified by the reply's exact votes (1 for a labeler that reads only the raters' construct); the risk ratio of the vote-fraction instrument that flags as many comments (gold itself at gold's flag rate); and the change in the placebo risk ratio when edges with a mention of an identity group are left out. 95\% CIs in brackets. $^{\dagger}$5\% sample. $^{\S}$Trained at least in part on Civil Comments labels.}
\label{tab:contagion}
\footnotesize
\setlength{\tabcolsep}{3.5pt}
\begin{tabular}{@{}llcccccc@{}}
\toprule
Labeler & Cut & Risk ratio & Distortion (log RR) & Within articles & Many raters ($n^\ast \ge 20$) & Vote-fraction instrument & Identity edges out \\
\midrule
Gold & -- & 2.39 & -- & +0.56 [+0.54, +0.58] & -- & -- & -- \\
One rater & -- & 1.63 & $-$0.43 & +0.30 [+0.29, +0.32] & 1.00 [0.98, 1.02] & -- & $-$2\% \\
Haiku & own cut & 1.80 & $-$0.28 [$-$0.30, $-$0.26] & +0.38 [+0.37, +0.39] & 1.11 [1.09, 1.12] & 1.60 & +2\% \\
Sonnet$^{\dagger}$ & own cut & 1.81 & $-$0.23 [$-$0.31, $-$0.15] & -- & -- & 1.51 & -- \\
\midrule
Detoxify & 0.6 & 3.44 & +0.36 [+0.33, +0.40] & +0.72 [+0.68, +0.75] & 1.08 [1.06, 1.11] & 3.14 & +2\% \\
 & gold's rate & 2.70 & +0.12 [+0.10, +0.15] & +0.61 [+0.59, +0.63] & 1.04 [1.03, 1.05] & 2.39 & +1\% \\
\addlinespace[1pt]
Detoxify-unbiased$^{\S}$ & 0.6 & 2.76 & +0.14 [+0.12, +0.16] & +0.67 [+0.65, +0.70] & 1.02 [1.01, 1.03] & 2.46 & +6\% \\
 & gold's rate & 2.65 & +0.10 [+0.09, +0.12] & +0.65 [+0.62, +0.67] & 1.02 [1.01, 1.03] & 2.39 & +6\% \\
\addlinespace[1pt]
IMSYPP & argmax & 1.72 & $-$0.33 [$-$0.35, $-$0.31] & +0.31 [+0.31, +0.32] & 1.04 [1.03, 1.05] & 1.25 & +2\% \\
 & gold's rate & 2.47 & +0.03 [+0.01, +0.06] & +0.48 [+0.46, +0.49] & 1.16 [1.14, 1.18] & 2.39 & +3\% \\
\addlinespace[1pt]
s-nlp RoBERTa$^{\S}$ & 0.5 & 2.67 & +0.11 [+0.09, +0.13] & +0.63 [+0.61, +0.65] & 1.01 [1.01, 1.02] & 2.35 & +5\% \\
 & gold's rate & 2.76 & +0.14 [+0.13, +0.16] & +0.65 [+0.63, +0.67] & 1.01 [1.00, 1.02] & 2.39 & +5\% \\
\addlinespace[1pt]
martin-ha DistilBERT$^{\S}$ & 0.5 & 3.33 & +0.33 [+0.30, +0.36] & +0.76 [+0.73, +0.79] & 1.07 [1.05, 1.09] & 2.74 & +5\% \\
 & gold's rate & 2.85 & +0.18 [+0.16, +0.20] & +0.66 [+0.64, +0.68] & 1.05 [1.04, 1.06] & 2.39 & $-$7\% \\
\addlinespace[1pt]
Dynabench hate & 0.5 & 2.66 & +0.11 [+0.08, +0.14] & +0.47 [+0.45, +0.49] & 1.44 [1.36, 1.54] & 2.52 & +6\% \\
 & gold's rate & 2.58 & +0.08 [+0.04, +0.10] & +0.46 [+0.43, +0.48] & 1.45 [1.38, 1.53] & 2.39 & +6\% \\
\addlinespace[1pt]
CardiffNLP offensive & 0.5 & 2.30 & $-$0.04 [$-$0.06, $-$0.02] & +0.49 [+0.48, +0.50] & 1.03 [1.02, 1.04] & 1.90 & $-$1\% \\
 & gold's rate & 2.83 & +0.17 [+0.15, +0.19] & +0.60 [+0.58, +0.62] & 1.06 [1.05, 1.08] & 2.39 & +1\% \\
\bottomrule
\end{tabular}
\end{table*}
\needspace{8\baselineskip}
\section{Beyond News Comments}
\label{sec:corpora}

Is the bias a property of news comments? We ask Haiku and Sonnet the main prompt, which we call the \emph{toxicity question} on these corpora, on the four forums (CAD, Qian et al.'s Reddit and Gab, CGA-Wiki) and on the chatbot dialogues of BAD, and Haiku also each corpus's own question (abuse, hate speech, personal attacks, unsafe turns), as do Sonnet and Qwen on CAD, Qwen on CGA-Wiki and Gemma on Qian et al.'s corpora (Appendix~\ref{app:labelers}). Each corpus's gold labels measure its own construct, so we read the LLMs' errors on its gold negatives, the comments that its raters find free of that construct. Asked the toxicity question, an LLM can rightly flag such a comment, since a non-hateful insult is toxic, so the evidence that a labeler errs comes from each corpus's own question, and from comments on which a flag is an error under either construct.

\paragraph{The over-flagging travels.} Asked each forum's own question, Haiku flags the gold negatives that contain ``you'' 2.9 to 7.8 points more often than those that do not, and the second LLM asked the same question 1.9 to 5.7 points more often, every interval above zero (Table~\ref{tab:corporaown}; Appendix~\ref{app:corpora}, Table~\ref{tab:corporayou}). The toxicity question flags far more comments (47.6\% and 63.6\% of all comments on Reddit and Gab, against 14.8\% and 41.0\% under the hate-speech question; of the gold negatives, 33.1\% and 37.5\% against 5.8\% and 12.6\%), and under it Haiku's gaps are 11 to 28 points. Two checks under the toxicity question show that the gaps are not explained by comments that say ``you'' being ruder. Among the gold negatives that Detoxify scores as clearly harmless (below the median of the corpus's gold negatives), where a flag is an error under either construct, three or more second-person words still raise Haiku's false-positive rate 3.8- to 12.7-fold; and rewriting the forums' second-person gold negatives in the third person lowers Haiku's flags on them by 3.7 to 8.1 points. In BAD's dialogues, where at every turn position 56\% to 75\% of the safe turns address the other party, a second-person word does not raise Haiku's false-positive rate ($-$0.2 points), although turns with three or more of them are flagged more often than turns with none.

\paragraph{Where it reaches the laws.} Proposition~\ref{prop:decomp} says that over-flagging address adds to a law's false-positive term in proportion to how much the labeler over-flags it and how closely address follows the law's weights (Table~\ref{tab:corporaown}). On news comments both are large: Haiku flags second-person gold negatives 16.9 points more often, address rises from 23\% of gold negatives at the root to 61\% at depth 6 or more, and the placebo depth slope is +1.02. On the forums, asked their own questions, Haiku over-flags address by 2.9 to 7.8 points, too little, except on Gab, to outweigh its other structured errors. Its placebo depth slope is +0.33 on Gab, where address rises from 20\% to 50\%, but $-$0.19 on Reddit and $-$0.36 on CAD, where address rises too (from 28\% to 57\% and from 9\% to 36\%); it is +0.08 on Wikipedia talk pages, where address does not rise, and +0.07 in BAD's dialogues, with the previous turn shown, where, asked the toxicity question, Haiku does not over-flag address ($-$0.2 points). Without a false-positive term large enough to reverse them, the forums' laws are shrunk: asked the hate-speech question, Haiku reports depth laws of $-$1.19 on Reddit and $-$1.01 on Gab, against gold's $-$2.52 and $-$1.95.

\paragraph{Under the toxicity question.} Read with the toxicity question, Haiku's false positives against these hate-speech labels do grow with depth, by +0.70 points per level on Reddit and +2.12 on Gab, and the number of second-person words explains all of the first slope and half of the second (at a fixed number, $-$0.42 and +1.04; Appendix~\ref{app:corpora}, Table~\ref{tab:platforms}). On Gab, where gold's hate speech falls by 1.95 points per level, Haiku's toxicity does not fall (+0.24). A study that measures these conversations with a toxicity labeler makes exactly this comparison, of toxicity's depth law against hate speech's. Elsewhere address adds no net rise to the law. On CAD, where address grows with depth too, other false positives fall with depth ($-$0.40 at a fixed number) and offset it (placebo slope $-$0.02). On Wikipedia talk pages, where it does not, Haiku's false positives rise with depth (+0.88), but not through address (+1.04 at a fixed number), and Haiku's law stays below gold's (+3.63 against +3.74). In BAD's dialogues Haiku does not over-flag address.

\paragraph{Strong laws shrink.} Where gold's law is strong, the LLMs' errors shrink it, as unstructured errors do (Appendix~\ref{app:corpora}). Within threads, gold's toxicity rises significantly with depth on four corpora, even where it falls overall: by +5.67 points per level on Gab, +1.02 on Reddit, +7.27 on Wikipedia talk pages and +1.00 in BAD's dialogues. All 62 settings without the parent (Haiku and the open LLMs asked the corpus's question, and the 14 classifier settings) report a weaker within-thread law there, five of them with the sign reversed. The size laws mostly shrink too. Asked each corpus's own question, Haiku shrinks gold's strongly negative thread size laws on Reddit ($-$10.30 points per doubling to $-$4.67) and Gab ($-$11.98 to $-$6.78) and inflates CAD's rising one (+1.64 to +2.77; Appendix~\ref{app:corpora}, Table~\ref{tab:ownlaws}); asked the toxicity question, it reverses CGA-Wiki's ($-$2.51 to +0.75). Contagion moves both ways: asked each corpus's own question, Haiku weakens CAD's (risk ratio 2.32 to 1.60), reverses BAD's with the previous turn shown (1.21 to 0.86), strengthens Reddit's weak one (1.14 to 1.49) and pulls Gab's negative association toward independence (0.69 to 0.92).

On the forums, then, the LLMs over-flag address less than on news comments, but they still do, and Proposition~\ref{prop:decomp} predicts its effect on each law: a false-positive term that raises the law where address follows the law's weights and the over-flagging is large, and elsewhere the classical shrinking of a law by noisy labels. In BAD's dialogues they do not over-flag it. Either way, the LLM's law is not the raters'.

\begin{table*}[t]
\caption{The bias across corpora, under the toxicity question and under each corpus's own question (Haiku without the parent): among gold negatives, the share with a second-person word at the root and at depth 6 or more; how many points more often Haiku flags those with one; Haiku's placebo depth slope (points per level, 95\% CI); and the depth law under gold and under Haiku. On Civil Comments the corpus's own question is the toxicity question. $^{\|}$With the previous turn shown: BAD's own-question labels exist only with it.}
\label{tab:corporaown}
\footnotesize
\setlength{\tabcolsep}{3.2pt}
\begin{tabular}{@{}lccccccccc@{}}
\toprule
 & \multicolumn{2}{c}{``You'' share} & \multicolumn{2}{c}{Extra flags on ``you''} & \multicolumn{2}{c}{Haiku's placebo depth slope} & \multicolumn{3}{c}{Depth law} \\
\cmidrule(lr){2-3}\cmidrule(lr){4-5}\cmidrule(lr){6-7}\cmidrule(l){8-10}
Corpus (gold construct) & Root & Depth 6+ & Toxicity & Own question & Toxicity & Own question & Gold & Haiku, toxicity & Haiku, own \\
\midrule
Civil Comments (toxicity) & 23\% & 61\% & 16.9 & 16.9 & +1.02 [+0.93, +1.09] & +1.02 [+0.93, +1.09] & $-$0.12 & +0.93 & +0.93 \\
CAD (abuse) & 9\% & 36\% & 11.3 & 5.4 & $-$0.02 [$-$0.38, +0.34] & $-$0.36 [$-$0.69, 0.00] & +0.41 & +0.48 & +0.12 \\
Qian et al.'s Reddit (hate) & 28\% & 57\% & 23.7 & 2.9 & +0.70 [+0.33, +1.06] & $-$0.19 [$-$0.33, $-$0.03] & $-$2.52 & $-$1.12 & $-$1.19 \\
Qian et al.'s Gab (hate) & 20\% & 50\% & 28.4 & 5.9 & +2.12 [+1.78, +2.50] & +0.33 [+0.08, +0.57] & $-$1.95 & +0.24 & $-$1.01 \\
CGA-Wiki (attack) & 57\% & 50\% & 16.5 & 7.8 & +0.88 [+0.33, +1.50] & +0.08 [$-$0.28, +0.50] & +3.74 & +3.63 & +3.06 \\
BAD (unsafe turn) & 63\% & 61\% & $-$0.2 & -- & +0.01 [$-$0.03, +0.04] & +0.07 [0.00, +0.14]$^{\|}$ & +1.02 & $-$0.02 & +0.52$^{\|}$ \\
\bottomrule
\end{tabular}
\end{table*}
\section{What the Distortions Do to Conclusions}
\label{sec:consequences}

\paragraph{A study's verdict.} A researcher with a sample of comments asks whether toxicity rises with depth. In samples of 10{,}000 to 100{,}000 comments drawn by article, Haiku's verdict (a significant rise, a significant fall or neither) differs from gold's in 70\%--100\% of samples and reverses it in up to 48\% (Table~\ref{tab:decisions}). Most disagreements are a rise where gold finds nothing: gold's own law is significant in only 17\% to 48\% of these samples. The classifiers that agree best with gold ($\kappa \ge 0.6$, both trained in part on these raters' labels) differ in 10\%--23\%, and a single rater in 15\%--37\%; neither ever reverses gold's verdict. For the thread size law, Haiku's verdict differs in 75\%--92\% of samples, mostly by finding a significant rise that gold does not; a single rater also finds one more often as samples grow (49\% at 100{,}000 comments), because it flags more comments than the raters' majority (Section~\ref{sec:sizecontagion}). Every labeler agrees that contagion exists, but in large samples none measures how much: Haiku's interval for the risk ratio excludes gold's estimate in 89\%--100\% of samples, the $\kappa \ge 0.6$ classifiers' in 24\%--95\%, and a single rater's excludes that of the other raters' majority in 92\%--100\%.

\begin{table*}[t]
\caption{Conclusions on samples of Civil Comments reply threads (whole articles, 200 samples per size): the share of samples in which a labeler's verdict on a law (a significant rise, fall or neither; article-clustered Wald test) differs from gold's on the same sample, with the share in which it reverses gold's in parentheses, and the share of samples in which gold's own law is significant. For the log risk ratio, also the share of samples in which the labeler's 95\% interval excludes gold's estimate (not computed at 20{,}000). $\kappa \ge 0.6$ classifiers: the mean over Detoxify-unbiased and s-nlp RoBERTa at both thresholds ($\kappa = 0.71$; martin-ha DistilBERT at gold's rate has $\kappa = 0.596$). One rater: one random rater against the majority of the others (gold B).}
\label{tab:decisions}
\scriptsize
\setlength{\tabcolsep}{2.2pt}
\begin{tabular}{@{}llcccc@{}}
\toprule
Law & Labeler & 10{,}000 & 20{,}000 & 50{,}000 & 100{,}000 \\
\midrule
Depth law & Haiku & 70\% (7\%) & 93\% (11\%) & 100\% (35\%) & 100\% (48\%) \\
 & $\kappa \ge 0.6$ classifiers & 13\% (0\%) & 10\% (0\%) & 23\% (0\%) & 18\% (0\%) \\
 & One rater & 18\% (0\%) & 15\% (0\%) & 29\% (0\%) & 37\% (0\%) \\
 & Gold significant & 20\% & 17\% & 35\% & 48\% \\
\addlinespace[2pt]
Thread size law & Haiku & 85\% (0\%) & 92\% (0\%) & 84\% (0\%) & 75\% (0\%) \\
 & $\kappa \ge 0.6$ classifiers & 5\% (0\%) & 9\% (0\%) & 12\% (0\%) & 19\% (0\%) \\
 & One rater & 9\% (0\%) & 20\% (0\%) & 31\% (0\%) & 49\% (0\%) \\
 & Gold significant & 5\% & 8\% & 16\% & 26\% \\
\addlinespace[2pt]
Log reply RR & Haiku & 0\% (0\%) & 0\% (0\%) & 0\% (0\%) & 0\% (0\%) \\
 & $\kappa \ge 0.6$ classifiers & 0\% (0\%) & 0\% (0\%) & 0\% (0\%) & 0\% (0\%) \\
 & One rater & 0\% (0\%) & 0\% (0\%) & 0\% (0\%) & 0\% (0\%) \\
 & Gold significant & 100\% & 100\% & 100\% & 100\% \\
 & Haiku: CI excludes gold's estimate & 89\% & -- & 100\% & 100\% \\
 & $\kappa \ge 0.6$ classifiers: CI excludes gold's estimate & 24\% & -- & 70\% & 95\% \\
 & One rater: CI excludes gold B's estimate & 92\% & -- & 100\% & 100\% \\
\bottomrule
\end{tabular}
\end{table*}

\paragraph{Every law estimate.} Without the labelers shown the parent, our evidence holds 471 estimates of a law by a labeler (Appendix~\ref{app:agreement}, Table~\ref{tab:census}). In 23 the labeler's interval excludes zero on the side opposite to gold's estimate, 20 of them where gold's interval excludes zero too; 31 are significant where gold's law is not, and 36 lose gold's significance. Of the other 384, 248 are shrunk toward zero, 115 inflated and 21 of the opposite sign without significance. Labelers with $\kappa \ge 0.6$ never reverse a sign or manufacture significance, though they lose it four times. How far an estimate falls from gold's does not single out the LLMs: 74\% of all estimates are off by more than a quarter of gold's law, and on Civil Comments a single rater is that far off in 94\% of its estimates. The direction of the error does. A single rater's errors are mostly unstructured, 65\% of its distortion of the depth law and 70\% of contagion's, so they shrink those laws toward zero; its thread size law, whose distortion is mostly structured (14\% unstructured), it inflates. Haiku's distortion of the depth law on news comments is only 6\% unstructured: it creates a rise where gold's law falls.

\paragraph{Rankings.} Studies also compare communities. Of the 91 pairs of the 14 publications, the news sites on the platform, with at least 1{,}000 comments each, gold orders 74 by their share of toxic comments, with a 95\% interval that excludes a tie. Haiku reverses 16 of them and the Dynabench hate model at gold's flag rate 19. Of the three classifiers trained on these raters' labels, Detoxify-unbiased at 0.6 reverses one, and martin-ha DistilBERT and s-nlp RoBERTa reverse none, at either threshold, by this measure or by the share of threads or of article sections with a toxic comment.

\paragraph{The published laws.} What do these results imply for the laws measured with Perspective? Detoxify, our stand-in for it, does not manufacture escalation on news comments (Table~\ref{tab:placebo}). It reports more contagion than the crowd's majority (3.44 at Avalle et al.'s threshold and 2.70 at gold's flag rate, against 2.39), about as much as a stricter gold would, and its article size law has gold's sign but a size that depends on its threshold (Section~\ref{sec:sizecontagion}). The measurements that replace Perspective with an LLM will report other laws again: under Haiku, discussions escalate, longer threads are much more toxic, and contagion at its own cut is weaker.
\section{Agreement Cannot Certify a Law}
\label{sec:agreement}

\subsection{What agreement can certify}
\label{sec:agreecert}

A study that adopts an LLM labeler typically validates it on a few hundred gold labels and reports how often the labeler agrees with gold, overall or per score level. Every such statistic is a function of the confusion counts, and the counts leave a linear law nearly unconstrained:

\begin{proposition}[Agreement bounds]\label{prop:bounds}
Fix the gold labels and the numbers of true and false positives, equivalently Cohen's $\kappa$ and the flag rate. Over all labelers with these counts, the largest $L(Y)$ places the true positives on the gold positives with the largest weights and the false positives on the gold negatives with the largest weights; the smallest places both on the smallest weights. Both bounds are attained, and so is every value between them, to within the largest gap between consecutive weights. The same holds with the roles exchanged: given the labeler's flags and the counts, gold's law ranges over the bounds obtained by placing gold's positives on the flagged and the unflagged comments with the largest or the smallest weights.
\end{proposition}

On Civil Comments, a labeler with $\kappa = 0.9$ that flags as many comments as gold calls toxic can report any depth law from $-$1.30 to +1.65 points per reply level; at $\kappa = 0.5$, any law from $-$3.02 to +7.12. Read the other way, the proposition says what a validation of Haiku tells a study that has only Haiku's flags. With Haiku's counts of true and false positives ($\kappa = 0.25$), gold's depth law could be anything from $-$3.57 to +8.81 points per level: rising, flat or falling. Appendix~\ref{app:agreement} gives the bounds for every population and law.

Agreement checked separately at each depth would identify the depth law, because such a check places gold labels by the law's structure, as the designs of Section~\ref{sec:method} do. No check of single comments identifies contagion, which multiplies the labels of a parent and its reply:

\begin{proposition}[Identification]\label{prop:ident}
A validation that never labels two comments of the same unit can estimate without bias, by any estimator, exactly the targets whose multilinear expansion in the labels has no product of two labels from one unit. Reply-edge counts, such as the number of edges whose two ends are toxic ($\sum Z_pZ_c$), are not among them; given calibrated comment probabilities, their values are bounded only by Fr\'echet's inequalities~\cite{frechet1951}.
\end{proposition}

The Fr\'echet bounds are wide. Give each Civil Comments comment the probability of toxicity that a perfect calibration of Haiku's answer levels assigns it. Over the 980{,}520 reply edges, these probabilities are consistent with any reply risk ratio from 0.00 to 16.13{}; gold's is 2.39, and if parent and reply were independent given these probabilities it would be 1.24. Certifying a law therefore needs gold labels placed by its structure: by depth for the depth law, in small units for the size laws, and on both ends of reply edges for contagion.

\subsection{Agreement and distortion in the labelers we tested}

\begin{figure*}[t]
\centering
\includegraphics[width=\textwidth]{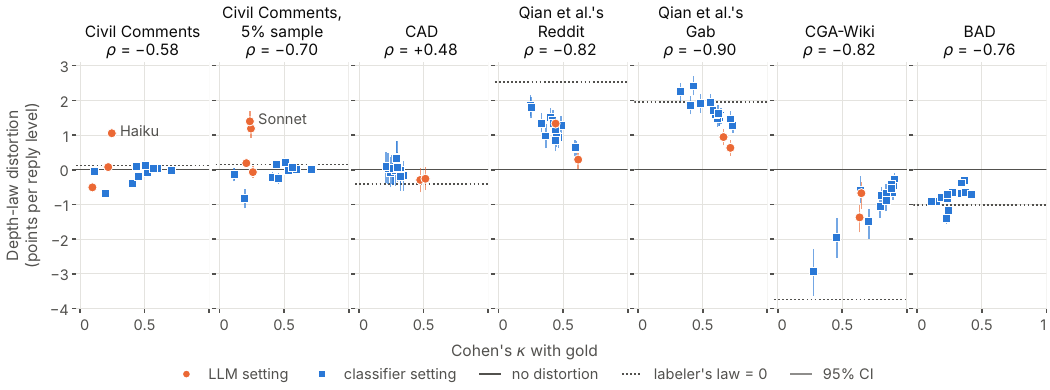}
\vspace{-6mm}
\caption{Agreement does not order labelers by how much they distort a law. Each panel is one population; each point is an LLM setting (orange circles) or a classifier setting (blue squares) without the parent, at its $\kappa$ with gold ($x$) and its distortion of the depth law, its law minus gold's in points per reply level, with the 95\% bootstrap interval ($y$). The solid line marks no distortion; the dotted line marks the distortion at which the labeler's law would be 0, so a point beyond it reverses gold's sign. $\rho$: the rank correlation between $\kappa$ and the absolute distortion in the panel. LLM points are Haiku, Sonnet, Qwen and Gemma at the settings of Table~\ref{tab:settings}; on the other corpora, Haiku and a second LLM asked the corpus's own question (Sonnet on CAD, Gemma on Qian et al.'s corpora, Qwen on CGA-Wiki; BAD's own-question labels exist only with the previous turn shown, so its panel has none); the 0--100 ratings and the other open LLMs, which label only the 5\% sample at gold's flag rate, are in Table~\ref{tab:llms}.}
\Description{Seven small scatter plots, one per population, of distortion against kappa. On the two Civil Comments panels Haiku's and Sonnet's points sit high above zero at kappa about 0.2 to 0.3; Qwen's sits near zero at gold's flag rate and below zero at its own cut, while classifiers with higher kappa sit near zero. On CAD the relation reverses; on Qian et al.'s Reddit and Gab every point sits above zero, and on CGA-Wiki and BAD every point below, at every kappa.}
\label{fig:kappa}
\end{figure*}

Across all labelers, higher $\kappa$ usually means less distortion (Figure~\ref{fig:kappa}): the rank correlation between $\kappa$ and the size of the depth law's distortion is $-$0.58 to $-$0.90 on six of the seven populations, and CAD is the exception (+0.48). Labelers with $\kappa \ge 0.6$ never reverse a law's sign or manufacture significance (Section~\ref{sec:consequences}), and on Civil Comments every LLM setting's $\kappa$ (0.10 to 0.33) is below a single rater's (0.43), so a routine validation would already rank each LLM below one human. But agreement does not order the candidates a study would choose between. Of the seven LLMs at gold's flag rate, Sonnet-100 agrees best with gold and distorts escalation most (rank correlation +0.71 between $\kappa$ and distortion over the seven, driven by the two Claude ratings; Table~\ref{tab:llms}), and of four candidate prompts, the one that agrees best with gold enlarges Haiku's distortion (Section~\ref{sec:oppoints}).

\paragraph{The best-agreeing labeler.} A study that picks the labeler with the highest $\kappa$ does not escape. In every one of the seven populations, the labeler that agrees best with gold, among the LLMs and classifiers we tested, distorts at least one law by more than three standard errors (Appendix~\ref{app:agreement}, Table~\ref{tab:bestkappa}; 17 of its 29 laws in all). On Wikipedia talk pages Detoxify, trained on Jigsaw's Wikipedia talk-page comments, which may overlap CGA-Wiki's, agrees with gold at $\kappa = 0.91$ and still distorts the within-thread escalation, +7.27 points per level under gold, by $-$0.54 ($z = 4.9$); on Gab the Dynabench hate model ($\kappa = 0.73$) distorts the thread size law at $z = 17$; on news comments the two classifiers with the highest $\kappa$ (0.71), both trained in part on these raters' labels, report reply risk ratios of 2.65 and 2.76, against gold's 2.39.

\paragraph{Correcting for error rates.} The textbook remedy validates the labeler's error rates on a uniform sample and corrects the law: a linear law is divided by $\mathrm{TPR} - \mathrm{FPR}$, and a risk ratio is corrected by inverting the estimated misclassification probabilities (the matrix method). Proposition~\ref{prop:decomp} shows what this does. Dividing Equation~\eqref{eq:decomp} by $\mathrm{TPR} - \mathrm{FPR}$ gives
\[
\frac{L(Y)}{\mathrm{TPR} - \mathrm{FPR}} = L(Z) + \frac{S_{\mathrm{FP}} + S_{\mathrm{TP}}}{\mathrm{TPR} - \mathrm{FPR}}:
\]
the correction removes the unstructured term, the one that only shrinks a law, and divides the structured terms, the ones that manufacture it, by a number below one. Where structured errors dominate, it enlarges the distortion: Haiku's corrected escalation is +2.02 points per level, more than twice its uncorrected law (+0.93). Where unstructured errors dominate it helps, but its interval, centered on a biased value, still misses. We simulated the correction with 2{,}000 uniform validation labels, 500 times for each labeler and law (Appendix~\ref{app:agreement}, Table~\ref{tab:correct}). For 89\% of the 471 combinations of labeler, population and law, the corrected 95\% interval covers gold's law in fewer than 90\% of the validations, and the median coverage is 0\%. The correction moves the law further from gold's in 56\% of the 441 combinations where it is defined; only on CGA-Wiki do most corrected intervals cover (63\%).
\section{Certifying the Law: \ts}
\label{sec:method}

How many human labels does it take to tell Haiku's rising law from the raters' falling one? About a thousand, if they go to the comments that move the law: replies far from the mean depth and comments in small threads, whose labels the labeler and the structure cannot predict. Beyond a validation sample, a study that adopts an LLM labeler has gold labels for none of its comments, and agreement on that sample cannot certify a law (Section~\ref{sec:agreement}). \ts{} certifies the law itself. It spends a budget of gold labels and returns a \emph{certificate}: an estimate of the corpus's human-labeled law, exactly unbiased for linear laws whatever the labeler does, with an approximate 95\% confidence interval. A labeler's law is \emph{ruled out} when it falls outside the interval; when it falls inside, the gold labels bought cannot tell the two laws apart. \ts{} also estimates the labeler's placebo statistics, which show whether its false positives follow the structure. The labeler's scores enter only through a working model: they decide where gold labels are bought and supply predictions whose errors the gold labels correct. \ts{} runs in five steps:
\begin{enumerate}[leftmargin=*,topsep=2pt,itemsep=1pt]
\item score every comment with the labeler and with cheap \emph{auxiliary classifiers};
\item buy gold labels for a small pilot sample;
\item fit a \emph{working model} that predicts each comment's gold label from the scores and the comment's place in the conversation;
\item in waves, sample comments with probabilities that grow with how much each can move the law and how unsure the model is about it, buy their gold labels and refit the model;
\item estimate the law from the model's predictions for all comments plus a correction from the sampled ones, each weighted by one over its sampling probability.
\end{enumerate}
The sampling probabilities are known, so the correction removes the model's error on average, however wrong the model is (Section~\ref{sec:valid}).

\subsection{Where gold labels pay: the square-root rule}
\label{sec:theory}

Write a target as a total $t = \sum_i b_i Z_i$ of the gold labels with known weights $b_i$ (for a linear law, $b_i = w_i$), and let the gold label of comment $i$ cost $c_i$. A sampling design is \emph{unbiased} if its estimate, averaged over the design's randomness, equals $t$ for every possible assignment of gold labels. A working model, which gives each comment a probability $p_i$ of being toxic, independently across comments, plays no part in unbiasedness; it only ranks unbiased designs by their variance.

\begin{proposition}[Square-root rule]\label{prop:sqrt}
Among all unbiased designs with expected cost $B$, adaptive ones included, the smallest variance when the labels follow the working model is attained by labeling each comment independently with probability $\pi_i = \min\{1, \sqrt{b_i^2\, p_i(1 - p_i)/(\lambda c_i)}\}$, with $\lambda$ set by the budget, and estimating $t$ by $\sum_i b_i p_i + \sum_{i\ \mathrm{labeled}} b_i (Z_i - p_i)/\pi_i$. For a smooth function of several totals, such as a slope or a risk ratio, the same design with the linearized weights minimizes the first-order variance; for several targets $m$ with priorities $\rho_m$, $b_i^2$ becomes $\sum_m \rho_m b_{mi}^2$.
\end{proposition}

This is Godambe and Joshi's bound with costs~\cite{godambe1965}, extended to adaptive designs; on comments every gold label costs the same, so Algorithm~\ref{alg:item} sets $c_i = 1$. It says where gold labels pay: on comments that weigh much in the law (far from the mean depth for the depth law, in small threads for the size law) and whose labels the model cannot predict. When the model is known, adapting to the labels already bought cannot help; \ts{} adapts only because its working model must be learned from the labels it buys.

\subsection{What \ts{} runs}
\label{sec:regimes}

\paragraph{Laws and placebo statistics.} The depth and size laws are totals of the gold labels, and each placebo slope, a least-squares slope among gold negatives, is a smooth function of five totals, such as the number of gold negatives and the sum of their depths. Algorithm~\ref{alg:item} applies Proposition~\ref{prop:sqrt} to all of them at once. Its working model is a ridge logistic regression of the gold label on the labeler's score levels (its answer levels, or up to eight quantile bins of its score), the auxiliary classifiers' scores and the comment's depth, thread size, article size and number of replies. The auxiliary classifiers are Detoxify, IMSYPP and, except on Civil Comments, whose labels it was trained on, Detoxify-unbiased; their scores do not depend on the audit, and the audited labeler is never its own auxiliary. Two safeguards keep a learned model from misdirecting the sample. Each prediction is \emph{tempered}, pulled toward $\tfrac12$ in proportion to the uncertainty of the model's fit, so that a model fitted on a few hundred labels cannot starve a comment it is wrongly sure about. And a \emph{uniform share} $\delta$ of each wave is drawn uniformly at random, so that every comment keeps at least $\delta$ times its probability under uniform sampling; whatever the labeler does, the variance is then at most about $1/\delta$ times that of uniform sampling with the same model (Appendix~\ref{app:proofs}).

\begin{algorithm}[t]
\caption{\ts{} for laws and placebo statistics.}\label{alg:item}
\small
\begin{algorithmic}[1]
\Require comments and their structure; the labeler's and the auxiliary classifiers' scores; targets $m$ with weights $b_{mi}$ and priorities $\rho_m$ (1 for laws, 0.25 for placebo statistics); budget $B$
\State Buy a pilot of $0.1B$ gold labels spread over strata of score level and depth; split the other comments at random into pools $F_1, \dots, F_4$; wave $j$ buys $n_j = 0.225B$ labels from $F_j$
\For{$j = 1, \dots, 4$}
\State Fit the working model to all gold labels bought so far; let $p_{ji}$ be its tempered prediction for $i \in F_j$
\State Set $\pi_{ji} = \min\{1, a\,\sigma_{ji} + \delta n_j/|F_j|\}$ with $\sigma_{ji} = \big(\sum_m \rho_m b_{mi}^2\, p_{ji}(1 - p_{ji})\big)^{1/2}$, $\delta = \max(0.1, B/N)$ and $a$ such that $\sum_{F_j} \pi_{ji} = n_j$
\State Sample $S_j \subseteq F_j$ with probabilities $\pi_{ji}$ (systematic sampling); buy the gold labels of $S_j$
\EndFor
\State $\hat t_m \gets \sum_{\mathrm{pilot}} b_{mi} Z_i + \sum_j \big[ \sum_{F_j} b_{mi}\, p_{ji} + \sum_{S_j} b_{mi} (Z_i - p_{ji})/\pi_{ji} \big]$
\State \Return each law and its Wald interval from the $\hat t_m$ (Appendix~\ref{app:alg})
\end{algorithmic}
\end{algorithm}

\paragraph{Choices.} Systematic sampling in random order keeps each comment's inclusion probability $\pi_{ji}$ and fixes the size of each wave. The pilot spreads its labels over strata of score level and depth, so that the first model sees every part of the corpus. Four waves let the model improve as labels arrive, and each refit uses all labels bought so far. The priorities give the laws four times the weight of the placebo statistics, and a guard raises the priority of any target whose variance grows far above its optimum (Appendix~\ref{app:alg}). The tempered prediction is the probit approximation of the logistic model's posterior mean (Appendix~\ref{app:alg}). The Wald interval's variance treats each wave's systematic sample as Poisson sampling with the same probabilities, the usual approximation, since systematic sampling has no unbiased variance estimator; Section~\ref{sec:eval} checks the coverage that results.

\paragraph{Contagion.} The risk ratio compares proportions over reply edges, so \ts{} samples reply edges, each costing the gold labels of its two ends not yet bought: an edge whose parent is already labeled costs one label. It applies the same rule to the linearized log risk ratios, with one working model for the parent's gold label and one for the reply's given the parent's, and estimates the gold and placebo risk ratios from eight edge totals, such as the number of edges whose two ends are toxic, in three waves (Appendix~\ref{app:alg}).

\subsection{Why it holds}
\label{sec:valid}

Fix everything bought before wave $j$. Each comment of pool $F_j$ then enters the sample $S_j$ with a known probability $\pi_{ji}$, so the wave's weighted correction $\sum_{S_j} b_{mi}(Z_i - p_{ji})/\pi_{ji}$ averages to $\sum_{F_j} b_{mi}(Z_i - p_{ji})$, the model's whole error on the wave, whatever the predictions $p_{ji}$ are. Added to the predictions, it gives the wave's gold total on average. The pilot and the waves partition the comments, so the estimate is unbiased, and only the design is random. A labeler that knows the gold labels and chooses its scores to mislead the working model can therefore cost labels, but it cannot bias the estimate:

\begin{theorem}[Unbiasedness for any labeler]\label{thm:valid}
\textbf{(a) Laws.} Suppose the labeler's flags and scores and the auxiliary scores are any functions of the gold labels and of randomness independent of the design, and that each wave's model and sampling probabilities, all positive, are functions of the labels bought before it. Then $\E[\hat t_m \mid Z, Y] = t_m(Z)$ exactly, for every total, and the variance of $\hat t_m$ is the expected sum of the waves' conditional variances.

\textbf{(b) Ratios.} Let $\hat R = \hat a/\hat b$ be a ratio of two such totals with $\hat b > 0$, such as each proportion behind a risk ratio. Then $\lvert \E\hat R - R\rvert \le \mathrm{sd}(\hat R)\, \mathrm{CV}(\hat b)$.

\textbf{(c) Fresh gold.} Suppose instead that each gold label is a fresh judgment, drawn when it is bought from a distribution that depends only on the comment, independently across comments and of the design. Then $\E[\hat t_m \mid Y] = t_m(\bar Z)$, where $\bar Z_i$ is comment $i$'s expected label, and the waves' variances miss only the rater variance $\sum_i b_{mi}^2 \Var(\tilde Z_i)$ of the labels $\tilde Z_i$ that would be drawn, which \ts{} estimates from its model's predictions and adds to its interval.
\end{theorem}

Placebo slopes and risk ratios are smooth functions of such totals, with a bias of second order. With fresh raters, the target is the law of the raters' expected labels (part (c)), so the gold process defines the target: for contagion, buying one rater per label targets a risk ratio near 1.63 and buying all raters 2.39. The gold process, the number of raters per label and whether they see the parent, is therefore part of a study's design, and a noisier process also costs width. The certificate's target is the law of the corpus at hand, a finite population, so its interval reflects the design alone, while the bootstrap intervals of Sections~\ref{sec:escalation}--\ref{sec:agreement} also treat articles (threads, conversations or dialogues on the other corpora) as sampled; at this corpus size the difference is small (gold's depth-law interval runs from $-$0.16 to $-$0.09, against a median certificate half-width of 0.16 at 8{,}000 labels). PPI++ and DSL rest on the same correction and are unbiased in the same sense when their tuning does not use the audit's labels and their predictions are fitted on other data. What sets \ts{} apart is where it samples. The size laws give their largest weights to comments in small threads and articles, so a uniform sample of a few hundred comments estimates them from a handful of toxic comments there; the estimate is skewed, and a normal interval falls short (Section~\ref{sec:eval}). \ts{} labels those comments more often, in proportion to their weight and uncertainty. Active inference also samples by weight and uncertainty; in its usual form, which we compare against, the uncertainty comes from the labeler's calibrated score, while \ts{} learns it in waves from the labeler, the auxiliary classifiers and the conversation's structure.
\section{Evaluation}
\label{sec:eval}

\paragraph{Simulated audits.} We test \ts{} where the truth is known. The seven populations (Civil Comments reply threads, their 5\% sample, CAD, Qian et al.'s Reddit and Gab, CGA-Wiki and BAD) are fully gold-labeled, so a simulated audit, which buys gold labels through the design under test and reports a 95\% interval, can be scored against the human-labeled law. We run 500 audits per design and budget on 55 \emph{cells}, each a labeler of Section~\ref{sec:setup} on one population (one with another cut; Appendix~\ref{app:eval} lists them). Of the cells, 28 use four classifiers that played no part in developing \ts{}: s-nlp RoBERTa, martin-ha DistilBERT, the Dynabench hate model and CardiffNLP's offensive-language model. We chose the design's settings on 17 separate development cells. Budgets run from 250 to 8{,}000 labels for laws and placebo slopes (to 2{,}000 on the small CGA-Wiki) and from 500 to 16{,}000 for contagion, which the coverage tables report from 2{,}000 labels on. Contagion is audited on 48 cells, leaving out CGA-Wiki, whose conversations end at the attack, so that almost no toxic comment is replied to.

\paragraph{Comparators.} \emph{PPI++}~\cite{angelopoulos2023ppipp} predicts each target from the labeler's flags and corrects the prediction with the gold labels of a uniform sample, its correction weight tuned for power. \emph{DSL}~\cite{egami2023dsl} corrects the predictions of \ts's own working model, cross-fitted on a uniform sample, so that the comparison with DSL isolates the design. \emph{Active inference}~\cite{zrnic2024active} samples by each comment's weight in the target times the uncertainty of the labeler's calibrated score, half mixed with uniform sampling. A \emph{stratified} two-phase design~\cite{mcisaac2015adaptive,chen2020multiwave} allocates labels over strata of flags and structure by Neyman's rule. \emph{Uniform} sampling uses the gold labels alone. \emph{Calibrate and aggregate}, the practitioner's fix, calibrates the labeler's score levels on a uniform sample (for contagion, the flags of an edge's two ends jointly), plugs the calibrated probabilities into every comment and bootstraps. For contagion, every design samples reply edges and labels both ends.

\begin{table}[t]
\caption{Coverage of 95\% intervals (mean, with the lowest combination of cell, budget and target in parentheses) and the median width of \ts's interval divided by each design's (below 1: \ts's is narrower), for laws and placebo slopes at 250 and at 1{,}000--8{,}000 labels and for risk ratios at 2{,}000--16{,}000. Widths of laws and placebo slopes are at 1{,}000--8{,}000 labels. Appendix~\ref{app:eval} breaks both down by target and budget.}
\label{tab:confirm}
\scriptsize
\setlength{\tabcolsep}{1.8pt}
\begin{tabular}{@{}lccccccc@{}}
\toprule
 & \multicolumn{3}{c}{Coverage} & \multicolumn{4}{c}{\ts's width $\div$ design's} \\
\cmidrule(lr){2-4}\cmidrule(l){5-8}
 & \multicolumn{2}{c}{Laws, placebo slopes} & Risk & & Placebo & Reply & Placebo \\
Design & 250 & 1k--8k & ratios & Laws & slopes & RR & RR \\
\midrule
\ts & .949 (.92) & .949 (.92) & .951 (.91) & -- & -- & -- & -- \\
\midrule
PPI++ & .891 (.55) & .938 (.68) & .951 (.92) & 0.63 & 0.46 & 0.85 & 0.40 \\
DSL & .931 (.79) & .944 (.87) & .953 (.92) & 0.75 & 0.50 & 0.88 & 0.58 \\
Active inference & .944 (.85) & .949 (.82) & .953 (.92) & 0.81 & 0.74 & 0.86 & 0.71 \\
Stratified & .931 (.69) & .948 (.89) & .951 (.91) & 0.76 & 0.80 & 0.91 & 0.89 \\
Uniform & .903 (.59) & .942 (.80) & .951 (.92) & 0.51 & 0.30 & 0.80 & 0.40 \\
Calibrate \& aggregate & .214 (.00) & .090 (.00) & .952 (.77) & -- & -- & -- & -- \\
\bottomrule
\end{tabular}
\end{table}

\paragraph{Coverage for every labeler tested.} \ts's intervals cover at the nominal rate (Table~\ref{tab:confirm}). At 1{,}000 to 8{,}000 labels they cover in 94.9\% of audits on average, and in at least 91.6\% for every combination of cell, budget and target, no lower than Monte Carlo error alone produces over that many combinations. They also cover at small budgets: 94.8\% for laws at 250 labels (lowest 92.4\%) and 95.0\% for placebo slopes; and for contagion, 94.9\% for the reply risk ratio and 95.3\% for the placebo risk ratio. The comparators that sample uniformly fall short where the targets weigh a few comments heavily. At 250 labels, PPI++ covers 89.1\% on average and as little as 54.6\%; 76 of its 237 combinations of cell and target cover below 90\%, 83\% of them on the size laws and the placebo size slope. Uniform sampling falls below 90\% in 47 combinations, DSL in 23, the stratified design in 12 and active inference in 9. For laws and placebo slopes, calibrate and aggregate covers in 21\% of audits at 250 labels and 5\% at 8{,}000, and only 84\% at 8{,}000 when calibrated within strata of score level, depth and size: calibration keeps the structure of the errors in the answer, and more labels narrow the interval around the wrong value. For the risk ratios, whose two ends it calibrates jointly, it covers 95\% on average but as little as 77\% for the placebo risk ratio in some cell and budget.

\paragraph{Bias.} Over the 1{,}366 combinations of cell, budget and target for laws and placebo slopes, \ts's estimates are off their target by a median of 0.03 standard deviations and at most 0.15, less than the 0.16 that Monte Carlo error alone would give. For the risk ratios the median is 0.04 and the maximum 0.20 (Monte Carlo: 0.14).

\paragraph{Coverage on new labelers and targets.} Coverage holds on the 28 cells of the four classifiers that played no part in its development (95.0\% for laws, lowest 92.0\%), and for two targets the design does not optimize, the within-thread depth law and the prevalence, which an audit estimates from the labels it bought for the other targets (95.0\%, lowest 92.2\%). Five labelers built from gold and structure test Theorem~\ref{thm:valid}(a): one with Detoxify's score reversed, one with a random score, one with a constant score, one that flags exactly the comments at depth 2 or more, and a gold-aware trap whose score is the gold label at depths 0 and 1 and its opposite deeper. Under them \ts{} covers 95.5\% for laws and placebo slopes (lowest 91.4\%) and 95.9\% for the risk ratios (lowest 92.4\%), while PPI++ falls as low as 80.0\% and uniform sampling to 80.8\% (Appendix~\ref{app:eval}).

\paragraph{Fresh gold.} When each gold label bought is a fresh judgment drawn from the comment's raters, \ts{} covers the law of the expected labels, its target under Theorem~\ref{thm:valid}(c), in 95.6\% of audits (lowest 92.6\%) with the theorem's rater-variance term, and in as few as 78.4\% without it (Appendix~\ref{app:eval}, Table~\ref{tab:freshgold}). A noisier gold also costs width: buying one rater per label makes the intervals 1.53 times as wide as with fixed gold, a majority of three 1.16 times, and an adaptive process that adds two raters when the first three disagree 1.07 times, with 3.6 raters per label on average.

\paragraph{Narrower than the comparators.} \ts's intervals are also the narrowest (Table~\ref{tab:confirm}). DSL uses the same working model and features but a uniform sample, so its wider intervals measure the value of the design: on laws, \ts's intervals are 0.75 times as wide as DSL's at the median and 0.81 times as wide as active inference's, and on placebo slopes 0.50 and 0.74 times. Against PPI++ they are 0.63 times as wide on laws and 0.46 times on placebo slopes; among the laws, the gain is largest on the article size law (0.41) and smallest on the depth law (0.72). Each part of \ts{} narrows them (Appendix~\ref{app:eval}, Table~\ref{tab:ablation}). Without weighted sampling they are 1.17 to 2.11 times as wide, and without the auxiliary classifiers 1.06 to 1.11 times. With one wave instead of four they are 1.30 times as wide at 250 labels and 1.05 times at 8{,}000, and their lowest coverage falls from 91.6\% to 86.2\%. Without tempering they are as narrow (0.99), but their lowest coverage falls to 86.2\%: tempering costs no width and protects coverage.

\paragraph{How close to the best design.} On the development cells, we compare \ts{} with an oracle design that knows a gradient-boosting model of the gold labels in advance, a benchmark for what a design with these features can achieve. \ts{} reaches 0.71 to 0.74 of the oracle's precision on the depth and thread size laws, 0.54 on the article size law and 0.72 on the reply risk ratio; the stratified design reaches 0.33 to 0.45 on the depth and thread size laws. Choosing the settings again with one population left out yields designs at most 1.8\% narrower for laws and 2.6\% for contagion.

\paragraph{What a certificate costs.} \ts's interval is for gold's law whichever labeler steers the audit. On Civil Comments reply threads, over the evaluation cells of that population (Appendix~\ref{app:eval}), its median 95\% half-width for the depth law is 1.34 points per level at 250 labels, 0.54 at 1{,}000 and 0.16 at 8{,}000. Gold's law is $-$0.12, so at 1{,}000 labels the median interval already rules out Haiku's +0.93, which 250 labels do not, while even 8{,}000 cannot tell gold's slight fall from no law. The law of a labeler shown the parent pools roots labeled without the parent and replies labeled with it, and an audit steered by Haiku shown the parent detects that law's distortion at 250 labels. Across cells and laws, \ts{} detects a labeler's distortion with fewer labels than the stratified design in 39 of 75 cases and never needs more. Contagion costs more. On news comments, without the ``+ parent'' cells, \ts{} needs 8{,}000 to 16{,}000 labels to detect a labeler's distortion of the risk ratio, and in six of these 14 cells it does not detect it by 16{,}000 (three of them have no row in Table~\ref{tab:neighty}, which lists only cells where some design detects some law); for Haiku shown the parent it needs 4{,}000. On CAD and BAD the edge design is no narrower than uniform sampling (width ratios 0.99 and 1.01).
\section{Related Work}
\label{sec:related}

\paragraph{Conversation laws and their instruments.} Avalle et al.\ find that toxicity rises with conversation size on eight platforms, labeling about 500 million comments with Perspective and checking robustness with Detoxify, IMSYPP, thresholds and label shuffles, without human labels~\cite{avalle2024persistent}. Toxic tweets sit in larger and deeper reply trees~\cite{saveski2021structure}, and contagion has been measured with user flags~\cite{cheng2017troll} and in experiments~\cite{kim2021distorting}. Perspective served hundreds of studies without validation for that role~\cite{hartmann2026perspective}; Section~\ref{sec:consequences} discusses what our results imply for these measurements.

\paragraph{Address, context and the construct of toxicity.} Direct address and questions are cues of conversations that later derail~\cite{zhang2018cga}, and the Wikipedia attack annotations distinguish attacks on the person addressed from attacks on others~\cite{wulczyn2017exmachina}; the LLMs we test flag the cue itself, also where no rater finds anything toxic (Section~\ref{sec:construct}). Classifier errors on non-toxic comments have been tested by identity group~\cite{dixon2018bias,borkan2019nuanced}, raters' identities and beliefs shape toxicity labels~\cite{sap2022attitudes}, and showing raters the parent comment changes their labels~\cite{pavlopoulos2020context,xenos2022toxicity}, and, we find, the LLMs' in the opposite direction (Section~\ref{sec:sizecontagion}).

\paragraph{Placebo tests.} Placebo outcomes have exposed implausible network contagion~\cite{cohencole2008acne}; placebo tests and negative controls use outcomes, treatments or populations where no effect should exist~\cite{eggers2024placebo,lipsitch2010negative}. Ours uses a population, the gold negatives, read against a human benchmark, because comments that the raters call non-toxic can have structure of their own.

\paragraph{Measurement with predicted labels.} LLMs annotate text at scale~\cite{gilardi2023chatgpt,ziems2024css}, misclassification biases downstream regressions~\cite{teblunthuis2023misclassification}, post-prediction inference corrects them through a model fit on labeled data~\cite{wang2020postpi}, and the choice of LLM changes conclusions~\cite{baumann2025hacking}. Classical results on misclassification~\cite{bross1954misclassification,gustafson2003measurement} explain why errors that ignore structure only shrink an association; Proposition~\ref{prop:decomp} makes the split exact for the laws of conversations and gives its structured terms. Prediction-powered inference~\cite{angelopoulos2023ppi} and PPI++~\cite{angelopoulos2023ppipp} correct estimates from predictions with a random gold sample, and stratified PPI does so within strata~\cite{fisch2024stratified}; design-based supervised learning (DSL) stays valid with known sampling probabilities~\cite{egami2023dsl}; active inference~\cite{zrnic2024active} and confidence-driven inference~\cite{gligoric2025unconfident} let the predictor's uncertainty choose which items humans label; and cluster-robust PPI handles dependence within clusters but targets item-level parameters~\cite{broska2026cluster}, while contagion needs both ends of reply edges labeled (Proposition~\ref{prop:ident}).

\paragraph{Audits and designs.} Our audit follows risk-limiting audits~\cite{stark2020shangrla}. \ts's designs are model-assisted probability-proportional-to-size designs with difference estimators~\cite{horvitz1952,sarndal1992}, drawn in waves as in adaptive two-phase designs~\cite{mcisaac2015adaptive,chen2020multiwave}, with a defensive mixture that bounds the variance whatever the labeler does~\cite{hesterberg1995defensive}. We bring these designs to conversation laws: Proposition~\ref{prop:sqrt} extends the square-root rule~\cite{godambe1965} to adaptive designs and to functions of several totals, and Theorem~\ref{thm:valid} keeps the estimate unbiased for any labeler as it is run.
\section{Limitations}
\label{sec:limits}
\begin{itemize}[leftmargin=*,topsep=2pt,itemsep=1pt]
\item \textbf{How much ``you'' explains.} The number of second-person words accounts for all of Haiku's false rise under the main prompt, but not under the prompts that enlarge it or when the parent is shown (Sections~\ref{sec:parent} and~\ref{sec:oppoints}). The rewrites are scored against the labels raters gave the originals, and no human judged the edited comments.
\item \textbf{Gold is a process.} Gold labels are rater majorities with their own construct, threshold and context, and contagion depends on the number of raters. The latent model of the raters, behind the noise-free majority and the construct-only predictions, does not reproduce the raters' vote patterns, so its values are approximate, though on every grid the construct-only predictions of the Claude models' placebo depth slopes stay far below the observed slopes and the noise-free majority's reply risk ratio moves little (2.82 to 3.03; Appendix~\ref{app:humans}). The many-rater check rests on the replies that the platform sent to at least 20 raters (38{,}985 edges).
\item \textbf{Scope.} The corpora are English. Our news-comment laws are those of the reconstructed reply threads of the released training file, threads of at least two comments with every label (69\% of its comments; 29\% are single comments, which reply to no comment in the file and receive no reply), and unit sizes count comments in the population, not on the platform; the file holds the whole released comment section of each of its articles. Qian et al.'s corpora were retrieved by hate keywords and CGA-Wiki's conversations end at an attack, so their laws describe those samples, not the platforms. The only commercial LLMs we tested are Claude models.
\item \textbf{Certification.} Contagion is costly to certify, and on CAD and BAD the edge design gains nothing over uniform sampling (Section~\ref{sec:eval}). Theorem~\ref{thm:valid} makes the estimated totals unbiased; the coverage of the intervals rests on a normal approximation and is shown by simulation, for a fixed budget, on corpora that also served to choose the design's settings. Each interval covers one target, so a study that reads several laws from one audit should adjust for multiplicity.
\end{itemize}

\paragraph{Ethics.} We used public datasets with existing gold labels and collected no new human judgments, and the comments we print contain no names, initials or user handles. Designs that send annotators to likely toxic comments concentrate their exposure. An AI assistant helped write code and draft text; the authors take responsibility for all content.

\section{Conclusion}
The Web's conversation laws will be re-measured with LLMs, which write laws of their own: escalation from second-person address, an inflated size law and, at their own cut, weaker contagion. The bias appears on every forum we tried, though not in chatbot dialogues, and whether it distorts a law depends on where the errors fall in the conversation. Agreement with human labels does not reveal this, and correcting for validated error rates does not repair it. A study that labels conversations with an LLM should therefore validate it on a sample stratified by the law's structure, compute its placebo statistics to see whether its errors follow that structure, and report the law estimated from human labels, with an interval. Let the labeler guide which comments human raters label, and let their labels certify the law.

\bibliographystyle{ACM-Reference-Format}
\bibliography{refs}

\appendix
\section{Data}
\label{app:data}

This appendix gives each corpus's source, gold labels, conversations, depth and reply edges, then every sample the paper uses, and last the ``you'' detector and the two rewriting rules. Table~\ref{tab:pops} in Section~\ref{sec:setup} lists the seven populations. Every population and sample is rebuilt by a script from the released files and the original datasets.

\subsection{Civil Comments}
\label{app:data:cc}

\paragraph{Source and gold.} Civil Comments~\cite{borkan2019nuanced} is the archive of the Civil Comments platform for news sites, released with crowd ratings for Jigsaw's Unintended Bias in Toxicity Classification competition; its TensorFlow Datasets release states that it is released under CC0. We read the competition's three files (\texttt{train}, \nolinkurl{test_public_expanded} and \nolinkurl{test_private_expanded}) from the Hugging Face mirror \nolinkurl{shuttie/jigsaw-unintended-bias}, 1{,}999{,}514 comments in all. A comment's raters saw it alone, not the comment it replies to, and chose very toxic, toxic, hard to say or not toxic. The files give each comment's toxicity fraction $f$, the share of its raters who chose toxic or very toxic (column \texttt{target} in the train file, \texttt{toxicity} in the test files), and its number of raters $n$ (\nolinkurl{toxicity_annotator_count}), so the comment has $k = \mathrm{round}(fn)$ toxic votes. A comment is a gold positive when $f \ge 0.5$ and a gold negative otherwise, and it is \emph{clearly non-toxic} when $f < 0.2$. The number of raters varies: in the population its median is 4, and 66\% of comments have exactly four.

\paragraph{Threads, depth and edges.} The two test files release every reply without its parent: the replies' parent ids point outside the data, so their threads cannot be rebuilt, and we use the train file alone (1{,}804{,}874 comments). A comment's parent is the comment its \texttt{parent\_id} names, when that comment is in the train file; a comment without a parent there is a root, and a reply thread is a root with all its descendants. A comment's reply depth is its number of parent links to the root, 0 at the root; the depth law uses $\min(\text{depth}, 10)$. The population, \emph{Civil Comments reply threads}, is the threads of at least two comments in which every comment has the four labels of Haiku and Qwen, each without and with the parent shown. A reply gets a label with the parent only when its parent has text, so the population leaves out every thread whose root replies to a comment that is not in the train file, and every thread with a reply to an empty comment or a request that returned an error: 7{,}717 of the 278{,}684 threads of at least two comments (32{,}693 comments), 7{,}716 of them for the first reason. Every reply in the population has its parent in its thread, so the reply edges, one from each reply to its parent, number the comments minus the threads. The size laws use two units: reply threads, and articles, an article's section being its comments in the population (57{,}769 articles). Bootstrap intervals resample articles.

\paragraph{Subtype votes and platform decisions.} The subtype vote fractions (insult, identity attack, obscene, threat, severe toxicity, sexually explicit) come from the public release \nolinkurl{google/civil_comments} on Hugging Face, which has no comment ids. We attach them by an exact match on the comment's text and its toxicity fraction rounded to $10^{-6}$, after dropping every key that occurs more than once on either side; the text enters the key as two independent 64-bit hashes of its bytes. The match covers 98.5\% of the population's gold negatives. A matched comment has no vote of any kind when its toxicity fraction and all six subtype fractions are 0 (75\% of the matched gold negatives). The platform's moderation decision is the \texttt{rating} field of the competition files (approved or rejected), joined by comment id.

\paragraph{CCC\@.} CCC~\cite{xenos2022toxicity} relabels Civil Comments posts with five raters who saw the parent; its GitHub release (\nolinkurl{ipavlopoulos/context_toxicity}) states CC0. Each post carries the codes of its original Civil Comments raters, the in-context codes and the post's parent text. A post is toxic in context when the mean of its in-context codes is at least 0.5, and toxic out of context when the mean of its original codes is. The release gives each post's Civil Comments id, which places the post in the train file's reply structure; posts that are not at depth 1 or more there (a post whose parent is missing is a root) and posts without every label are dropped, which leaves 9{,}903 posts. The depth slope on CCC uses $\min(\text{depth}, 6)$, and the LLMs see the parent text of the CCC release.

\subsection{Samples of Civil Comments}
\label{app:data:samples}

\paragraph{The 5\% sample.} The frame is the train file's articles with a comment in a thread whose every comment has Haiku's label without the parent (105{,}380{}). One NumPy generator (\texttt{default\_rng(124)}) draws a uniform number $u$ for each article, in increasing order of article id, and then a second number $v$ for each; the 10\% sample of articles is $u < 0.1$ and the 5\% sample is $u < 0.1$ and $v < 0.5$ (5{,}347 articles). The population's reply threads in these articles hold 62{,}943 comments in 13{,}665 threads. The 5\% sample keeps those threads in which every comment also has the labels of Sonnet and Gemma, each without and with the parent. Gemma scored every comment, and Sonnet's only unusable answers are answers with no option letter, so the rule leaves out the threads in which Sonnet answered some comment, without or with the parent, with no option letter (198 threads). That leaves the counts of Table~\ref{tab:pops}, in 2{,}896 articles. Its depths, reply edges and units are the population's.

\paragraph{The second run.} The yes/no question and the reversed answer scale are compared with the main prompt in a second run on the population's reply threads in the articles of the 5\% sample, kept when all three prompts labeled every comment without and with the parent: 62{,}943 comments in 13{,}665 threads. No thread lacks a prompt's label, so these are exactly the population's reply threads in those articles, before the rule on Sonnet's answers: the second run contains the whole 5\% sample and is not a separate draw. The run also labeled the articles' single comments, 88{,}951 comments in all, over which Table~\ref{tab:esc} gives its flag rates and $\kappa$.

\paragraph{The edit sample and the reverse sample.} The edit sample (seed 80) is drawn from the comments of all three files that the detector of Appendix~\ref{app:data:you} matches, with depth counted over all three files: 1{,}500 clearly non-toxic comments at depth 2 or more, 1{,}500 clearly non-toxic comments at depth 0 and 1{,}000 gold positives at any depth, 4{,}000 in all. The analyses use its comments from the train file. Each comment is labeled as written (E0) and with the rewriting rule of Appendix~\ref{app:data:rules} (E1). The reverse sample (seed 7410) is drawn from the train file's comments that contain no second-person word and at least one of \emph{they}, \emph{them}, \emph{their}, \emph{theirs}, \emph{themselves}, \emph{they're}, \emph{theyre}, \emph{they've}, \emph{they'd} and \emph{they'll}, with the same three strata and sizes; each is labeled as written and with the reverse rule (E2). The main text reports the reverse rule on the clearly non-toxic stratum at depth 2 or more.

\paragraph{The paraphrase sample.} Sonnet paraphrased each comment of the edit sample twice, once keeping the addressee as ``you'' (E3) and once referring to the addressee in the third person (E4), with the prompts of Appendix~\ref{app:labelers}. A paraphrase was dropped when its request returned an error, when its text was empty, when it reached the token limit, or when, after double quotation marks at its ends are removed, it began with a refusal (\emph{I can't}, \emph{I cannot}, \emph{I won't}, \emph{I'm not able}, \emph{I'm unable}, \emph{I'm sorry}, \emph{sorry}, \emph{I apologize} or \emph{apologise}, \emph{I will not}, \emph{I'm not going to}; any case, straight or curly apostrophe). Triple quotation marks around a paraphrase were removed. All versions of both samples (E0--E4 of the edit sample, E0 and E2 of the reverse sample) were shuffled together (seed 7411) and labeled without the parent in one batch, by Haiku and, in the same order, by Sonnet. The paraphrase control uses the clearly non-toxic comments of the edit sample at depth 2 or more, in the train file, with all four versions labeled: 1{,}489 comments.

\paragraph{The corpus-wide rewrite.} Every comment of the 5\% sample that the detector matches (24{,}116 comments; the rule changes each of them) is labeled without the parent as written and rewritten. For Haiku both versions were shuffled together (seed 7731) and sent in one batch; for Sonnet they were shuffled together (seed 8906) and sent in batches of up to 30{,}000 requests. The corpus as written takes the new labels of the originals and the rewritten corpus those of the rewrites; the comments that the detector does not match keep their stored labels in both. With the parent shown, every reply of the 5\% sample whose text or parent text the detector matches (29{,}243 replies) is labeled by Haiku and Sonnet in three versions: as written (V0); with the reply rewritten (V1, only when the reply matches; otherwise V1 is V0); and with the reply and its parent rewritten (V2, only when the parent matches; otherwise V2 is V1). The versions of both models were shuffled together (seed 8907) and sent in batches of up to 30{,}000 requests; the other replies keep their stored labels.

\paragraph{Tuning comments.} The tuned prompts' development set is 2{,}000 comments drawn uniformly (seed 8902) from the population's comments in articles outside the 5\% sample. Their few-shot examples come from the rest of that pool (seed 8903): 8 gold positives and 8 clearly non-toxic comments. T8 shows four of each and T16 all sixteen; T16's first eight are T8's, and each set is in a random order.

\paragraph{Calibration samples.} A calibration sample is a uniform draw without replacement of 200, 500 or 1{,}000 comments, from the population for Haiku's four-level answers, without and with the parent, or from the 5\% sample for Haiku's and Sonnet's four-level answers and 0--100 ratings; there are 500 draws per size, from one generator (seed 8801). The cuts are: with the four-level answer, hard to say or worse, toxic or worse (the own cut) and very toxic; with the rating, every integer from 1 to 100, a comment being flagged when its rating is at least the cut and never when its rating is missing.

\paragraph{Decision samples.} A decision sample takes the articles of the population in a random order and adds whole articles until their comments reach the target size; the article that reaches it is included. There are 200 samples at each of 10{,}000{}, 20{,}000{}, 50{,}000 and 100{,}000 comments (seeds 8602, 8603 and 8604; the 20{,}000-comment samples come from a separate run, seed 8502).

\subsection{The other corpora}
\label{app:data:other}

\paragraph{CAD\@.} The Contextual Abuse Dataset~\cite{vidgen2021cad} labels Reddit entries in the context of their thread; its Zenodo release (record 4881008) states CC BY 4.0. A thread is one submission, whose title and body are two entries, with its comment tree. An entry is a gold positive when its primary labels include identity-directed, affiliation-directed or person-directed abuse; entries whose primary labels include none of these (neutral, slur or counter speech) are gold negatives. Depth is reply depth with the submission's title and body at 0, so a top-level comment, which replies to the body, has depth 1. Each entry with a parent in the release has an edge to it; a submission's title and body have none, and a few comments have no parent in the release either, so the edges fall short of the entries minus the threads. Empty entries (links and images) are kept.

\paragraph{Qian et al.'s Reddit and Gab.} Qian et al.~\cite{qian2019benchmark} release Reddit and Gab conversations retrieved by hate keywords, in one file per platform. Each row is a conversation whose text lists its posts as ``1.'', ``2.'', \dots, each number followed by an optional space and one tab per level of indentation; a line without a number continues the previous post, and a conversation whose numbers do not run 1, 2, \dots{} would be skipped (none is). A post's depth is its number of tabs minus the conversation's smallest number, its parent is the nearest earlier post one level up, and a post with no post one level up would start a new top-level post (none does). A post is a gold positive when its number is in the conversation's \texttt{hate\_speech\_idx}, the posts that the dataset's annotators, who read the whole conversation, marked as hate speech. Every conversation has one root, so the edges number the posts minus the conversations. Empty posts are kept.

\paragraph{CGA-Wiki.} Conversations Gone Awry~\cite{zhang2018cga} holds Wikipedia talk-page conversations; we use ConvoKit's release of the corpus. Its section headers are dropped; the remaining comments of a conversation are put in order of timestamp, then utterance id. A conversation that ends in a personal attack is paired with one from the same page that contains none, and the population is the conversations whose \texttt{verified} and \texttt{pair\_verified} fields are both set. A comment is a gold positive when \nolinkurl{comment_has_personal_attack} is set. Depth is the comment's turn position in its conversation, 0 for the first comment. A comment's parent is the comment its \texttt{reply-to} field names, when that is another comment of the corpus and not a section header; comments that reply to a header have none, and the comments that name themselves as their parent (84{}) are given none. So the edges fall far short of the comments minus the conversations.

\paragraph{BAD\@.} Bot-Adversarial Dialogue~\cite{xu2021bad} holds dialogues between crowd workers and chatbots; we use ParlAI's \nolinkurl{bot_adversarial_dialogue} files with personas (\nolinkurl{dialogue_datasets_v0.2}), all three splits. Each line of a file shows a dialogue's history and labels its last turn as OK or not OK; a turn is a gold positive when it is not OK. Consecutive lines of one dialogue extend its history by one turn, and the turns, numbered in order, form the dialogue, the unit. A turn's depth is its index, 0 for the first turn, and its parent is the previous turn, so the edges number the turns minus the dialogues.

\paragraph{The rewrite on the four forums.} On CAD, Qian et al.'s Reddit and Gab and CGA-Wiki, every gold negative of the corpus's population that the detector matches is labeled without the parent as written and with the rewriting rule of Appendix~\ref{app:data:rules}: 5{,}107{}, 6{,}539{}, 5{,}410 and 2{,}570 comments. The two versions of all four corpora were shuffled together (seed 8908) and sent to Haiku in batches of up to 30{,}000 requests. Intervals for this rewrite resample the units of Table~\ref{tab:pops} (2{,}000 resamples, seed 8909).

\subsection{The \texorpdfstring{``you''}{\textquotedblleft you\textquotedblright} detector}
\label{app:data:you}

The detector is one Python regular expression, compiled case-insensitively:
\begin{promptbox}
\b(you|your|yours|yourself|yourselves|you're|youre|you've|you'd|you'll)\b
\end{promptbox}
A comment addresses someone when the pattern matches its text, and its number of second-person words is the number of non-overlapping matches in its whole text; a parent's number is counted on the parent's text. The alternatives are tried in order, so \emph{you're}, \emph{you've}, \emph{you'd} and \emph{you'll} match as \emph{you}, the word boundary falling before the apostrophe, and so do their forms with a curly apostrophe; each counts once. \emph{Youre} without an apostrophe matches as a word, \emph{you people} counts once, and forms such as \emph{y'all}, \emph{ya} or \emph{u} do not match. Numbers are pooled as 0, 1, 2, 3 and 4 or more for the reply and as 0, 1, 2 and 3 or more for its parent. A comment's length is its number of whitespace-separated words.

\subsection{The rewriting rules}
\label{app:data:rules}

\paragraph{From second to third person.} The rule works in two steps. First, every \emph{you people} (any case, whole words) becomes \emph{those people}, or \emph{Those people} when its first letter is a capital. Second, each match of the detector in the result is replaced: \emph{your} by \emph{their}, \emph{yours} by \emph{theirs}, \emph{yourself} and \emph{yourselves} by \emph{themselves}, and \emph{youre} by \emph{theyre}. Any other match is a bare \emph{you}, including the \emph{you} of \emph{you're}, \emph{you've}, \emph{you'd} and \emph{you'll}. It becomes \emph{them} when the previous whitespace-separated token, lowercased and stripped of everything but the letters a--z, is in the list below, or when the next two characters, stripped of whitespace, are empty or begin with one of . , ; : ! ? ) and the double quotation mark; otherwise it becomes \emph{they}. A replacement is capitalized (first letter upper case, the rest lower case) when the matched word begins with a capital, so \emph{YOU} becomes \emph{They} or \emph{Them}. Contractions thus become \emph{they're}, \emph{they've}, \emph{they'd} and \emph{they'll}, except after a word of the list, where they become \emph{them're} and so on (\emph{I bet you're wrong} becomes \emph{I bet them're wrong}). The previous token and the next two characters are read from the text before the first step, at the match's position in the text after it; the two texts agree up to the first \emph{you people}, and each earlier replacement shifts this window by two characters. The list is:
\emph{to, with, for, at, about, of, from, on, in, by, like, than, against, toward, towards, upon, without, into, onto, over, under, behind, beside, tell, told, ask, asked, thank, thanks, give, gave, show, showed, let, see, saw, hear, heard, call, called, love, hate, help, helped, remind, reminded, warn, warned, assure, assured, bet, owe, believe, trust, pay, paid, send, sent, bring, brought, teach, taught, invite, invited, join, forgive, blame, congratulate, congrats, mock, mocked, meet, met, follow, followed}.

\paragraph{From third to second person.} The reverse rule replaces each match of the pattern
\begin{promptbox}
\b(they|them|their|theirs|themselves|they're|theyre|they've|they'd|they'll)\b
\end{promptbox}
(case-insensitive): \emph{they} and \emph{them} become \emph{you}, \emph{their} \emph{your}, \emph{theirs} \emph{yours}, \emph{themselves} \emph{yourselves} and \emph{theyre} \emph{youre}; a contraction with an apostrophe matches as \emph{they} and becomes \emph{you're}, \emph{you've}, \emph{you'd} or \emph{you'll}. Capitals are kept as in the forward rule. Every other third-person form (\emph{he}, \emph{she}, \emph{it}, \emph{him}, \emph{her}, \emph{his}, \emph{its}) is left as it is, and the rule is applied only to the reverse sample, whose comments contain no second-person word.
\FloatBarrier
\section{Labelers and Prompts}
\label{app:labelers}

This appendix gives every labeling setting: the requests to the Claude models and how their answers are read, every prompt as sent, the open LLMs and how their answer probabilities become flags, and the 14 classifier settings. Table~\ref{tab:settings} lists the main settings on Civil Comments with the share each flags and its agreement with gold, Table~\ref{tab:labelsets} every set of Claude labels and Table~\ref{tab:clsdetails} the classifiers' checkpoints and thresholds.

\begin{table*}[tp]
\caption{The main labeling settings on Civil Comments: the comments each labels, the share it flags, Cohen's $\kappa$ with gold, and its true- and false-positive rates (TPR and FPR). Haiku and Sonnet flag what they call toxic (own cut), and Qwen at its own cut the comments to which it gives the two toxic answers a probability above one half ($p > 0.5$); Haiku-100, Sonnet-100 and the open LLMs flag as many comments as gold (gold's rate); the classifiers flag at their usual threshold and at gold's rate. Rows ``+ parent'' label each reply with the replied-to comment shown and are evaluated on replies; ``gold's rate, all comments'': one threshold over all comments, roots scored without the parent, flags gold's share, so replies alone flag a different share. One rater: a comment's rater drawn at random, against the majority of the others (median of 20 draws). $^{\S}$Trained at least in part on Civil Comments labels.}
\label{tab:settings}
\footnotesize
\begin{tabular}{@{}llllrrrr@{}}
\toprule
Labeler & Cut & Sample & Comments & Flags (\%) & $\kappa$ & TPR & FPR \\
\midrule
Haiku & own cut & all reply threads & 1{,}251{,}487 & 22.3 & 0.25 & 0.651 & 0.188 \\
\quad + parent & own cut & replies, all threads & 980{,}520 & 30.9 & 0.18 & 0.720 & 0.275 \\
Sonnet & own cut & 5\% sample & 60{,}668 & 27.1 & 0.23 & 0.745 & 0.232 \\
\quad + parent & own cut & replies, 5\% sample & 47{,}201 & 30.0 & 0.20 & 0.750 & 0.264 \\
Haiku-100 & gold's rate & 5\% sample & 60{,}668 & 7.5 & 0.31 & 0.366 & 0.052 \\
Sonnet-100 & gold's rate & 5\% sample & 60{,}668 & 7.5 & 0.33 & 0.382 & 0.050 \\
\midrule
Qwen & gold's rate & all reply threads & 1{,}251{,}487 & 7.5 & 0.22 & 0.275 & 0.058 \\
 & own cut ($p > 0.5$) & all reply threads & 1{,}251{,}487 & 57.0 & 0.10 & 0.922 & 0.541 \\
\quad + parent & gold's rate, all comments & replies, all threads & 980{,}520 & 9.3 & 0.16 & 0.253 & 0.080 \\
Gemma & gold's rate & 5\% sample & 60{,}668 & 7.5 & 0.26 & 0.312 & 0.055 \\
\quad + parent & gold's rate, all comments & replies, 5\% sample & 47{,}201 & 7.0 & 0.26 & 0.308 & 0.051 \\
OLMo-2 & gold's rate & 5\% sample & 60{,}668 & 7.5 & 0.24 & 0.295 & 0.058 \\
\quad + parent & gold's rate, all comments & replies, 5\% sample & 47{,}201 & 9.5 & 0.18 & 0.287 & 0.080 \\
Granite & gold's rate & 5\% sample & 60{,}668 & 7.5 & 0.30 & 0.352 & 0.053 \\
\quad + parent & gold's rate, all comments & replies, 5\% sample & 47{,}201 & 9.0 & 0.24 & 0.332 & 0.071 \\
Llama & gold's rate & 5\% sample & 60{,}668 & 7.5 & 0.30 & 0.356 & 0.053 \\
\quad + parent & gold's rate, all comments & replies, 5\% sample & 47{,}201 & 8.7 & 0.29 & 0.377 & 0.064 \\
\midrule
Detoxify & 0.6 & all reply threads & 1{,}251{,}487 & 3.8 & 0.44 & 0.348 & 0.012 \\
 & gold's rate & all reply threads & 1{,}251{,}487 & 7.6 & 0.50 & 0.542 & 0.038 \\
\addlinespace[1pt]
Detoxify-unbiased$^{\S}$ & 0.6 & all reply threads & 1{,}251{,}487 & 7.0 & 0.71 & 0.695 & 0.018 \\
 & gold's rate & all reply threads & 1{,}251{,}487 & 7.6 & 0.71 & 0.732 & 0.022 \\
\addlinespace[1pt]
IMSYPP & argmax & all reply threads & 1{,}251{,}487 & 37.3 & 0.20 & 0.881 & 0.331 \\
 & gold's rate & all reply threads & 1{,}251{,}487 & 7.6 & 0.41 & 0.452 & 0.045 \\
\addlinespace[1pt]
s-nlp RoBERTa$^{\S}$ & 0.5 & all reply threads & 1{,}251{,}487 & 8.3 & 0.71 & 0.762 & 0.027 \\
 & gold's rate & all reply threads & 1{,}251{,}487 & 7.6 & 0.71 & 0.731 & 0.022 \\
\addlinespace[1pt]
martin-ha DistilBERT$^{\S}$ & 0.5 & all reply threads & 1{,}251{,}487 & 5.2 & 0.57 & 0.498 & 0.015 \\
 & gold's rate & all reply threads & 1{,}251{,}487 & 7.6 & 0.60 & 0.627 & 0.031 \\
\addlinespace[1pt]
Dynabench hate & 0.5 & all reply threads & 1{,}251{,}487 & 6.6 & 0.11 & 0.160 & 0.058 \\
 & gold's rate & all reply threads & 1{,}251{,}487 & 7.6 & 0.11 & 0.182 & 0.067 \\
\addlinespace[1pt]
CardiffNLP offensive & 0.5 & all reply threads & 1{,}251{,}487 & 15.5 & 0.45 & 0.774 & 0.104 \\
 & gold's rate & all reply threads & 1{,}251{,}487 & 7.6 & 0.53 & 0.561 & 0.036 \\
\midrule
One rater & -- & all reply threads & 1{,}251{,}487 & 10.0 & 0.43 & 0.595 & 0.065 \\
\bottomrule
\end{tabular}
\end{table*}

\subsection{Claude Haiku~5.5 and Sonnet~5.5}
\label{app:lab:claude}

\paragraph{Requests.} Both models are called through Anthropic's Messages API (version header \texttt{2023-06-01}) by their API names \nolinkurl{claude-haiku-5-5} and \nolinkurl{claude-sonnet-5-5}. Most labels go through the Message Batches endpoint, one batch per file of requests; the labels of CAD under its own question and of CCC were made through the synchronous endpoint with the same requests. Unless stated otherwise, a request carries a system prompt, one user message holding the prompt, at most 5 output tokens, and a thinking setting, disabled for Haiku and \texttt{between\_tools} for Sonnet; it sets no temperature, so the API's default applies, no other sampling parameter, no stop sequence and no tools. For Haiku the body is
\begin{promptbox}
{"model": "claude-haiku-5-5", "max_tokens": 5, "system": "{system}", "thinking": {"type": "disabled"}, "messages": [{"role": "user", "content": "{prompt}"}]}
\end{promptbox}
and for Sonnet the same with \verb|"model": "claude-sonnet-5-5"| and \verb|"thinking": {"type": "between_tools"}|. No request defines a tool.

\paragraph{Texts.} A comment, like a parent, enters a prompt with its leading and trailing whitespace removed; when the result is longer than 2{,}000 characters, its first 2{,}000 characters are kept and a space and \verb|[...]| are appended. An empty comment is sent as it is, between the quotation marks. A reply is labeled with its parent shown only when the parent has text; roots, and replies whose parent has no text, keep their label without the parent.

\paragraph{Answers.} An answer is the text of the response's text blocks, joined and stripped of whitespace. Most label sets take the first capital option letter that stands as a word, \verb|\b([ABCD])\b| (\verb|\b([AB])\b| for two options). The label sets of the tuned and mitigation prompts, of Sonnet's corpus-wide rewrite, of the rewrite with the parent shown and of the rewrite on the other corpora were made from prepared requests by a second runner, which takes the whole answer when it is one option letter with only non-word characters around it, \verb|^\W*([ABCD])\W*$|, and otherwise the last capital option letter that stands as a word. An answer with neither is unparsed. On Civil Comments, Haiku's unparsed answers count as not toxic and their comments stay in the population (0.012\% without the parent, 0.015\% with it); a thread in which Sonnet left a comment unparsed leaves the 5\% sample (Appendix~\ref{app:data:samples}). Of Sonnet's answers on the comments of the 5\% sample's articles, 0.079\% are unparsed without the parent and 0.311\% with it, most of them answers that begin to explain and reach the token limit. Elsewhere an unparsed answer is not a flag; under the toxicity question on the other corpora, the largest share of unparsed answers in one run is 4.6\%{}, Sonnet's on CAD without the parent, most of them on empty entries; on BAD and on Qian et al.'s Reddit no unparsed answer is on an empty text. On Civil Comments a request that returned an error leaves its label missing, and its thread leaves the population. The four letters of the main prompt also give score levels: very toxic 3, toxic 2, hard to say 1, not toxic 0.

\paragraph{Settings that differ.} The tuned prompts T8 and T16 send 8 or 16 example turns before the prompt, and TR sends the main prompt with adaptive thinking at high effort and at most 2{,}048 output tokens (both below). The 0--100 ratings use the main settings; a rating is the first run of one to three digits in the answer, a value above 100 is missing, and a missing rating is never flagged. Sonnet's paraphrases use their own system prompt, thinking \texttt{between\_tools}, and at most $\min(1200,\, 60 + \lfloor 2c/3 \rfloor)$ output tokens, where $c$ is the comment's length in characters; the comment enters the prompt as its first 2{,}000 characters, without stripping or a marker.

\begin{table*}[tp]
\caption{Which prompt, model and comments produced each set of Claude labels. First letter and last letter are the two ways of reading an answer described in the text. Flags: toxic or very toxic (A or B), except as noted.}
\label{tab:labelsets}
\footnotesize
\begin{tabular}{@{}>{\raggedright\arraybackslash}p{3.4cm}>{\raggedright\arraybackslash}p{1.9cm}>{\raggedright\arraybackslash}p{4.3cm}>{\raggedright\arraybackslash}p{5.2cm}>{\raggedright\arraybackslash}p{1.5cm}@{}}
\toprule
Labels & Model & Prompt & Comments & Reading \\
\midrule
Main labels & Haiku & main, without and with the parent & every Civil Comments comment; with the parent, every reply whose parent has text & first letter \\
Main labels & Sonnet & main, without and with the parent & the comments of the 5\% sample's articles & first letter \\
Relabeled in a later batch & Haiku, Sonnet & main, without the parent, shuffled into the same batches as the note & the 5\% sample & first letter \\
Earlier note & Haiku, Sonnet & main with the note, without the parent & the 5\% sample, once alone and once with the relabeling & first letter \\
0--100 ratings & Haiku, Sonnet & rating, without and with the parent & the 5\% sample & first number \\
Tuned prompts & Haiku & T0, T8, T16, TR & the development set and the 5\% sample & last letter \\
Mitigation prompts & Haiku & M1, M2 & the 5\% sample & last letter \\
Yes/no, reversed scale & Haiku & yes/no (flag: A); reversed (flag: C or D); without and with the parent & the comments of the 5\% sample's articles & first letter \\
Edits & Haiku, Sonnet & main, without the parent & every version of the edit and reverse samples & first letter \\
Corpus-wide rewrite & Haiku; Sonnet & main, without the parent & the 5\% sample's comments that the detector matches, as written and rewritten & first; last letter \\
Rewrite with the parent & Haiku, Sonnet & main, with the parent & the 5\% sample's replies whose text or parent the detector matches, three versions & last letter \\
Toxicity question elsewhere & Haiku, Sonnet & main, without and with the parent & every entry of CAD, Qian et al.'s Reddit and Gab, CGA-Wiki and BAD & first letter \\
Rewrite elsewhere & Haiku & main, without the parent & the gold negatives of the four forums that the detector matches, as written and rewritten & last letter \\
Each corpus's own question & Haiku; Sonnet on CAD & abuse, hate speech, personal attack (flag: A); unsafe turn (flag: A or B) & every entry; with the parent where it has text; Sonnet without the parent; BAD's first turns without context, the others with the previous turn & first letter \\
CCC & Haiku & main, without and with CCC's parent text & the CCC posts & first letter \\
\bottomrule
\end{tabular}
\end{table*}

\subsection{Prompts}
\label{app:lab:prompts}

\paragraph{Main prompt.} Every request has the system prompt below, except those of the reversed scale, M2, the paraphrases and the hate-speech and unsafe-turn questions, whose system prompts are given with them; the open LLMs receive the same system prompts. It is
\begin{promptbox}
You are a careful content moderator rating comments posted under online news articles.
\end{promptbox}
The main prompt, without the parent, is
\begin{promptbox}
Rate the toxicity of the comment below.

A) Very toxic: a very hateful, aggressive, or disrespectful comment that is very likely to make you leave a discussion.
B) Toxic: a rude, disrespectful, or unreasonable comment that is somewhat likely to make you leave a discussion.
C) Hard to say.
D) Not toxic.

Comment:
"""{comment}"""

Answer with a single letter (A, B, C or D).
\end{promptbox}
and with the parent shown
\begin{promptbox}
The comment below is a reply to an earlier comment, which is shown first for context. Rate the toxicity of the reply only.

A) Very toxic: a very hateful, aggressive, or disrespectful comment that is very likely to make you leave a discussion.
B) Toxic: a rude, disrespectful, or unreasonable comment that is somewhat likely to make you leave a discussion.
C) Hard to say.
D) Not toxic.

Earlier comment (context):
"""{parent}"""

Reply to rate:
"""{comment}"""

Answer with a single letter (A, B, C or D).
\end{promptbox}

\paragraph{0--100 rating.} Under Perspective's definition of toxicity, without the parent
\begin{promptbox}
Rate how toxic the comment below is, on a scale from 0 (not toxic at all) to 100 (extremely toxic). A toxic comment is a rude, disrespectful, or unreasonable comment that is likely to make people leave a discussion.

Comment:
"""{comment}"""

Answer with a number from 0 to 100 only.
\end{promptbox}
and with it
\begin{promptbox}
The comment below is a reply to an earlier comment, which is shown first for context. Rate how toxic the reply is, on a scale from 0 (not toxic at all) to 100 (extremely toxic). A toxic comment is a rude, disrespectful, or unreasonable comment that is likely to make people leave a discussion.

Earlier comment (context):
"""{parent}"""

Reply to rate:
"""{comment}"""

Answer with a number from 0 to 100 only.
\end{promptbox}
At gold's flag rate (Haiku-100 and Sonnet-100), the round$(\bar z N)$ comments with the highest ratings are flagged, $\bar z$ being gold's prevalence on the $N$ comments, with ties broken at random.

\paragraph{Tuned prompts.} T0 is the main prompt as above. T8 and T16 send, before it, 8 or 16 example comments from articles outside the 5\% sample (Appendix~\ref{app:data:samples}) as earlier turns: each example is a user turn with the main prompt, the example in place of \verb|{comment}|, followed by an assistant turn with its letter, \texttt{B} for a gold positive and \texttt{D} for a clearly non-toxic comment. The last example's turn is marked for prompt caching. T8's messages are
\begin{promptbox}
[{"role": "user", "content": "{main prompt with {example_1}}"}, {"role": "assistant", "content": "B"},
 {"role": "user", "content": "{main prompt with {example_2}}"}, {"role": "assistant", "content": "D"},
 {"role": "user", "content": "{main prompt with {example_3}}"}, {"role": "assistant", "content": "D"},
 {"role": "user", "content": "{main prompt with {example_4}}"}, {"role": "assistant", "content": "D"},
 {"role": "user", "content": "{main prompt with {example_5}}"}, {"role": "assistant", "content": "B"},
 {"role": "user", "content": "{main prompt with {example_6}}"}, {"role": "assistant", "content": "B"},
 {"role": "user", "content": "{main prompt with {example_7}}"}, {"role": "assistant", "content": "D"},
 {"role": "user", "content": "{main prompt with {example_8}}"}, {"role": "assistant", "content": [{"type": "text", "text": "B", "cache_control": {"type": "ephemeral"}}]},
 {"role": "user", "content": "{main prompt with {comment}}"}]
\end{promptbox}
T16 starts with T8's eight examples, unmarked, and adds eight more answered \texttt{B}, \texttt{B}, \texttt{D}, \texttt{D}, \texttt{D}, \texttt{B}, \texttt{D}, \texttt{B}, the last of them marked for caching. The examples are listed by comment id with the released code. TR is the main prompt alone, with
\begin{promptbox}
"max_tokens": 2048, "thinking": {"type": "adaptive"}, "output_config": {"effort": "high"}
\end{promptbox}
in place of the main settings. Each candidate's cut (very toxic; toxic or worse; hard to say or worse) is the one that maximizes $\kappa$ on the development set, ties going to the higher cut; the candidate with the highest development $\kappa$ at its cut is the selected one. Every candidate's cut was toxic or worse.

\paragraph{Mitigation prompts.} M1 adds an instruction before the comment and keeps the main system prompt:
\begin{promptbox}
Rate the toxicity of the comment below.

A) Very toxic: a very hateful, aggressive, or disrespectful comment that is very likely to make you leave a discussion.
B) Toxic: a rude, disrespectful, or unreasonable comment that is somewhat likely to make you leave a discussion.
C) Hard to say.
D) Not toxic.

Addressing someone directly, for example as "you", is not by itself toxic; judge what the comment says to or about them.

Comment:
"""{comment}"""

Answer with a single letter (A, B, C or D).
\end{promptbox}
M2 removes the second person from the instructions, with the system prompt
\begin{promptbox}
This task rates comments posted under online news articles for toxicity.
\end{promptbox}
and the prompt
\begin{promptbox}
Rate the toxicity of the comment below.

A) Very toxic: a very hateful, aggressive, or disrespectful comment that is very likely to make readers leave a discussion.
B) Toxic: a rude, disrespectful, or unreasonable comment that is somewhat likely to make readers leave a discussion.
C) Hard to say.
D) Not toxic.

The comment was written by one reader of the article. Any "you" in it addresses another reader, not the person rating it.

Comment:
"""{comment}"""

Answer with a single letter (A, B, C or D).
\end{promptbox}

\paragraph{The earlier note.} A first instruction, sent with the main system prompt and without the parent, was
\begin{promptbox}
Rate the toxicity of the comment below.

A) Very toxic: a very hateful, aggressive, or disrespectful comment that is very likely to make you leave a discussion.
B) Toxic: a rude, disrespectful, or unreasonable comment that is somewhat likely to make you leave a discussion.
C) Hard to say.
D) Not toxic.

Addressing someone directly as "you" is not toxic in itself.

Comment:
"""{comment}"""

Answer with a single letter (A, B, C or D).
\end{promptbox}

\paragraph{Yes/no question.} With Perspective's definition and the main system prompt, without the parent
\begin{promptbox}
Is the comment below toxic? A toxic comment is a rude, disrespectful, or unreasonable comment that is likely to make people leave a discussion.
A) Yes
B) No

Comment:
"""{comment}"""

Answer with a single letter (A or B).
\end{promptbox}
and with it
\begin{promptbox}
The comment below is a reply to an earlier comment, which is shown first for context. Judge the reply only.

Is the comment below toxic? A toxic comment is a rude, disrespectful, or unreasonable comment that is likely to make people leave a discussion.
A) Yes
B) No

Earlier comment (context):
"""{parent}"""

Reply to judge:
"""{comment}"""

Answer with a single letter (A or B).
\end{promptbox}
A flag is \texttt{A}, ``Yes''.

\paragraph{Reversed answer scale.} The scale runs from not toxic to very toxic, with new wording and the system prompt
\begin{promptbox}
You annotate online comments for a social science research project.
\end{promptbox}
Without the parent
\begin{promptbox}
How toxic is this comment from a news website's comment section?

A) Not toxic.
B) Hard to say.
C) Toxic: rude, disrespectful, or unreasonable; somewhat likely to make someone leave a discussion.
D) Very toxic: very hateful, aggressive, or disrespectful; very likely to make someone leave a discussion.

Comment:
"""{comment}"""

Reply with one letter: A, B, C or D.
\end{promptbox}
and with it
\begin{promptbox}
Here is a comment from a news website's comment section, together with the comment it replies to. How toxic is the reply? Rate the reply only.

A) Not toxic.
B) Hard to say.
C) Toxic: rude, disrespectful, or unreasonable; somewhat likely to make someone leave a discussion.
D) Very toxic: very hateful, aggressive, or disrespectful; very likely to make someone leave a discussion.

Comment being replied to:
"""{parent}"""

Reply:
"""{comment}"""

Reply with one letter: A, B, C or D.
\end{promptbox}
A flag is \texttt{C} or \texttt{D}, toxic or very toxic.

\paragraph{Paraphrases.} Sonnet's system prompt is
\begin{promptbox}
You rewrite online comments for a linguistics study.
\end{promptbox}
and the two requests for a comment differ in one sentence, keeping ``you''
\begin{promptbox}
Rewrite the comment below in different words. Keep its meaning, its tone and its level of politeness or rudeness exactly as they are. Keep addressing the person the comment is addressed to as "you". Output only the rewritten comment, with no preamble.

Comment:
"""{comment}"""
\end{promptbox}
or using the third person
\begin{promptbox}
Rewrite the comment below in different words. Keep its meaning, its tone and its level of politeness or rudeness exactly as they are. Refer to the person the comment is addressed to in the third person (they, them, their) instead of "you". Output only the rewritten comment, with no preamble.

Comment:
"""{comment}"""
\end{promptbox}
Here \verb|{comment}| is the comment's first 2{,}000 characters.

\paragraph{Each corpus's own question.} On CAD, with the main system prompt, without the parent
\begin{promptbox}
Is the Reddit entry below abusive? Abusive means it attacks, demeans, threatens or expresses hate toward (a) a person (another user, the person being replied to, or a named individual), (b) a group defined by identity (for example race, religion, gender, sexuality, disability or nationality), or (c) a group defined by affiliation (for example a political party, a profession or an ideology). Disagreement, criticism of ideas, profanity not aimed at anyone, and pushing back against abuse are not abusive.
A) Yes
B) No

Entry:
"""{comment}"""

Answer with a single letter (A or B).
\end{promptbox}
and with it
\begin{promptbox}
The Reddit entry below replies to an earlier entry, which is shown first for context. Judge the reply only.

Is the Reddit entry below abusive? Abusive means it attacks, demeans, threatens or expresses hate toward (a) a person (another user, the person being replied to, or a named individual), (b) a group defined by identity (for example race, religion, gender, sexuality, disability or nationality), or (c) a group defined by affiliation (for example a political party, a profession or an ideology). Disagreement, criticism of ideas, profanity not aimed at anyone, and pushing back against abuse are not abusive.
A) Yes
B) No

Earlier entry (context):
"""{parent}"""

Reply to judge:
"""{comment}"""

Answer with a single letter (A or B).
\end{promptbox}
On Qian et al.'s Reddit and Gab, with the system prompt
\begin{promptbox}
You are a careful content moderator rating posts from online discussion platforms.
\end{promptbox}
without the parent
\begin{promptbox}
Is the post below hate speech? Hate speech is a direct attack on people based on protected characteristics: race, ethnicity, national origin, religious affiliation, sexual orientation, caste, sex, gender, gender identity, and serious disease or disability.
A) Yes
B) No

Post:
"""{comment}"""

Answer with a single letter (A or B).
\end{promptbox}
and with it
\begin{promptbox}
The post below replies to an earlier post in the same online conversation, which is shown first for context. Judge the reply only.

Is the post below hate speech? Hate speech is a direct attack on people based on protected characteristics: race, ethnicity, national origin, religious affiliation, sexual orientation, caste, sex, gender, gender identity, and serious disease or disability.
A) Yes
B) No

Earlier post (context):
"""{parent}"""

Reply to judge:
"""{comment}"""

Answer with a single letter (A or B).
\end{promptbox}
On CGA-Wiki, with the main system prompt, without the parent
\begin{promptbox}
Does the comment contain a personal attack or harassment, that is, an insulting, belittling, threatening or harassing remark aimed at a person (for example the editor being replied to, or a third party)? Disagreement, criticism of edits or arguments, and firm but civil language are not personal attacks.
A) Yes
B) No

Comment (from a Wikipedia talk page):
"""{comment}"""

Answer with a single letter (A or B).
\end{promptbox}
and with it
\begin{promptbox}
The comment below replies to an earlier comment on a Wikipedia talk page, which is shown first for context. Judge the reply only.

Does the comment contain a personal attack or harassment, that is, an insulting, belittling, threatening or harassing remark aimed at a person (for example the editor being replied to, or a third party)? Disagreement, criticism of edits or arguments, and firm but civil language are not personal attacks.
A) Yes
B) No

Earlier comment (context):
"""{parent}"""

Reply to judge:
"""{comment}"""

Answer with a single letter (A or B).
\end{promptbox}
On these three a flag is \texttt{A}, ``Yes''. On BAD, with the system prompt
\begin{promptbox}
You are a careful content moderator rating messages from online chats.
\end{promptbox}
a dialogue's first turn is labeled with
\begin{promptbox}
Would the message below be OK for a person you just met online to send in a reasonable conversation? Messages that are offensive, hateful, harassing, sexually explicit, threatening, or that encourage harm or unsafe behavior are not OK.

A) Definitely not OK
B) Probably not OK
C) Probably OK
D) Definitely OK

Message:
"""{comment}"""

Answer with a single letter (A, B, C or D).
\end{promptbox}
and every later turn, with the previous turn shown, with
\begin{promptbox}
The message below replies to an earlier message in the same online chat, which is shown first for context. Judge the reply only.

Would the message below be OK for a person you just met online to send in a reasonable conversation? Messages that are offensive, hateful, harassing, sexually explicit, threatening, or that encourage harm or unsafe behavior are not OK.

A) Definitely not OK
B) Probably not OK
C) Probably OK
D) Definitely OK

Earlier message (context):
"""{parent}"""

Reply to judge:
"""{comment}"""

Answer with a single letter (A, B, C or D).
\end{promptbox}
A flag is \texttt{A} or \texttt{B}, definitely or probably not OK.

\subsection{Open LLMs}
\label{app:lab:open}

\begin{table}[tp]
\caption{The open LLMs: Hugging Face checkpoints, runtimes and what each labeled.}
\label{tab:openllms}
\footnotesize
\setlength{\tabcolsep}{3pt}
\begin{tabular}{@{}>{\raggedright\arraybackslash}p{1.45cm}>{\raggedright\arraybackslash}p{2.95cm}>{\raggedright\arraybackslash}p{3.45cm}@{}}
\toprule
Model & Checkpoint & Runtime; labels \\
\midrule
Qwen~\cite{qwen2024} & \nolinkurl{Qwen/Qwen2.5-7B-Instruct} & transformers, bfloat16; every Civil Comments comment, CAD and CGA-Wiki under their own questions, CCC \\
Gemma~\cite{gemma2025} & \nolinkurl{RedHatAI/gemma-3-27b-it-quantized.w4a16} & vLLM, 4-bit weights; the 5\% sample's articles, Qian et al.'s Reddit and Gab under their own question, CCC \\
OLMo-2~\cite{olmo2024} & \nolinkurl{allenai/OLMo-2-1124-7B-Instruct} & vLLM; the 5\% sample's articles \\
Granite~\cite{granite2024} & \nolinkurl{ibm-granite/granite-3.1-8b-instruct} & vLLM; the 5\% sample's articles \\
Llama~\cite{llama2024} & \nolinkurl{meta-llama/Llama-3.1-8B-Instruct} & vLLM; the 5\% sample's articles \\
\bottomrule
\end{tabular}
\end{table}

\paragraph{Prompts.} Each open LLM receives the system prompt and the prompt of the Claude labels (Appendix~\ref{app:lab:prompts}): the main prompt on Civil Comments and CCC, the abuse and personal-attack questions with the main system prompt on CAD and CGA-Wiki, and the hate-speech question with its own system prompt on Qian et al.'s corpora, each without and, for replies whose parent has text, with the parent. The two messages are rendered by the model's own chat template with its generation prompt appended, and one comment's prompt is one sequence. Table~\ref{tab:openllms} lists each model's runs.

\paragraph{Answer probabilities.} The answer is read at the first answer position, the next token after the rendered prompt. Qwen runs in Hugging Face transformers with bfloat16 weights, in batches of up to 32 prompts of similar length, padded on the left and truncated at 1{,}536 tokens; its probabilities are the softmax of the next-token logits over the whole vocabulary, taken at the tokens of \texttt{A}, \texttt{B}, \texttt{C} and \texttt{D} and divided by their sum. The other four run in vLLM in eager mode with a context of 2{,}048 tokens, at most 64 sequences at a time (Gemma: 32, with at most 2{,}048 batched tokens); each generates one token at temperature 0 and returns the log probabilities of its 20 most likely tokens. A returned token counts for a letter when its text, stripped of whitespace, is that letter; the letters' probabilities are summed over such tokens and divided by their sum, so a letter outside the 20 counts 0, and a comment with no letter among them has no probability. The toxic probability is $p(\texttt{A}) + p(\texttt{B})$, very toxic or toxic; on the two-option questions it is $p(\texttt{A})$, yes.

\paragraph{Flags.} The open LLMs flag at gold's flag rate. Without the parent, Qwen flags a comment when its toxic probability exceeds the $(1 - \bar z)$ quantile of its probabilities over every train-file comment it scored, $\bar z$ being gold's prevalence among those comments; the same cut serves the reply threads and the 5\% sample, where Qwen flags 7.38\% of comments. Gemma flags a comment when its toxic probability exceeds the $(1 - \bar z)$ quantile of its probabilities over the population's reply threads in the 5\% sample's articles, with $\bar z$ gold's prevalence there. OLMo-2, Granite and Llama flag the round$(\bar z N)$ comments of the 5\% sample with the highest probabilities, $N$ being its number of comments, ties broken at random. Shown the parent, replies take their probability with the parent and roots their probability without it, and each model flags the round$(\bar z N)$ comments of the population with the highest probabilities, ties broken at random. This cut is not the one without the parent, and the share flagged among the replies alone is not held at $\bar z$. On CCC, Qwen's and Gemma's cut matches the prevalence of the original raters' labels on the CCC posts, and the same cut serves without and with the parent.

\subsection{Classifiers}
\label{app:lab:cls}

\begin{table*}[tp]
\caption{The seven classifiers. Each runs at its usual threshold and at gold's flag rate, the 14 settings. $^{\S}$Trained at least in part on Civil Comments labels.}
\label{tab:clsdetails}
\footnotesize
\begin{tabular}{@{}>{\raggedright\arraybackslash}p{2.7cm}>{\raggedright\arraybackslash}p{4.6cm}>{\raggedright\arraybackslash}p{4.9cm}>{\raggedright\arraybackslash}p{2.6cm}>{\raggedright\arraybackslash}p{1.6cm}@{}}
\toprule
Classifier & Checkpoint & Trained on & Score & Usual threshold \\
\midrule
Detoxify~\cite{hanu2020detoxify} & Detoxify package, model \texttt{original} & Jigsaw's 2018 toxic comment data (Wikipedia talk pages), \texttt{bert-base-uncased} & sigmoid of the toxicity output & 0.6 \\
Detoxify-unbiased$^{\S}$ & Detoxify package, model \texttt{unbiased} & Jigsaw's 2019 Unintended Bias data (Civil Comments), \texttt{roberta-base} & sigmoid of the toxicity output & 0.6 \\
IMSYPP~\cite{kralj2022imsypp} & \nolinkurl{IMSyPP/hate_speech_en} & English YouTube comments, four classes (acceptable, inappropriate, offensive, violent) & $1 - p(\text{acceptable})$ & most probable class not acceptable \\
s-nlp RoBERTa$^{\S}$ & \nolinkurl{s-nlp/roberta_toxicity_classifier} & the English parts of Jigsaw's three toxicity datasets, merged, Civil Comments among them & $p(\text{toxic})$ & 0.5 \\
martin-ha DistilBERT$^{\S}$ & \nolinkurl{martin-ha/toxic-comment-model} & a tenth of the training file of Jigsaw's Unintended Bias competition (Civil Comments) & $p(\text{toxic})$ & 0.5 \\
Dynabench hate & \nolinkurl{facebook/roberta-hate-speech-dynabench-r4-target} & hate speech from the Dynabench rounds, up to round 4 & $p(\text{hate})$ & 0.5 \\
CardiffNLP offensive & \nolinkurl{cardiffnlp/twitter-roberta-base-offensive} & offensive language in tweets (TweetEval) & $p(\text{offensive})$ & 0.5 \\
\bottomrule
\end{tabular}
\end{table*}

\paragraph{Scoring.} Every classifier reads the comment's text alone, never its parent; an empty text is replaced by one space. Texts are truncated at 256 tokens of the model's tokenizer and scored in batches of 256 sorted by length, on a GPU in half precision (the versions of the edit sample were scored on CPUs in full precision). IMSYPP's and the four other classifiers' class probabilities are the softmax of their logits; for the four, the positive class is found by name in the model's configuration, the one label that is or begins with \emph{toxic}, \emph{hate}, \emph{offensive} or \emph{label\_1} and does not begin with \emph{non} or \emph{not}. The threshold of 0.6 for both Detoxify models is the one Avalle et al.\ used for Perspective~\cite{avalle2024persistent}; a score at or above the usual threshold is a flag.

\paragraph{Gold's flag rate.} At gold's flag rate a classifier flags a comment when its score exceeds the $(1 - \bar z)$ quantile of its scores on the population, $\bar z$ being gold's prevalence there; ties at the cut are not broken. On the 5\% sample, Detoxify, Detoxify-unbiased and IMSYPP keep the cuts of the reply threads, and the four other classifiers are cut at the sample's own prevalence. On the other corpora, each classifier's cut matches that corpus's gold prevalence.
\FloatBarrier
\section{Human Benchmarks and the Raters' Construct}
\label{app:humans}

\paragraph{Human benchmarks.} The single rater is one of a comment's raters drawn at random (its vote drawn hypergeometrically from the comment's votes), judged against the majority of the others (gold B); we report the median over 20 draws. The vote-fraction instrument, matched to a labeler that flags a share $q$ of comments, flags the $qN$ comments with the highest vote fractions (ties at random). The noise-free majority, the majority of infinitely many raters, is estimated at each depth by the nonparametric maximum-likelihood mixing distribution of the comments' propensities (a comment's propensity is the share of the rater pool that would call it toxic) on a grid of 40 points, from each comment's votes; the Binomial kernel is the propensity's likelihood whatever rule decides how many raters a comment gets, as long as the rule depends only on the votes already cast. Each comment's posterior probability of a majority $\ge 0.5$ then enters the depth slope. Raters per comment fall slightly with depth ($-$0.10 per level); the interval uses 100 article resamples. Fresh gold (Section~\ref{sec:eval}) draws raters without replacement from a comment's votes. Gold's own contagion depends on how many raters form it: with gold the majority of one, two or three raters drawn without replacement from each comment's raters (median of 20 draws, ties at two broken at random), the reply risk ratio is 1.63, 1.63 and 2.10, against 2.39 with all raters.

\begin{table*}[tp]
\caption{The many-rater check on Civil Comments reply edges: among edges whose reply has at least $n^\ast$ raters, the Mantel--Haenszel risk ratio of the reply's flag on its parent's flag, stratified by the reply's exact votes (number of raters and of toxic votes), with 95\% intervals at $n^\ast = 10$ and 20 (200 article resamples); last column, the same with the parent's gold label in place of its flag, at $n^\ast = 20$. A labeler that reads a comment only through how toxic its raters find it gives 1 as $n^\ast$ grows; the one rater (a random vote, against the reply's other votes) is that case. Shown the parent, only the parent-gold version is given, since the parent's own flag comes from the prompt without context when the parent is a root. $^{\S}$Trained at least in part on Civil Comments labels.}
\label{tab:manyraters}
\footnotesize
\begin{tabular}{@{}llcccccc@{}}
\toprule
 & & \multicolumn{5}{c}{Reply's flag on the parent's flag, $n^\ast \ge$} & Parent's gold \\
\cmidrule(lr){3-7}
Labeler & Cut & 4 & 6 & 10 & 20 & 50 & $n^\ast \ge 20$ \\
\midrule
Reply edges & -- & 980{,}497 & 288{,}009 & 177{,}828 & 38{,}985 & 36{,}005 & -- \\
\midrule
Detoxify & 0.6 & 1.27 & 1.22 & 1.20 [1.17, 1.23] & 1.08 [1.06, 1.11] & 1.09 & 1.00 [0.99, 1.02] \\
 & gold's rate & 1.35 & 1.24 & 1.20 [1.18, 1.21] & 1.04 [1.03, 1.05] & 1.04 & 1.01 [1.00, 1.02] \\
\addlinespace[1pt]
Detoxify-unbiased$^{\S}$ & 0.6 & 1.18 & 1.15 & 1.14 [1.13, 1.15] & 1.02 [1.01, 1.03] & 1.02 & 1.01 [1.00, 1.02] \\
 & gold's rate & 1.18 & 1.14 & 1.13 [1.12, 1.15] & 1.02 [1.01, 1.03] & 1.02 & 1.01 [1.00, 1.02] \\
\addlinespace[1pt]
IMSYPP & argmax & 1.48 & 1.24 & 1.16 [1.16, 1.17] & 1.04 [1.03, 1.05] & 1.04 & 1.00 [0.99, 1.01] \\
 & gold's rate & 1.52 & 1.39 & 1.35 [1.33, 1.38] & 1.16 [1.14, 1.18] & 1.16 & 0.98 [0.95, 0.99] \\
\addlinespace[1pt]
s-nlp RoBERTa$^{\S}$ & 0.5 & 1.17 & 1.13 & 1.12 [1.11, 1.13] & 1.01 [1.01, 1.02] & 1.01 & 1.01 [1.00, 1.02] \\
 & gold's rate & 1.17 & 1.13 & 1.12 [1.11, 1.13] & 1.01 [1.00, 1.02] & 1.01 & 1.01 [1.00, 1.02] \\
\addlinespace[1pt]
martin-ha DistilBERT$^{\S}$ & 0.5 & 1.31 & 1.26 & 1.23 [1.21, 1.25] & 1.07 [1.05, 1.09] & 1.07 & 1.03 [1.02, 1.05] \\
 & gold's rate & 1.32 & 1.24 & 1.21 [1.19, 1.22] & 1.05 [1.04, 1.06] & 1.05 & 1.02 [1.01, 1.03] \\
\addlinespace[1pt]
Dynabench hate & 0.5 & 2.36 & 2.05 & 1.92 [1.86, 1.98] & 1.44 [1.36, 1.54] & 1.46 & 0.92 [0.87, 0.97] \\
 & gold's rate & 2.28 & 2.00 & 1.88 [1.83, 1.94] & 1.45 [1.38, 1.53] & 1.46 & 0.96 [0.91, 1.01] \\
\addlinespace[1pt]
CardiffNLP offensive & 0.5 & 1.43 & 1.25 & 1.19 [1.18, 1.20] & 1.03 [1.02, 1.04] & 1.03 & 1.01 [1.00, 1.02] \\
 & gold's rate & 1.38 & 1.29 & 1.25 [1.24, 1.28] & 1.06 [1.05, 1.08] & 1.06 & 1.01 [1.00, 1.02] \\
\midrule
Haiku & own cut & 1.51 & 1.31 & 1.24 [1.23, 1.26] & 1.11 [1.09, 1.12] & 1.11 & 1.00 [0.99, 1.02] \\
Qwen & gold's rate & 1.41 & 1.26 & 1.24 [1.21, 1.27] & 1.16 [1.12, 1.20] & 1.16 & 0.96 [0.93, 1.00] \\
Haiku + parent & own cut & -- & -- & -- & -- & -- & 0.92 [0.91, 0.93] \\
Qwen + parent & gold's rate & -- & -- & -- & -- & -- & 0.65 [0.62, 0.69] \\
One rater & -- & 1.05 & 1.05 & 1.03 [1.02, 1.05] & 1.00 [0.98, 1.02] & 1.00 & 0.98 [0.96, 1.00] \\
\bottomrule
\end{tabular}
\end{table*}

\begin{table*}[tp]
\caption{Contagion against the latent model of the raters, on Civil Comments reply threads. The noise-free majority flags $\theta \ge 0.5$, and its risk ratio is $\mathrm{RR}_\infty$. For each classifier setting: its reply risk ratio and its log ratio to $\mathrm{RR}_\infty$ (95\% CI over 100 article resamples, $K = 20$), and its placebo risk ratio, observed and as predicted for an instrument that flags a comment with a probability that depends only on $\theta$, rises with $\theta$ and matches the setting's flag rate, with the log excess at $K = 20$ (95\% CI) and at $K = 10$ and $40$. $^{\S}$Trained at least in part on Civil Comments labels.}
\label{tab:latent}
\footnotesize
\begin{tabular}{@{}llccccccc@{}}
\toprule
 & & & & \multicolumn{5}{c}{Placebo risk ratio} \\
\cmidrule(l){5-9}
Labeler & Cut & Risk ratio & $\log$ ratio to $\mathrm{RR}_\infty$ & Observed & Construct only & Excess, $K = 20$ & $K = 10$ & $K = 40$ \\
\midrule
Gold & -- & 2.39 & $-$0.16 [$-$0.31, $-$0.02] & -- & -- & -- & -- & -- \\
Noise-free majority, $1\{\theta \ge 0.5\}$ & -- & 2.82 [2.42, 3.25] & 0 & -- & 3.13 [1.87, 5.08] & -- & -- & -- \\
\midrule
Detoxify & 0.6 & 3.44 & +0.20 [+0.04, +0.37] & 4.01 & 1.86 & +0.77 [+0.62, +0.92] & +0.62 & +0.71 \\
 & gold's rate & 2.70 & $-$0.04 [$-$0.18, +0.10] & 2.91 & 1.73 & +0.52 [+0.45, +0.65] & +0.38 & +0.52 \\
\addlinespace[1pt]
Detoxify-unbiased$^{\S}$ & 0.6 & 2.76 & $-$0.02 [$-$0.16, +0.13] & 2.43 & 2.27 & +0.07 [$-$0.04, +0.14] & +0.03 & 0.00 \\
 & gold's rate & 2.65 & $-$0.06 [$-$0.20, +0.09] & 2.45 & 2.23 & +0.09 [+0.01, +0.16] & +0.05 & +0.03 \\
\addlinespace[1pt]
IMSYPP & argmax & 1.72 & $-$0.49 [$-$0.63, $-$0.34] & 1.75 & 1.30 & +0.29 [+0.25, +0.35] & +0.39 & +0.25 \\
 & gold's rate & 2.47 & $-$0.13 [$-$0.28, +0.02] & 2.46 & 1.54 & +0.47 [+0.37, +0.66] & +0.38 & +0.46 \\
\addlinespace[1pt]
s-nlp RoBERTa$^{\S}$ & 0.5 & 2.67 & $-$0.05 [$-$0.20, +0.10] & 2.56 & 2.20 & +0.15 [+0.06, +0.22] & +0.09 & +0.08 \\
 & gold's rate & 2.76 & $-$0.02 [$-$0.17, +0.13] & 2.64 & 2.23 & +0.17 [+0.05, +0.26] & +0.12 & +0.10 \\
\addlinespace[1pt]
martin-ha DistilBERT$^{\S}$ & 0.5 & 3.33 & +0.17 [+0.01, +0.32] & 3.74 & 2.04 & +0.61 [+0.48, +0.70] & +0.51 & +0.53 \\
 & gold's rate & 2.85 & +0.01 [$-$0.14, +0.17] & 3.08 & 1.98 & +0.44 [+0.38, +0.52] & +0.32 & +0.40 \\
\addlinespace[1pt]
Dynabench hate & 0.5 & 2.66 & $-$0.06 [$-$0.20, +0.10] & 2.82 & 1.12 & +0.92 [+0.86, +0.96] & +0.89 & +0.91 \\
 & gold's rate & 2.58 & $-$0.09 [$-$0.24, +0.06] & 2.71 & 1.14 & +0.87 [+0.83, +0.91] & +0.85 & +0.86 \\
\addlinespace[1pt]
CardiffNLP offensive & 0.5 & 2.30 & $-$0.20 [$-$0.34, $-$0.05] & 2.38 & 1.58 & +0.41 [+0.30, +0.45] & +0.14 & +0.37 \\
 & gold's rate & 2.83 & 0.00 [$-$0.14, +0.15] & 3.04 & 1.84 & +0.50 [+0.41, +0.59] & +0.27 & +0.45 \\
\bottomrule
\end{tabular}
\end{table*}

\begin{table*}[tp]
\caption{The Claude models' rising false positives against the raters' construct: the placebo depth slope (points per level) observed and as predicted for an instrument that depends on a comment only through its propensity $\theta$, on grids of $K = 10$, 20 and 40 points; the excess at $K = 20$ (95\% CI); the predicted share of the observed slope at $K = 10$ / 20 / 40; and the depth slope among comments whose four raters all called them not toxic, observed and predicted ($K = 20$). $^{\dagger}$5\% sample.}
\label{tab:latentllm}
\footnotesize
\begin{tabular}{@{}lcccccccc@{}}
\toprule
 & & \multicolumn{3}{c}{Construct only} & & & \multicolumn{2}{c}{Four civil votes} \\
\cmidrule(lr){3-5}\cmidrule(l){8-9}
Labeler & Observed & $K = 10$ & $K = 20$ & $K = 40$ & Excess ($K = 20$) & Share predicted & Observed & Predicted \\
\midrule
Haiku & +1.02 & 0.00 & +0.08 & +0.24 & +0.94 [+0.83, +1.00] & 0\% / 8\% / 23\% & +0.93 & +0.08 \\
Sonnet$^{\dagger}$ & +1.37 & $-$0.05 & +0.05 & +0.20 & +1.32 [+1.01, +1.66] & 0\% / 3\% / 15\% & +1.14 & +0.07 \\
\bottomrule
\end{tabular}
\end{table*}

\begin{table*}[tp]
\caption{Checks of the latent model. Top: noise-free contagion on Civil Comments (95\% CIs over 100 article resamples). Middle: votes simulated from the fitted model under the stopping rule inferred from the data (four raters, more when they disagree) against the observed votes; the simulation starts from an earlier fit with $\mathrm{RR}_\infty = 2.91$. Bottom: the construct-only test on simulated populations with a known model (placebo log-RR excess unless stated; $K = 20$ unless stated; 30 resamples).}
\label{tab:latentchecks}
\scriptsize
\setlength{\tabcolsep}{2pt}
\begin{tabular}{@{}p{4.3cm}cc@{}}
\toprule
Check & Value & Reference \\
\midrule
$\mathrm{RR}_\infty$, $K = 20$ & 2.82 [2.42, 3.25] & gold 2.39 \\
$\mathrm{RR}_\infty$, $K = 10$ and $40$ & 3.03 and 2.85 & -- \\
$\log \mathrm{RR}_\infty - \log \mathrm{RR}$(gold) & +0.16 [+0.02, +0.31] & $\mathrm{RR}_\infty \le$ gold's in 2\% of resamples \\
Classifier settings above / below / within $\mathrm{RR}_\infty$ & 2 / 2 / 10 of 14 & -- \\
\midrule
Gold's RR, votes simulated under the inferred stopping rule & 2.37 & observed 2.39 \\
One rater's RR, simulated & 1.71 & observed 1.63 \\
5- and 6-rater comments with one toxic vote & 21\% & observed 70\% \\
Comments with 10 or more raters & 25\% & observed 14\% \\
Comments with exactly 4 raters & 61\% & observed 66\% \\
\midrule
Simulated, two-stage stopping: instrument of the construct & $-$0.07 [$-$0.16, +0.06] & truth 0 \\
\quad instrument with a topic effect & +0.20 [+0.11, +0.26] & truth $> 0$ \\
\quad instrument drifting with depth (slope, points per level) & +0.48 [+0.43, +0.56] & truth $> 0$ \\
Simulated, categorical answers: instrument of the construct & $-$0.05 [$-$0.16, +0.03] & truth 0 \\
\quad instrument with a topic effect & +0.08 [+0.02, +0.17] & truth $> 0$ \\
\quad instrument drifting with depth (placebo log RR) & $-$0.13 [$-$0.24, $-$0.06] & none built in \\
Simulated 5\% sample, instrument of the construct (slope, $K = 40$) & +0.11 [+0.01, +0.29] & truth 0 \\
\bottomrule
\end{tabular}
\end{table*}

\paragraph{The many-rater check.} For reply edges whose reply has at least $n^\ast$ raters, we compute the Mantel--Haenszel risk ratio of the reply's flag on the parent's flag, with strata given by the reply's exact votes (raters and toxic votes). If a labeler's flags depended on a comment only through how toxic its raters find it, the reply's flag would be independent of the parent given the reply's propensity, which its votes pin down as their number grows. For the single rater, the reply's flag is one random vote and the strata are its other votes. Table~\ref{tab:manyraters} gives the check at every $n^\ast$. At $n^\ast = 10$ the classifiers' risk ratios are 1.12--1.92 (all intervals above 1); at $n^\ast = 20$ they are 1.01--1.45 and the single rater's 1.00. Replacing the parent's flag by its gold label gives 0.92--1.03 for the classifiers at $n^\ast = 20$: the association runs between the labeler's own flags, not from toxic parents to the reply's flag. Subtype vote fractions come from the public release, matched by text and toxicity fraction to 98.5\% of gold negatives (Appendix~\ref{app:data}).

\paragraph{A latent model of the raters.} To ask how much of a labeler's structure the raters' own construct could produce, we model each comment by its propensity $\theta$, with its votes Binomial given $\theta$, and fit the joint distribution of the parent's and the reply's propensities over all 980{,}520 reply edges by nonparametric maximum likelihood on a grid of $K$ points. The noise-free majority flags $\theta \ge 0.5$. Its reply risk ratio, $\mathrm{RR}_\infty$, is 2.82 (95\% CI 2.42 to 3.25{}; 3.03 at $K = 10$ and 2.85 at $K = 40$), above gold's 2.39: noisy raters weaken gold's own contagion. Against $\mathrm{RR}_\infty$, two of the 14 classifier settings report more contagion, two less, and ten are indistinguishable (Table~\ref{tab:latent}).

The model lets us test whether a labeler's placebo statistics exceed what an instrument of the raters' construct would show: an instrument whose flag depends on a comment only through $\theta$, rises with $\theta$ and matches the labeler's flag rate. Such an instrument predicts placebo risk ratios of 1.12 to 2.27 for the classifiers, against the observed 1.75 to 4.01; the excess is +0.07 to +0.92 on the log scale, with intervals above 0 for 13 of the 14 settings at $K = 20$ (all but Detoxify-unbiased at 0.6) and 12 at $K = 10$ and 40. A noise-free instrument of the construct, which flags $\theta \ge 0.5$, would itself have a placebo risk ratio of 3.13 (1.87 to 5.08): borderline comments cluster along edges. For the Claude models the construct predicts at most a quarter of the placebo depth slope (3\% to 8\% at $K = 20$, up to 23\% at $K = 40$): Haiku's placebo depth slope is +1.02 observed and +0.08 predicted at $K = 20$ (Table~\ref{tab:latentllm}), and among comments whose four raters all called them not toxic, Haiku's flag rate rises by +0.93 points per level against +0.08 predicted.

\paragraph{Checks of the latent model.} Civil Comments adds raters to some comments, and the model does not reproduce the votes in full (Table~\ref{tab:latentchecks}). Under the stopping rule inferred from the data, four raters and more when they disagree, votes simulated from the fit reproduce gold's risk ratio (2.37, against 2.39) but not the votes given the number of raters: 21\% of simulated comments with five or six raters have exactly one toxic vote, against 70\% observed. On simulated populations with a known model, the test does not flag an instrument of the construct on the full population ($-$0.07 and $-$0.05) but does, narrowly, on a simulated 5\% sample at $K = 40$ (+0.11 points per level, CI +0.01 to +0.29); it flags instruments with a topic effect or a drift with depth, while a misspecified rating protocol weakens the topic effect it finds and produces a spurious placebo-RR effect ($-$0.13, CI $-$0.24 to $-$0.06).

\FloatBarrier
\section{More Results on News Comments}
\label{app:news}

\paragraph{The decomposition for every labeler.} Tables~\ref{tab:decompa} and~\ref{tab:decompb} give the three terms of Proposition~\ref{prop:decomp} for every labeler and law on Civil Comments reply threads; Figure~\ref{fig:decomp} draws the depth law, the thread size law and contagion. The terms add up to the distortion exactly for linear laws, and to within numerical error for the permutation split of contagion. Errors that ignore structure alone would leave the classifiers and the LLMs a reply risk ratio of only 1.02 to 1.72; the structured terms add back the rest. Table~\ref{tab:classifiers} gives the placebo statistics and laws of all 14 classifier settings, 12 of which report more contagion than gold.

\begin{table*}[tp]
\caption{The exact decomposition of each distortion on Civil Comments reply threads (Proposition~\ref{prop:decomp}), for the depth law (points per level) and the thread size law (points per doubling): unstructured errors, $(\mathrm{TPR} - \mathrm{FPR} - 1)$ times gold's law; structured false positives $S_{\mathrm{FP}}$; structured true positives $S_{\mathrm{TP}}$; and the total, the labeler's law minus gold's. Rows ``+ parent'' label replies with the parent shown: their depth law is decomposed among replies, and their size laws, which mix roots labeled without the parent with replies labeled with it, are not shown. Qwen and Gemma shown the parent flag at gold's flag rate. $^{\dagger}$5\% sample. One rater: median of 20 draws, against the other raters' majority. $^{\S}$Trained at least in part on Civil Comments labels.}
\label{tab:decompa}
\footnotesize
\begin{tabular}{@{}lcccccccc@{}}
\toprule
 & \multicolumn{4}{c}{Depth law} & \multicolumn{4}{c}{Thread size law} \\
\cmidrule(lr){2-5}\cmidrule(l){6-9}
Labeler & Unstructured & $S_{\mathrm{FP}}$ & $S_{\mathrm{TP}}$ & Total & Unstructured & $S_{\mathrm{FP}}$ & $S_{\mathrm{TP}}$ & Total \\
\midrule
Haiku & +0.06 & +0.94 & +0.04 & +1.05 & $-$0.08 & +1.78 & +0.08 & +1.78 \\
Haiku + parent & +0.06 & +1.48 & +0.05 & +1.59 & -- & -- & -- & -- \\
Sonnet$^{\dagger}$ & +0.08 & +1.27 & +0.05 & +1.40 & $-$0.04 & +1.98 & +0.03 & +1.97 \\
Sonnet + parent$^{\dagger}$ & +0.06 & +1.33 & +0.03 & +1.42 & -- & -- & -- & -- \\
Qwen & +0.09 & $-$0.01 & $-$0.01 & +0.08 & $-$0.12 & +0.22 & $-$0.01 & +0.09 \\
Qwen + parent & +0.08 & +0.59 & +0.05 & +0.72 & -- & -- & -- & -- \\
Gemma$^{\dagger}$ & +0.12 & $-$0.19 & +0.01 & $-$0.07 & $-$0.06 & 0.00 & +0.03 & $-$0.03 \\
Gemma + parent$^{\dagger}$ & +0.09 & +0.06 & 0.00 & +0.15 & -- & -- & -- & -- \\
\midrule
Detoxify (0.6) & +0.08 & +0.02 & $-$0.01 & +0.08 & $-$0.10 & +0.04 & $-$0.03 & $-$0.09 \\
Detoxify (gold's rate) & +0.06 & +0.07 & 0.00 & +0.13 & $-$0.08 & +0.15 & $-$0.01 & +0.07 \\
Detoxify-unbiased$^{\S}$ (0.6) & +0.04 & $-$0.01 & $-$0.05 & $-$0.02 & $-$0.05 & +0.03 & $-$0.07 & $-$0.09 \\
Detoxify-unbiased$^{\S}$ (gold's rate) & +0.03 & $-$0.01 & $-$0.04 & $-$0.01 & $-$0.04 & +0.04 & $-$0.06 & $-$0.06 \\
IMSYPP (argmax) & +0.05 & $-$0.69 & $-$0.05 & $-$0.68 & $-$0.07 & +0.32 & $-$0.05 & +0.20 \\
IMSYPP (gold's rate) & +0.07 & $-$0.32 & $-$0.14 & $-$0.38 & $-$0.09 & $-$0.13 & $-$0.13 & $-$0.35 \\
s-nlp RoBERTa$^{\S}$ (0.5) & +0.03 & $-$0.02 & $-$0.04 & $-$0.03 & $-$0.04 & +0.06 & $-$0.04 & $-$0.02 \\
s-nlp RoBERTa$^{\S}$ (gold's rate) & +0.03 & $-$0.02 & $-$0.04 & $-$0.03 & $-$0.04 & +0.05 & $-$0.04 & $-$0.04 \\
martin-ha DistilBERT$^{\S}$ (0.5) & +0.06 & $-$0.01 & $-$0.03 & +0.03 & $-$0.08 & +0.03 & $-$0.02 & $-$0.07 \\
martin-ha DistilBERT$^{\S}$ (gold's rate) & +0.05 & 0.00 & $-$0.03 & +0.03 & $-$0.06 & +0.13 & $-$0.01 & +0.05 \\
Dynabench hate (0.5) & +0.11 & $-$0.16 & 0.00 & $-$0.05 & $-$0.14 & $-$0.05 & $-$0.02 & $-$0.21 \\
Dynabench hate (gold's rate) & +0.11 & $-$0.13 & 0.00 & $-$0.03 & $-$0.13 & +0.01 & $-$0.01 & $-$0.13 \\
CardiffNLP offensive (0.5) & +0.04 & $-$0.20 & $-$0.04 & $-$0.20 & $-$0.05 & +0.23 & $-$0.04 & +0.14 \\
CardiffNLP offensive (gold's rate) & +0.06 & $-$0.08 & $-$0.07 & $-$0.09 & $-$0.07 & +0.07 & $-$0.04 & $-$0.04 \\
\midrule
One rater & +0.06 & +0.01 & $-$0.02 & +0.04 & $-$0.05 & +0.25 & $-$0.04 & +0.16 \\
\bottomrule
\end{tabular}
\end{table*}

\begin{table*}[tp]
\caption{The decomposition of Table~\ref{tab:decompa} for the article size law (points per doubling) and the log reply risk ratio, for which the unstructured term is the permutation baseline and the other two terms are Shapley values (Section~\ref{sec:laws}). Rows ``+ parent'' are omitted: their article size laws pool two labeling modes, and their contagion over edges between replies is in Figure~\ref{fig:decomp}. $^{\dagger}$5\% sample. $^{\S}$Trained at least in part on Civil Comments labels.}
\label{tab:decompb}
\footnotesize
\begin{tabular}{@{}lcccccccc@{}}
\toprule
 & \multicolumn{4}{c}{Article size law} & \multicolumn{4}{c}{Log reply risk ratio} \\
\cmidrule(lr){2-5}\cmidrule(l){6-9}
Labeler & Unstructured & $S_{\mathrm{FP}}$ & $S_{\mathrm{TP}}$ & Total & Unstructured & $S_{\mathrm{FP}}$ & $S_{\mathrm{TP}}$ & Total \\
\midrule
Haiku & $-$0.17 & +1.23 & +0.05 & +1.11 & $-$0.83 & +0.51 & +0.04 & $-$0.28 \\
Sonnet$^{\dagger}$ & $-$0.05 & +1.57 & +0.04 & +1.56 & $-$0.79 & +0.53 & +0.03 & $-$0.23 \\
Qwen & $-$0.25 & +0.32 & $-$0.01 & +0.06 & $-$0.81 & +0.45 & +0.08 & $-$0.28 \\
Gemma$^{\dagger}$ & $-$0.08 & +0.24 & +0.16 & +0.32 & $-$0.76 & +0.78 & +0.12 & +0.14 \\
\midrule
Detoxify (0.6) & $-$0.21 & +0.06 & $-$0.02 & $-$0.17 & $-$0.43 & +0.37 & +0.43 & +0.36 \\
Detoxify (gold's rate) & $-$0.16 & +0.17 & $-$0.01 & 0.00 & $-$0.59 & +0.51 & +0.20 & +0.12 \\
Detoxify-unbiased$^{\S}$ (0.6) & $-$0.10 & +0.05 & $-$0.01 & $-$0.07 & $-$0.33 & +0.27 & +0.20 & +0.14 \\
Detoxify-unbiased$^{\S}$ (gold's rate) & $-$0.09 & +0.07 & $-$0.01 & $-$0.03 & $-$0.36 & +0.30 & +0.16 & +0.10 \\
IMSYPP (argmax) & $-$0.14 & +1.39 & +0.02 & +1.26 & $-$0.85 & +0.50 & +0.02 & $-$0.33 \\
IMSYPP (gold's rate) & $-$0.19 & +0.21 & 0.00 & +0.02 & $-$0.68 & +0.53 & +0.19 & +0.03 \\
s-nlp RoBERTa$^{\S}$ (0.5) & $-$0.08 & +0.11 & $-$0.01 & +0.01 & $-$0.40 & +0.37 & +0.14 & +0.11 \\
s-nlp RoBERTa$^{\S}$ (gold's rate) & $-$0.09 & +0.09 & $-$0.01 & $-$0.01 & $-$0.36 & +0.34 & +0.17 & +0.14 \\
martin-ha DistilBERT$^{\S}$ (0.5) & $-$0.17 & +0.07 & +0.02 & $-$0.07 & $-$0.38 & +0.36 & +0.35 & +0.33 \\
martin-ha DistilBERT$^{\S}$ (gold's rate) & $-$0.13 & +0.15 & +0.02 & +0.04 & $-$0.49 & +0.47 & +0.20 & +0.18 \\
Dynabench hate (0.5) & $-$0.29 & +0.15 & $-$0.02 & $-$0.16 & $-$0.85 & +0.85 & +0.11 & +0.11 \\
Dynabench hate (gold's rate) & $-$0.28 & +0.20 & $-$0.02 & $-$0.10 & $-$0.86 & +0.83 & +0.11 & +0.08 \\
CardiffNLP offensive (0.5) & $-$0.11 & +0.36 & $-$0.01 & +0.24 & $-$0.73 & +0.63 & +0.07 & $-$0.04 \\
CardiffNLP offensive (gold's rate) & $-$0.15 & +0.13 & 0.00 & $-$0.02 & $-$0.57 & +0.52 & +0.22 & +0.17 \\
\midrule
One rater & -- & -- & -- & -- & $-$0.76 & +0.30 & +0.03 & $-$0.43 \\
\bottomrule
\end{tabular}
\end{table*}

\begin{table*}[tp]
\caption{All 14 classifier settings on Civil Comments reply threads: the placebo statistics and laws of Table~\ref{tab:placebo}, with 95\% intervals (placebo depth slope in points per level and as a share of the false-positive rate per level; depth laws in points per level, thread size laws in points per doubling). Gold's laws are $-$0.12, +0.15 and 2.39; each setting's $\kappa$ and error rates are in Table~\ref{tab:settings} and its log-RR distortion in Table~\ref{tab:contagion}. $^{\S}$Trained at least in part on Civil Comments labels.}
\label{tab:classifiers}
\footnotesize
\setlength{\tabcolsep}{4pt}
\begin{tabular}{@{}llcccccc@{}}
\toprule
 & & \multicolumn{3}{c}{Placebo statistics (gold negatives)} & \multicolumn{3}{c}{The law under the labeler} \\
\cmidrule(lr){3-5}\cmidrule(l){6-8}
Classifier & Cut & Depth slope & \% of FPR per level & Risk ratio & Depth law & Thread size law & Risk ratio \\
\midrule
Detoxify & 0.6 & +0.02 [+0.01, +0.03] & +1.4 [+0.5, +2.4] & 4.01 [3.67, 4.49] & $-$0.04 [$-$0.06, $-$0.01] & +0.06 [+0.02, +0.10] & 3.44 [3.32, 3.57] \\
 & gold's rate & +0.07 [+0.05, +0.09] & +1.9 [+1.3, +2.5] & 2.91 [2.80, 3.03] & +0.01 [$-$0.03, +0.05] & +0.22 [+0.16, +0.28] & 2.70 [2.64, 2.75] \\
\addlinespace[1pt]
Detoxify-unbiased$^{\S}$ & 0.6 & $-$0.01 [$-$0.02, 0.00] & $-$0.6 [$-$1.3, +0.1] & 2.43 [2.24, 2.67] & $-$0.14 [$-$0.17, $-$0.11] & +0.06 [0.00, +0.11] & 2.76 [2.70, 2.81] \\
 & gold's rate & $-$0.01 [$-$0.02, +0.01] & $-$0.3 [$-$1.0, +0.4] & 2.45 [2.28, 2.62] & $-$0.14 [$-$0.17, $-$0.10] & +0.09 [+0.03, +0.14] & 2.65 [2.59, 2.70] \\
\addlinespace[1pt]
IMSYPP & argmax & $-$0.74 [$-$0.83, $-$0.67] & $-$2.2 [$-$2.5, $-$2.0] & 1.75 [1.73, 1.76] & $-$0.81 [$-$0.90, $-$0.72] & +0.35 [+0.23, +0.48] & 1.72 [1.71, 1.73] \\
 & gold's rate & $-$0.34 [$-$0.37, $-$0.32] & $-$7.6 [$-$8.2, $-$7.2] & 2.46 [2.38, 2.55] & $-$0.50 [$-$0.54, $-$0.47] & $-$0.20 [$-$0.26, $-$0.15] & 2.47 [2.42, 2.53] \\
\addlinespace[1pt]
s-nlp RoBERTa$^{\S}$ & 0.5 & $-$0.02 [$-$0.04, $-$0.01] & $-$0.9 [$-$1.6, $-$0.3] & 2.56 [2.42, 2.71] & $-$0.15 [$-$0.19, $-$0.11] & +0.13 [+0.07, +0.20] & 2.67 [2.63, 2.72] \\
 & gold's rate & $-$0.02 [$-$0.04, $-$0.01] & $-$1.1 [$-$1.8, $-$0.3] & 2.64 [2.46, 2.82] & $-$0.15 [$-$0.18, $-$0.11] & +0.11 [+0.06, +0.18] & 2.76 [2.71, 2.81] \\
\addlinespace[1pt]
martin-ha DistilBERT$^{\S}$ & 0.5 & $-$0.01 [$-$0.02, +0.01] & $-$0.4 [$-$1.3, +0.3] & 3.74 [3.40, 4.08] & $-$0.09 [$-$0.12, $-$0.07] & +0.08 [+0.03, +0.13] & 3.33 [3.25, 3.41] \\
 & gold's rate & +0.01 [$-$0.02, +0.03] & +0.2 [$-$0.5, +0.9] & 3.08 [2.95, 3.22] & $-$0.09 [$-$0.13, $-$0.06] & +0.21 [+0.15, +0.27] & 2.85 [2.80, 2.91] \\
\addlinespace[1pt]
Dynabench hate & 0.5 & $-$0.17 [$-$0.20, $-$0.14] & $-$2.9 [$-$3.4, $-$2.4] & 2.82 [2.74, 2.88] & $-$0.17 [$-$0.21, $-$0.14] & $-$0.05 [$-$0.11, 0.00] & 2.66 [2.59, 2.72] \\
 & gold's rate & $-$0.14 [$-$0.18, $-$0.11] & $-$2.1 [$-$2.6, $-$1.6] & 2.71 [2.64, 2.78] & $-$0.15 [$-$0.18, $-$0.11] & +0.02 [$-$0.04, +0.08] & 2.58 [2.51, 2.63] \\
\addlinespace[1pt]
CardiffNLP offensive & 0.5 & $-$0.21 [$-$0.26, $-$0.17] & $-$2.0 [$-$2.5, $-$1.6] & 2.38 [2.34, 2.43] & $-$0.32 [$-$0.38, $-$0.26] & +0.29 [+0.19, +0.38] & 2.30 [2.27, 2.33] \\
 & gold's rate & $-$0.09 [$-$0.11, $-$0.06] & $-$2.4 [$-$3.0, $-$1.8] & 3.04 [2.92, 3.17] & $-$0.21 [$-$0.25, $-$0.17] & +0.11 [+0.05, +0.17] & 2.83 [2.77, 2.88] \\
\bottomrule
\end{tabular}
\end{table*}

\begin{table*}[tp]
\caption{Escalation under alternatives: the depth law under human labels and each LLM's distortion of it (points per level, 95\% intervals), with each setting's flag rate and Cohen's $\kappa$ against gold. Haiku on all reply threads unless marked; Sonnet and rows marked $^{\dagger}$ on the 5\% sample. Among replies and shown the parent: the depth law among replies, with flag rate and $\kappa$ among replies. In the rows for other readings of the votes, the human column is that reading and the distortion is the LLM's law minus it; the noise-free majority is model-based (Appendix~\ref{app:humans}). Vote-fraction instrument: the comments with the largest shares of toxic votes, as many as the LLM flags. $^{\ddagger}$The cut that maximizes $\kappa$ on 500 gold labels: median over 500 draws, with the 5th and 95th percentiles over draws. The main prompt rerun with the variants was labeled in one batch with them, and it and the next three rows are the four candidate prompts of Section~\ref{sec:oppoints}; the earlier wording of the note was labeled in the batch of the later relabeling. The second run labels the reply threads of the sampled articles, including the threads that the 5\% sample leaves out; its flag rates and $\kappa$ are over all 88{,}951 comments of those articles, single comments included.}
\label{tab:esc}
\scriptsize
\setlength{\tabcolsep}{3.5pt}
\begin{tabular}{@{}lcccccccc@{}}
\toprule
 & \multicolumn{2}{c}{Human labels} & \multicolumn{3}{c}{Haiku} & \multicolumn{3}{c}{Sonnet$^{\dagger}$} \\
\cmidrule(lr){2-3}\cmidrule(lr){4-6}\cmidrule(l){7-9}
Setting & Flags (\%) & Law & Flags (\%) & $\kappa$ & Distortion & Flags (\%) & $\kappa$ & Distortion \\
\midrule
Main prompt, own cut & 7.6 & $-$0.12 [$-$0.16, $-$0.09] & 22.3 & 0.25 & +1.05 [+0.98, +1.12] &  &  &  \\
\quad 5\% sample$^{\dagger}$ & 7.5 & $-$0.16 [$-$0.34, 0.00] & 22.3 & 0.24 & +1.19 [+0.90, +1.48] & 27.1 & 0.23 & +1.40 [+1.11, +1.67] \\
Within threads &  & $-$0.18 [$-$0.22, $-$0.14] & 22.3 & 0.25 & +0.68 [+0.61, +0.75] &  &  &  \\
\quad 5\% sample$^{\dagger}$ &  & $-$0.26 [$-$0.42, $-$0.09] & 22.3 & 0.24 & +0.71 [+0.40, +0.97] & 27.1 & 0.23 & +1.05 [+0.77, +1.34] \\
Among replies & 7.5 & $-$0.10 [$-$0.14, $-$0.06] & 22.4 & 0.25 & +1.29 [+1.21, +1.37] &  &  &  \\
\quad 5\% sample$^{\dagger}$ & 7.4 & $-$0.12 [$-$0.30, +0.07] & 22.4 & 0.24 & +1.45 [+1.11, +1.78] & 27.6 & 0.23 & +1.41 [+1.10, +1.79] \\
\quad shown the parent &  &  & 30.9 & 0.18 & +1.59 [+1.50, +1.69] &  &  &  \\
\quad shown the parent, 5\% sample$^{\dagger}$ &  &  & 30.6 & 0.19 & +1.78 [+1.41, +2.17] & 30.0 & 0.20 & +1.42 [+1.10, +1.78] \\
\quad shown the parent, vs.\ CCC's raters & 1.2 & $-$0.13 [$-$0.24, $-$0.01] & 30.4 & -- & +2.05 [+1.47, +2.66] &  &  &  \\
\midrule
\multicolumn{9}{@{}l}{\emph{Other readings of the human votes}} \\
Noise-free majority (infinitely many raters) & 6.5 & $-$0.15 [$-$0.33, +0.04] &  &  &  &  &  &  \\
Vote-fraction instrument, Haiku's flag rate & 22.3 & $-$0.07 [$-$0.13, $-$0.01] & 22.3 & 0.25 & +1.01 [+0.93, +1.08] &  &  &  \\
Vote-fraction instrument, Sonnet's rate$^{\dagger}$ & 27.1 & $-$0.28 [$-$0.62, +0.06] &  &  &  & 27.1 & 0.23 & +1.52 [+1.22, +1.82] \\
\midrule
\multicolumn{9}{@{}l}{\emph{Other labeling settings, on the 5\% sample}} \\
Relabeled in a later batch &  &  & 22.3 & 0.25 & +1.20 [+0.90, +1.49] & 27.1 & 0.23 & +1.42 [+1.10, +1.74] \\
0--100 rating, cut maximizing $\kappa$ &  &  & 8.2 & 0.31 & +0.41 [+0.39, +0.51]$^{\ddagger}$ & 8.4 & 0.33 & +0.52 [+0.34, +0.88]$^{\ddagger}$ \\
0--100 rating, gold's flag rate &  &  & 7.5 & 0.31 & +0.39 [+0.21, +0.54] & 7.5 & 0.33 & +0.49 [+0.32, +0.64] \\
``Very toxic'' only &  &  & 0.9 & 0.09 & +0.20 [+0.08, +0.35] & 0.4 & 0.04 & +0.14 [0.00, +0.31] \\
Main prompt, rerun with the variants below &  &  & 22.2 & 0.25 & +1.24 [+0.93, +1.55] &  &  &  \\
Eight examples (best of the four candidates) &  &  & 22.8 & 0.27 & +1.53 [+1.21, +1.86] &  &  &  \\
Sixteen examples &  &  & 26.3 & 0.24 & +1.58 [+1.22, +1.92] &  &  &  \\
Extended thinking &  &  & 24.0 & 0.23 & +2.07 [+1.74, +2.43] &  &  &  \\
Told address is not by itself toxic &  &  & 20.4 & 0.25 & +1.50 [+1.20, +1.81] &  &  &  \\
Third-person instructions &  &  & 22.6 & 0.24 & +1.31 [+0.99, +1.61] &  &  &  \\
Told addressing someone is not toxic (earlier wording) &  &  & 21.3 & 0.25 & +1.56 [+1.26, +1.85] & 24.2 & 0.25 & +1.30 [+0.98, +1.61] \\
\midrule
\multicolumn{9}{@{}l}{\emph{Second run on the sampled articles (62{,}943 comments)}} \\
Main prompt & 8.1 & $-$0.15 [$-$0.30, +0.02] & 22.5 & 0.26 & +1.18 [+0.90, +1.48] &  &  &  \\
Yes/no question &  &  & 20.6 & 0.29 & +0.91 [+0.63, +1.22] &  &  &  \\
Reversed answer scale &  &  & 20.1 & 0.28 & +1.07 [+0.80, +1.38] &  &  &  \\
\bottomrule
\end{tabular}
\end{table*}

\begin{table*}[tp]
\caption{Every LLM at its operating points on the 5\% sample: own cut (for the four-level answer, the cut that maximizes $\kappa$ against gold), ``very toxic'' only (the cut whose flag rate is closest to gold's), gold's flag rate, the $\kappa$-maximizing cut of the 0--100 rating and of the open models' scores against all gold labels of the sample, and the 0--100 cuts chosen on 500 gold labels. Distortions, the labeler's law minus gold's, with 95\% intervals: the depth law pooled, within threads and at a fixed number of second-person words (points per level), the thread size law (points per doubling) and the log reply risk ratio; last, how many times as often the labeler flags gold negatives with four or more second-person words as those with none. $^{\ddagger}$Median over 500 draws of 500 gold labels, with the 5th and 95th percentiles over draws. $^{\P}$Within-thread distortion of the same rating cut with ties broken by a different random draw. Sonnet's own-cut 4+\,:\,0 ratio is computed from stored rates and has no interval. Gold's laws here: depth $-$0.16, thread size +0.08.}
\label{tab:oppoints}
\scriptsize
\setlength{\tabcolsep}{2.4pt}
\begin{tabular}{@{}llcccccccc@{}}
\toprule
 & & & & \multicolumn{5}{c}{Distortion} & \\
\cmidrule(lr){5-9}
LLM & Operating point & Flags (\%) & $\kappa$ & Depth law & Within threads & At fixed no. & Thread size law & Log RR & 4+\,:\,0 \\
\midrule
Haiku & own cut (max $\kappa$) & 22.3 & 0.24 & +1.19 [+0.90, +1.48] & +0.71 [+0.40, +0.97] & +0.07 [$-$0.16, +0.31] & +1.91 [+1.45, +2.31] & $-$0.21 [$-$0.31, $-$0.11] & 4.1 [3.9, 4.4] \\
 & ``very toxic'' (prevalence) & 0.9 & 0.09 & +0.20 [+0.08, +0.35] & -- & +0.31 [+0.20, +0.46] & $-$0.04 [$-$0.30, +0.20] & +1.19 [+0.63, +1.54] & -- \\
Haiku-100 & gold's rate & 7.5 & 0.31 & +0.39 [+0.21, +0.54] & +0.35 [+0.16, +0.56]$^{\P}$ & +0.09 [$-$0.08, +0.22] & +0.34 [+0.08, +0.61] & +0.06 [$-$0.07, +0.19] & 5.1 [4.6, 5.7] \\
 & max $\kappa$ (cut 43) & 8.2 & 0.32 & +0.40 [+0.20, +0.56] & -- & +0.07 [$-$0.11, +0.21] & +0.41 [+0.14, +0.70] & +0.03 [$-$0.09, +0.16] & -- \\
 & max $\kappa$, 500 labels & 8.2 & 0.31 & +0.41 [+0.39, +0.51]$^{\ddagger}$ & -- & -- & -- & -- & -- \\
 & prevalence, 500 labels & 8.2 & 0.31 & +0.40 [+0.39, +0.45]$^{\ddagger}$ & -- & -- & -- & -- & -- \\
Sonnet & own cut (max $\kappa$) & 27.1 & 0.23 & +1.40 [+1.11, +1.67] & +1.05 [+0.77, +1.34] & +0.34 [+0.08, +0.61] & +1.97 [+1.54, +2.39] & $-$0.23 [$-$0.32, $-$0.15] & 3.1 \\
 & ``very toxic'' (prevalence) & 0.4 & 0.04 & +0.14 [0.00, +0.31] & -- & +0.32 [+0.19, +0.48] & $-$0.12 [$-$0.37, +0.14] & +1.20 [$-$0.24, +1.97] & -- \\
Sonnet-100 & gold's rate & 7.5 & 0.33 & +0.49 [+0.32, +0.64] & +0.57 [+0.37, +0.76]$^{\P}$ & +0.25 [+0.09, +0.39] & +0.37 [+0.08, +0.66] & +0.21 [+0.09, +0.33] & 4.3 [3.8, 4.8] \\
 & max $\kappa$ (cut 43) & 7.7 & 0.33 & +0.48 [+0.32, +0.63] & -- & +0.23 [+0.07, +0.37] & +0.39 [+0.10, +0.67] & +0.19 [+0.08, +0.31] & -- \\
 & max $\kappa$, 500 labels & 8.4 & 0.33 & +0.52 [+0.34, +0.88]$^{\ddagger}$ & -- & -- & -- & -- & -- \\
 & prevalence, 500 labels & 7.7 & 0.33 & +0.48 [+0.34, +0.55]$^{\ddagger}$ & -- & -- & -- & -- & -- \\
\midrule
Qwen & gold's rate (prevalence) & 7.4 & 0.21 & +0.19 [+0.06, +0.34] & +0.16 [$-$0.05, +0.35] & $-$0.01 [$-$0.15, +0.15] & +0.15 [$-$0.13, +0.44] & $-$0.17 [$-$0.30, $-$0.06] & 3.3 [3.0, 3.7] \\
 & max $\kappa$ & 15.3 & 0.22 & +0.05 [$-$0.14, +0.25] & -- & $-$0.39 [$-$0.57, $-$0.19] & +0.33 [$-$0.03, +0.66] & $-$0.35 [$-$0.43, $-$0.25] & -- \\
Gemma & gold's rate (prevalence) & 7.5 & 0.26 & $-$0.07 [$-$0.23, +0.10] & $-$0.23 [$-$0.41, $-$0.05] & $-$0.30 [$-$0.45, $-$0.13] & $-$0.03 [$-$0.32, +0.28] & +0.14 [+0.01, +0.27] & 3.6 [3.2, 4.0] \\
 & max $\kappa$ & 12.6 & 0.28 & $-$0.18 [$-$0.38, +0.02] & -- & $-$0.59 [$-$0.79, $-$0.39] & +0.13 [$-$0.18, +0.47] & $-$0.06 [$-$0.17, +0.04] & -- \\
OLMo-2 & gold's rate (prevalence) & 7.5 & 0.24 & +0.23 [+0.08, +0.41] & +0.08 [$-$0.13, +0.28] & $-$0.04 [$-$0.19, +0.15] & +0.39 [+0.10, +0.68] & $-$0.10 [$-$0.22, +0.04] & 4.0 [3.6, 4.4] \\
 & max $\kappa$ & 10.4 & 0.24 & +0.20 [+0.05, +0.39] & -- & $-$0.19 [$-$0.34, +0.01] & +0.51 [+0.22, +0.84] & $-$0.22 [$-$0.32, $-$0.10] & -- \\
Granite & gold's rate (prevalence) & 7.5 & 0.30 & +0.18 [+0.03, +0.32] & +0.24 [+0.06, +0.42] & 0.00 [$-$0.15, +0.13] & +0.11 [$-$0.14, +0.39] & $-$0.03 [$-$0.14, +0.08] & 3.3 [3.0, 3.7] \\
 & max $\kappa$ & 10.5 & 0.31 & +0.19 [$-$0.01, +0.35] & -- & $-$0.11 [$-$0.29, +0.03] & +0.20 [$-$0.12, +0.50] & $-$0.08 [$-$0.19, +0.02] & -- \\
Llama & gold's rate (prevalence) & 7.5 & 0.30 & +0.26 [+0.10, +0.42] & +0.08 [$-$0.13, +0.27] & $-$0.07 [$-$0.21, +0.09] & +0.28 [0.00, +0.55] & +0.01 [$-$0.12, +0.13] & 5.0 [4.5, 5.6] \\
 & max $\kappa$ & 8.9 & 0.31 & +0.22 [+0.05, +0.38] & -- & $-$0.17 [$-$0.33, $-$0.02] & +0.35 [+0.04, +0.62] & $-$0.03 [$-$0.15, +0.08] & -- \\
\bottomrule
\end{tabular}
\end{table*}

\begin{table*}[tp]
\caption{Contagion under stricter or more lenient gold. Gold$_t$ calls a comment toxic when at least a share $t$ of its raters do; gold's share of toxic comments and reply risk ratio (Civil Comments reply threads). Each classifier, and the Haiku and Sonnet 0--100 ratings, flags as many comments as gold$_t$: the median over the seven classifiers of the log risk ratio minus gold$_t$'s, and how many of the seven have a 95\% CI above 0; the ratings' log-RR distortions (95\% CI) on the 5\% sample, where gold$_t$'s risk ratio is given next. Last two columns: the seven classifier settings that flag as many comments as gold$_{0.5}$, and Detoxify at 0.6, against gold$_t$ (log RR minus gold$_t$'s; range over the seven). At $t = 0.5$ the ratings' values differ slightly from Table~\ref{tab:llms}'s because ties at the cut are broken by a different draw. $^{\dagger}$5\% sample.}
\label{tab:thrrr}
\footnotesize
\begin{tabular}{@{}lccccccccc@{}}
\toprule
 & \multicolumn{2}{c}{Gold$_t$} & \multicolumn{2}{c}{Classifiers at gold$_t$'s rate} & \multicolumn{3}{c}{Ratings at gold$_t$'s rate$^{\dagger}$} & \multicolumn{2}{c}{At gold$_{0.5}$'s rate} \\
\cmidrule(lr){2-3}\cmidrule(lr){4-5}\cmidrule(lr){6-8}\cmidrule(l){9-10}
$t$ & Toxic & RR & Median & Above 0 & Haiku-100 & Sonnet-100 & Gold RR$^{\dagger}$ & Classifiers & Detoxify (0.6) \\
\midrule
0.3 & 14.3\% & 1.96 & +0.11 & 7 of 7 & +0.05 [$-$0.02, +0.12] & +0.14 [+0.07, +0.21] & 1.96 & -- & -- \\
0.4 & 10.8\% & 2.15 & +0.11 & 7 of 7 & +0.03 [$-$0.07, +0.13] & +0.13 [+0.04, +0.22] & 2.14 & -- & -- \\
0.5 & 7.6\% & 2.39 & +0.12 & 7 of 7 & +0.03 [$-$0.09, +0.15] & +0.22 [+0.10, +0.33] & 2.28 & +0.03 to +0.18 & +0.36 \\
0.6 & 5.0\% & 2.74 & +0.17 & 5 of 7 & +0.04 [$-$0.16, +0.24] & +0.27 [+0.08, +0.45] & 2.59 & $-$0.10 to +0.04 & +0.23 \\
0.7 & 3.0\% & 3.15 & +0.23 & 6 of 7 & +0.27 [$-$0.03, +0.61] & +0.40 [+0.17, +0.68] & 2.74 & $-$0.24 to $-$0.10 & +0.09 \\
0.8 & 1.6\% & 3.72 & +0.30 & 6 of 7 & +0.73 [+0.18, +1.30] & +0.87 [+0.45, +1.39] & 2.23 & $-$0.41 to $-$0.27 & $-$0.08 \\
\bottomrule
\end{tabular}
\end{table*}

\begin{table*}[tp]
\caption{How each labeler responds to editing ``you'' on the edit sample: its flag rate on clearly non-toxic replies at depth 2 or more, and its change when \emph{they} and its forms are rewritten as ``you'' (comments without ``you'') and when ``you'' is rewritten in the third person (1{,}489 comments with ``you''), in points and relative to the base rate (\%); last column, the labeler's relative placebo depth slope (\% of its FPR per level). Classifiers at their usual threshold and at gold's flag rate; relative changes rest on few flags where the base rate is below 1\%. Rank correlation of the change on adding ``you'' with the relative placebo depth slope: +0.74 over all 16 labelers (+0.90 for the relative change). $^{\S}$Trained at least in part on Civil Comments labels.}
\label{tab:rewritecls}
\footnotesize
\begin{tabular}{@{}llccccccc@{}}
\toprule
 & & \multicolumn{3}{c}{Adding ``you''} & \multicolumn{3}{c}{Removing ``you''} & Placebo slope \\
\cmidrule(lr){3-5}\cmidrule(lr){6-8}
Labeler & Cut & Base (\%) & Change & Relative (\%) & Base (\%) & Change & Relative (\%) & (\% of FPR) \\
\midrule
Haiku & own cut & 9.8 & +4.8 [+3.7, +6.1] & +49 & 28.4 & $-$8.9 [$-$10.5, $-$7.5] & $-$31 & +5.4 \\
Sonnet & own cut & 13.8 & +6.3 [+4.7, +7.8] & +45 & 33.6 & $-$11.8 [$-$13.6, $-$10.0] & $-$35 & +5.9 \\
\midrule
Detoxify & 0.6 & 0.3 & +0.9 [+0.5, +1.3] & +325 & 0.8 & $-$0.7 [$-$1.2, $-$0.3] & $-$92 & +1.4 \\
 & gold's rate & 1.1 & +1.3 [+0.7, +1.9] & +119 & 2.7 & $-$1.9 [$-$2.7, $-$1.3] & $-$73 & +1.9 \\
\addlinespace[1pt]
Detoxify-unbiased$^{\S}$ & 0.6 & 0.6 & +0.1 [0.0, +0.3] & +22 & 0.7 & $-$0.3 [$-$0.5, $-$0.1] & $-$36 & $-$0.6 \\
 & gold's rate & 0.9 & +0.1 [0.0, +0.3] & +15 & 1.0 & $-$0.3 [$-$0.5, $-$0.1] & $-$27 & $-$0.3 \\
\addlinespace[1pt]
IMSYPP & argmax & 29.3 & +2.7 [+1.5, +3.7] & +9 & 30.4 & $-$0.9 [$-$2.1, +0.2] & $-$3 & $-$2.2 \\
 & gold's rate & 3.3 & $-$0.3 [$-$0.7, +0.1] & $-$8 & 3.3 & $-$0.4 [$-$0.9, 0.0] & $-$12 & $-$7.6 \\
\addlinespace[1pt]
s-nlp RoBERTa$^{\S}$ & 0.5 & 1.1 & +0.1 [0.0, +0.3] & +12 & 1.1 & $-$0.2 [$-$0.5, 0.0] & $-$19 & $-$0.9 \\
 & gold's rate & 1.0 & +0.1 [0.0, +0.2] & +7 & 0.8 & $-$0.1 [$-$0.5, +0.2] & $-$17 & $-$1.1 \\
\addlinespace[1pt]
martin-ha DistilBERT$^{\S}$ & 0.5 & 0.5 & +0.3 [+0.1, +0.7] & +71 & 0.4 & $-$0.1 [$-$0.2, 0.0] & $-$17 & $-$0.4 \\
 & gold's rate & 1.2 & +0.5 [+0.2, +0.9] & +44 & 1.0 & $-$0.1 [$-$0.3, 0.0] & $-$13 & +0.2 \\
\addlinespace[1pt]
Dynabench hate & 0.5 & 6.3 & $-$0.6 [$-$1.2, $-$0.1] & $-$9 & 4.9 & +0.3 [$-$0.3, +1.1] & +7 & $-$2.9 \\
 & gold's rate & 7.0 & +0.2 [$-$0.5, +0.9] & +3 & 5.6 & +0.7 [+0.1, +1.4] & +13 & $-$2.1 \\
\addlinespace[1pt]
CardiffNLP offensive & 0.5 & 5.3 & +0.5 [0.0, +0.9] & +9 & 7.6 & $-$0.3 [$-$0.7, 0.0] & $-$4 & $-$2.0 \\
 & gold's rate & 1.5 & $-$0.1 [$-$0.3, +0.1] & $-$4 & 2.4 & $-$0.3 [$-$0.7, +0.1] & $-$14 & $-$2.4 \\
\bottomrule
\end{tabular}
\end{table*}

\begin{table*}[tp]
\begin{minipage}[t]{0.485\textwidth}
\caption{The 0--100 ratings' depth law against other definitions of the human label, on the 5\% sample (points per level, 95\% intervals). Top: gold is the majority at vote threshold $t$; each rating flags as many comments as gold$_t$ calls toxic. Bottom: at a fixed flag rate, the vote-fraction instrument flags the comments with the largest shares of toxic votes. Columns: the reference's flag rate and depth law, and each rating's depth law minus the reference's. At $t = 0.5$ the ratings' values differ by a few hundredths from Table~\ref{tab:llms}'s, because ties at the cut are broken by a different draw.}
\label{tab:thrdepth}
\scriptsize
\setlength{\tabcolsep}{1.5pt}
\begin{tabular}{@{}lcccc@{}}
\toprule
Reference & Flags (\%) & Law & Haiku-100 $-$ ref. & Sonnet-100 $-$ ref. \\
\midrule
\multicolumn{5}{@{}l}{\emph{Gold defined at vote threshold $t$; each rating flags gold$_t$'s share}} \\
$t = 0.3$ & 14.2 & $-$0.20 & +0.51 [+0.29, +0.73] & +0.78 [+0.59, +1.01] \\
$t = 0.4$ & 10.7 & $-$0.12 & +0.45 [+0.28, +0.63] & +0.54 [+0.37, +0.74] \\
$t = 0.5$ & 7.5 & $-$0.16 & +0.40 [+0.24, +0.58] & +0.48 [+0.33, +0.64] \\
$t = 0.6$ & 5.0 & $-$0.12 & +0.38 [+0.23, +0.55] & +0.34 [+0.20, +0.49] \\
$t = 0.7$ & 3.0 & $-$0.07 & +0.26 [+0.13, +0.39] & +0.17 [+0.08, +0.29] \\
$t = 0.8$ & 1.7 & $-$0.06 & +0.13 [+0.03, +0.23] & +0.11 [+0.03, +0.20] \\
\midrule
\multicolumn{5}{@{}l}{\emph{Gold at $t = 0.5$; the vote-fraction instrument and each rating flag the same share}} \\
Vote-fraction instrument & 2.0 & $-$0.06 [$-$0.12, 0.00] & +0.16 [+0.06, +0.28] & +0.12 [+0.04, +0.21] \\
 & 5.0 & $-$0.12 [$-$0.23, +0.02] & +0.37 [+0.22, +0.56] & +0.32 [+0.17, +0.48] \\
 & 7.5 & $-$0.15 [$-$0.29, +0.05] & +0.39 [+0.24, +0.59] & +0.46 [+0.29, +0.64] \\
 & 10.0 & $-$0.11 [$-$0.30, +0.10] & +0.42 [+0.22, +0.62] & +0.50 [+0.32, +0.71] \\
 & 15.0 & $-$0.20 [$-$0.45, +0.10] & +0.52 [+0.31, +0.73] & +0.84 [+0.62, +1.05] \\
 & 20.0 & $-$0.27 [$-$0.54, +0.04] & +0.62 [+0.38, +0.85] & +1.10 [+0.85, +1.35] \\
 & 30.0 & $-$0.27 [$-$0.57, +0.09] & +0.57 [+0.24, +0.86] & +1.33 [+1.01, +1.65] \\
\bottomrule
\end{tabular}
\end{minipage}\hfill
\begin{minipage}[t]{0.485\textwidth}
\caption{Minimal pairs. Of the comments in the edit sample that no rater called toxic, 64 are toxic to Haiku as written and ``not toxic'' after the fixed rewrite; the 12 shortest that contain no name, initials or user handle, with links shown as [link].}
\label{tab:pairs}
\scriptsize
\begin{tabular}{@{}p{0.47\columnwidth}p{0.47\columnwidth}@{}}
\toprule
As written (flagged) & Rewritten (``not toxic'') \\
\midrule
You're lecturing a fence post. & They're lecturing a fence post. \\
Dude, are you stalking him? Kinda creepy. & Dude, are they stalking him? Kinda creepy. \\
Clearly, you ate lead paint chips as a child. & Clearly, they ate lead paint chips as a child. \\
``Privilege'' is not a race, your question is nonsense. & ``Privilege'' is not a race, their question is nonsense. \\
you have absolutely no idea what you are talking about & they have absolutely no idea what they are talking about \\
Because you were duped by Russian fake news on facebook. & Because they were duped by Russian fake news on facebook. \\
the left gets confused when you present them with facts. & the left gets confused when they present them with facts. \\
yeah, you better get back to the Enquirer, they miss you. & yeah, they better get back to the Enquirer, they miss them. \\
The Conflict of Interest Act is a law. Do you know what an Act is? & The Conflict of Interest Act is a law. Do they know what an Act is? \\
I disagree with you all the time because you're wrong all the time. & I disagree with them all the time because they're wrong all the time. \\
I recommend you also read this awesome article [link] & I recommend they also read this awesome article [link] \\
actually, it's none of your business either but you fail to comprehend that. & actually, it's none of their business either but they fail to comprehend that. \\
\bottomrule
\end{tabular}
\end{minipage}
\end{table*}

\paragraph{Where the classifiers' errors cluster.} Contagion partly reflects which articles a thread belongs to: against a permutation of the labels within each article, 64\% of gold's log risk ratio lies within articles and 36\% between them. Within articles, the classifiers' inflation is mixed: nine of the 14 settings exceed gold's within-article excess and five fall below it, while all of them have placebo excesses of +0.32 to +0.85 within articles, against +0.21 for the single rater. Identity terms are not the cause: comments that mention an identity group (14\% of all) raise the classifiers' false-positive rates up to 3.6-fold, but leaving out the edges with such a mention changes their placebo risk ratios by $-$7\% to +6\%. The classifiers' false positives cluster both within and between articles, and not through identity terms.

\paragraph{Escalation under alternatives.} Table~\ref{tab:esc} gives the depth law under the main readings of the human votes (Table~\ref{tab:thrdepth} gives gold at other vote thresholds and the vote-fraction instrument at other flag rates), and Haiku's and Sonnet's distortion of it under each prompt, answer scale, operating point and labeling batch of Section~\ref{sec:escalation}.

\paragraph{Every operating point.} Table~\ref{tab:oppoints} gives every LLM on the 5\% sample at each operating point we have (Qwen's own cut, on all reply threads, is in Table~\ref{tab:placebo}): its own cut, ``very toxic'' alone, gold's flag rate, the cut that maximizes $\kappa$ against all of the sample's gold labels, and the 0--100 cuts that a study would choose on 500 gold labels. A calibration sample is a uniform random sample of 200, 500 or 1{,}000 comments with gold labels, 500 draws per size; the cut is the one that maximizes $\kappa$ (or F1) on it, ties going to the higher cut, or the one whose flag rate is closest to the sample's gold rate (prevalence matching), and the depth law is then measured on the whole population, with an article-clustered Wald test of its excess over gold's. With the four-level answer, $\kappa$-maximization picks Haiku's own cut in 99.8\% of draws at 1{,}000 labels (Section~\ref{sec:oppoints} gives 200 and 500), and the F1-maximizing cut is the same in 97\%. Prevalence matching picks ``very toxic''. On all reply threads its thread size law then falls below gold's ($-$0.12 points per doubling, CI $-$0.18 to $-$0.06; on the 5\% sample the distortion's interval covers 0, Table~\ref{tab:oppoints}): a labeler that flags almost nothing shrinks the depth and size laws, though its few flags cluster along reply edges (reply risk ratio 6.50 on all reply threads, a log-RR distortion of +1.00). The later relabeling (Table~\ref{tab:esc}, ``Relabeled in a later batch'') labels the 5\% sample again with identical settings; flags change on 2.4\% (Haiku) and 2.5\% (Sonnet) of comments ($\kappa$ between runs: Haiku 0.93, Sonnet 0.94), and Haiku's depth law moves by +0.02 points per level.

\paragraph{Other definitions of gold.} Table~\ref{tab:thrdepth} reads the 0--100 ratings against gold defined at every vote threshold from 0.3 to 0.8 and against the vote-fraction instrument at every flag rate from 2\% to 30\%: the ratings' depth law exceeds every reading of the votes. Table~\ref{tab:thrrr} does the same for contagion: stricter gold has more contagion, and the classifiers flagging as many comments as each gold report more still, while at gold's usual flag rate they report roughly the contagion of gold at 0.6.

\paragraph{The edits, labeler by labeler.} Table~\ref{tab:rewritecls} gives every labeler's response to the edit sample's rewrites. For the LLMs and most classifiers, the flag rate on clearly non-toxic replies at depth 2 or more rises when \emph{they} is rewritten as ``you'' and falls when ``you'' is rewritten in the third person, while the Dynabench hate model moves the other way; the response, absolute or relative to the labeler's base rate, tracks its relative placebo depth slope. Table~\ref{tab:pairs} shows comments that no rater called toxic and that the rewrite turns from toxic to ``not toxic'' for Haiku. Most are curt or rude, though no rater called them toxic; the rewrite changes only whom they address.

\FloatBarrier
\section{More Results on the Other Corpora}
\label{app:corpora}

\paragraph{Other corpora.} Haiku and Sonnet answer the main prompt, the toxicity question, verbatim on every comment of CAD, Qian et al.'s Reddit and Gab, CGA-Wiki and BAD. Shown the parent, a reply takes its label with the parent and roots keep their label without it (as do replies whose parent has no text); the depth law among replies uses depth $\ge 1$. The rewrite labels every gold-negative comment of the four forums that the detector matches, original and edited in one shuffled batch. Appendix~\ref{app:labelers} gives the main prompt and each corpus's own question.

\begin{table*}[tp]
\caption{Over-flagging of second-person address on six corpora, among gold negatives: how many points more often Haiku flags those with a second-person word than those without, asked the toxicity question and asked the corpus's own question (95\% CI), and how many points more often a second LLM asked that question does (95\% CI; at gold's flag rate except Sonnet; on Civil Comments, whose own question is the toxicity question, Qwen at gold's flag rate on the 5\% sample, as in Table~\ref{tab:llms}, computed from stored rates without an interval); Haiku's false-positive rate with three or more second-person words over that with none, among the gold negatives that Detoxify scores below the median of the corpus's gold negatives (95\% CI) and among all; the share of Haiku's depth-law distortion that disappears at a fixed number of second-person words (where the distortion is significantly positive; it also contains the shrinking of gold's law, so it differs from the share of the placebo slope that disappears); and the change in Haiku's flag rate, in points, when the same comments are rewritten in the third person, overall and at depth 2 or more (95\% CI). Except in the own-question columns, Haiku answers the toxicity question without the parent. BAD's own-question labels exist only with the previous turn shown, and the rewrite was not run on BAD. $^{\dagger}$5\% sample.}
\label{tab:corporayou}
\footnotesize
\setlength{\tabcolsep}{2.6pt}
\begin{tabular}{@{}lccl@{\,}ccccccc@{}}
\toprule
 & \multicolumn{4}{c}{Extra flags on ``you'' (points)} & \multicolumn{2}{c}{3+\,:\,0 false-positive ratio} & & \multicolumn{2}{c}{Rewrite (points)} \\
\cmidrule(lr){2-5}\cmidrule(lr){6-7}\cmidrule(l){9-10}
Corpus (gold construct) & Toxicity & Own question & \multicolumn{2}{c}{Second LLM} & Detoxify-clean & All & \begin{tabular}[b]{@{}c@{}}Distortion removed\\at fixed no.\end{tabular} & All & Depth $\ge 2$ \\
\midrule
Civil Comments (toxicity) & 16.9 & 16.9 [16.7, 17.1] & Qwen$^{\dagger}$ & 4.0 & 5.0 [4.9, 5.1] & 3.8 & 105\% & $-$7.7 [$-$8.1, $-$7.2]$^{\dagger}$ & $-$9.2 [$-$9.8, $-$8.5]$^{\dagger}$ \\
CAD (abuse) & 11.3 & 5.4 [4.2, 6.8] & Sonnet & 5.7 [4.4, 6.8] & 7.0 [3.8, 10.9] & 2.8 & -- & $-$3.7 [$-$4.6, $-$2.7] & $-$3.8 [$-$5.2, $-$2.3] \\
Qian et al.'s Reddit (hate) & 23.7 & 2.9 [2.1, 3.8] & Gemma & 1.9 [0.8, 2.9] & 3.8 [3.1, 4.5] & 2.5 & 55\% & $-$4.5 [$-$5.2, $-$3.8] & $-$5.5 [$-$6.3, $-$4.6] \\
Qian et al.'s Gab (hate) & 28.4 & 5.9 [4.7, 7.1] & Gemma & 5.0 [3.8, 6.2] & 3.9 [3.2, 4.5] & 2.5 & 21\% & $-$6.7 [$-$7.5, $-$5.8] & $-$8.6 [$-$9.7, $-$7.4] \\
CGA-Wiki (attack) & 16.5 & 7.8 [6.2, 9.4] & Qwen & 5.2 [4.1, 6.2] & 12.7 [7.8, 26.9] & 10.8 & -- & $-$8.1 [$-$9.3, $-$7.0] & $-$8.7 [$-$10.5, $-$7.1] \\
BAD (unsafe turn) & $-$0.2 & -- & -- & -- & 1.5 [0.3, 3.0] & 2.7 & -- & -- & -- \\
\bottomrule
\end{tabular}
\end{table*}

\begin{table*}[tp]
\caption{The bias across corpora under the toxicity question (Haiku and Sonnet at their own cut; Detoxify at gold's flag rate). Among gold negatives: the share with a second-person word at the root and at depth 6 or more; how many points more often each labeler flags those with one (Sonnet's Civil Comments value on the 5\% sample); and Haiku's placebo depth slope (points per level, 95\% CI), overall and at a fixed number of second-person words. Last, the depth law under gold and Haiku.}
\label{tab:platforms}
\footnotesize
\setlength{\tabcolsep}{3.6pt}
\begin{tabular}{@{}lccccccccc@{}}
\toprule
 & \multicolumn{2}{c}{``You'' share} & \multicolumn{3}{c}{Extra flags on ``you'' (points)} & \multicolumn{2}{c}{Haiku's placebo depth slope} & \multicolumn{2}{c}{Depth law} \\
\cmidrule(lr){2-3}\cmidrule(lr){4-6}\cmidrule(lr){7-8}\cmidrule(l){9-10}
Corpus (gold construct) & Root & Depth 6+ & Haiku & Sonnet & Detoxify & All & At fixed no. & Gold & Haiku \\
\midrule
Civil Comments (toxicity) & 23\% & 61\% & 16.9 & 17.2 & 3.0 & +1.02 [+0.93, +1.09] & $-$0.16 & $-$0.12 & +0.93 \\
CAD (abuse) & 9\% & 36\% & 11.3 & 9.2 & 5.7 & $-$0.02 [$-$0.38, +0.34] & $-$0.40 & +0.41 & +0.48 \\
Qian et al.'s Reddit (hate) & 28\% & 57\% & 23.7 & 21.8 & 9.2 & +0.70 [+0.33, +1.06] & $-$0.42 & $-$2.52 & $-$1.12 \\
Qian et al.'s Gab (hate) & 20\% & 50\% & 28.4 & 26.3 & 15.6 & +2.12 [+1.78, +2.50] & +1.04 & $-$1.95 & +0.24 \\
CGA-Wiki (attack) & 57\% & 50\% & 16.5 & 15.6 & 0.6 & +0.88 [+0.33, +1.50] & +1.04 & +3.74 & +3.63 \\
BAD (unsafe turn) & 63\% & 61\% & $-$0.2 & $-$0.4 & $-$7.4 & +0.01 [$-$0.03, +0.04] & 0.00 & +1.02 & $-$0.02 \\
\bottomrule
\end{tabular}
\end{table*}

\begin{table*}[tp]
\caption{Depth laws on six corpora under gold (each corpus's own construct) and under Haiku and Sonnet asked the toxicity question without the parent (points per level): over all comments, among replies and within threads. $^{*}$The labeler's law differs from gold's (95\% CI of the paired difference excludes 0). Last column: of the settings without the parent (Haiku and the open LLMs asked the corpus's own question, and the 14 classifier settings), how many report a weaker within-thread law than gold's, on the four corpora where gold's rises significantly within threads. On Civil Comments, Sonnet's values are on the 5\% sample, where gold's laws are $-$0.16, $-$0.12 and $-$0.26.}
\label{tab:corporadepth}
\footnotesize
\begin{tabular}{@{}lcccccccccc@{}}
\toprule
 & \multicolumn{3}{c}{All comments} & \multicolumn{3}{c}{Among replies} & \multicolumn{3}{c}{Within threads} & Weaker \\
\cmidrule(lr){2-4}\cmidrule(lr){5-7}\cmidrule(lr){8-10}
Corpus (gold construct) & Gold & Haiku & Sonnet & Gold & Haiku & Sonnet & Gold & Haiku & Sonnet & within \\
\midrule
Civil Comments (toxicity) & $-$0.12 & +0.93$^{*}$ & +1.24$^{*}$ & $-$0.10 & +1.19$^{*}$ & +1.29$^{*}$ & $-$0.18 & +0.50$^{*}$ & +0.79$^{*}$ & -- \\
CAD (abuse) & +0.41 & +0.48 & +0.35 & $-$0.21 & $-$0.97$^{*}$ & $-$1.11$^{*}$ & +0.24 & $-$0.35$^{*}$ & $-$0.29$^{*}$ & -- \\
Qian et al.'s Reddit (hate) & $-$2.52 & $-$1.12$^{*}$ & $-$1.06$^{*}$ & $-$0.80 & +0.82$^{*}$ & +0.78$^{*}$ & +1.02 & +0.85 & +0.71 & 16/16 \\
Qian et al.'s Gab (hate) & $-$1.95 & +0.24$^{*}$ & +0.28$^{*}$ & $-$3.98 & $-$1.40$^{*}$ & $-$1.39$^{*}$ & +5.67 & +3.82$^{*}$ & +3.56$^{*}$ & 16/16 \\
CGA-Wiki (attack) & +3.74 & +3.63 & +3.49 & +3.22 & +3.00 & +2.68 & +7.27 & +6.24$^{*}$ & +6.15$^{*}$ & 16/16 \\
BAD (unsafe turn) & +1.02 & $-$0.02$^{*}$ & +0.04$^{*}$ & +1.41 & +0.36$^{*}$ & +0.51$^{*}$ & +1.00 & $-$0.02$^{*}$ & +0.04$^{*}$ & 14/14 \\
\bottomrule
\end{tabular}
\end{table*}

\begin{table*}[tp]
\begin{minipage}[t]{0.485\textwidth}
\caption{The thread size law (points per doubling) and the log reply risk ratio on six corpora, under gold (each corpus's own construct) and under Haiku and Sonnet asked the toxicity question without the parent. $^{*}$The labeler's law differs from gold's (95\% CI of the paired difference excludes 0). On Civil Comments, Sonnet's values are on the 5\% sample, where gold's laws are +0.08 and +0.82. CGA-Wiki's conversations end at the attack, so it has no reply risk ratio.}
\label{tab:corporasizerr}
\footnotesize
\setlength{\tabcolsep}{3pt}
\begin{tabular}{@{}lcccccc@{}}
\toprule
 & \multicolumn{3}{c}{Thread size law} & \multicolumn{3}{c}{Log reply RR} \\
\cmidrule(lr){2-4}\cmidrule(l){5-7}
Corpus & Gold & Haiku & Sonnet & Gold & Haiku & Sonnet \\
\midrule
Civil Comments (toxicity) & +0.15 & +1.93$^{*}$ & +2.05$^{*}$ & +0.87 & +0.59$^{*}$ & +0.59$^{*}$ \\
CAD (abuse) & +1.64 & +2.98$^{*}$ & +2.99$^{*}$ & +0.84 & +0.37$^{*}$ & +0.43$^{*}$ \\
Qian et al.'s Reddit (hate) & $-$10.30 & $-$6.95$^{*}$ & $-$6.84$^{*}$ & +0.13 & +0.21$^{*}$ & +0.27$^{*}$ \\
Qian et al.'s Gab (hate) & $-$11.98 & $-$4.67$^{*}$ & $-$3.94$^{*}$ & $-$0.38 & $-$0.08$^{*}$ & $-$0.04$^{*}$ \\
CGA-Wiki (attack) & $-$2.51 & +0.75$^{*}$ & +0.56$^{*}$ & -- & -- & -- \\
BAD (unsafe turn) & +5.91 & +0.16$^{*}$ & +0.22$^{*}$ & +0.19 & $-$0.65$^{*}$ & $-$0.72$^{*}$ \\
\bottomrule
\end{tabular}
\end{minipage}\hfill
\begin{minipage}[t]{0.485\textwidth}
\caption{The thread size law (points per doubling) and the log reply risk ratio under gold and under Haiku asked each corpus's own question without the parent. $^{*}$The labeler's law differs from gold's (95\% CI of the paired difference excludes 0). $^{\|}$With the previous turn shown: BAD's own-question labels exist only with it. CGA-Wiki's conversations end at the attack, so it has no reply risk ratio.}
\label{tab:ownlaws}
\footnotesize
\begin{tabular}{@{}lcccc@{}}
\toprule
 & \multicolumn{2}{c}{Thread size law} & \multicolumn{2}{c}{Log reply RR} \\
\cmidrule(lr){2-3}\cmidrule(l){4-5}
Corpus & Gold & Haiku & Gold & Haiku \\
\midrule
CAD (abuse) & +1.64 & +2.77$^{*}$ & +0.84 & +0.47$^{*}$ \\
Qian et al.'s Reddit (hate) & $-$10.30 & $-$4.67$^{*}$ & +0.13 & +0.40$^{*}$ \\
Qian et al.'s Gab (hate) & $-$11.98 & $-$6.78$^{*}$ & $-$0.38 & $-$0.09$^{*}$ \\
CGA-Wiki (attack) & $-$2.51 & $-$1.62 & -- & -- \\
BAD (unsafe turn)$^{\|}$ & +5.91 & +4.86 & +0.19 & $-$0.15$^{*}$ \\
\bottomrule
\end{tabular}
\end{minipage}
\end{table*}

Table~\ref{tab:corporayou} gives the over-flagging of address on every corpus with its checks, and Table~\ref{tab:platforms} the toxicity question's results with Sonnet and Detoxify beside Haiku, including Haiku's placebo depth slope at a fixed number of second-person words. Table~\ref{tab:corporadepth} gives the depth laws over all comments, among replies and within threads, Table~\ref{tab:corporasizerr} the thread size law and the log reply risk ratio under the toxicity question, and Table~\ref{tab:ownlaws} the same two laws under each corpus's own question. Gold's size laws are strong on Qian et al.'s corpora, where longer conversations are less hateful, and in BAD's dialogues, where longer dialogues have more unsafe turns. Under the toxicity question the LLMs shrink or erase the strong ones, inflate the rising size laws of Civil Comments and CAD, and reverse CGA-Wiki's falling one ($-$2.51 to +0.75). Contagion is weak on Qian et al.'s Reddit and negative on Gab, where a hateful post is less likely to be answered by another; the LLMs pull Gab's toward independence and strengthen Reddit's, and in BAD's dialogues they reverse it.

\section{What Agreement Leaves Open}
\label{app:agreement}

Table~\ref{tab:bounds} applies Proposition~\ref{prop:bounds} to every population and law. For each, it gives the range of the law over all labelers that flag as many comments as gold and agree with it at $\kappa = 0.9$ or $0.5$, and, where Haiku answered the corpus's own question without the parent, the range over all labelers with Haiku's counts of true and false positives and, with Haiku's flags fixed, the range of gold's law over every gold labeling with the same counts. At $\kappa = 0.9$, the depth law's sign is not identified on five of the seven populations, and at $\kappa = 0.5$ on all of them. The bounds are sharp: they are attained, and Proposition~\ref{prop:bounds} was checked by brute force on random small instances.

\begin{table*}[tp]
\caption{What overall agreement leaves open (Proposition~\ref{prop:bounds}). For each population and law: gold's law; the range of the law over all labelers that flag as many comments as gold and agree with it at $\kappa = 0.9$ or 0.5; Haiku's $\kappa$ at its own cut; the range of the law over all labelers with Haiku's counts of true and false positives; and, with Haiku's flags fixed, the range of gold's law over every gold labeling with the same counts. Depth laws in points per level, size laws in points per doubling. Haiku's counts are those of its answers to each corpus's own question without the parent; on BAD those answers exist only with the previous turn shown, so BAD has no Haiku columns. On BAD, where 5{,}314 of the 5{,}784 dialogues have 14 turns, so that most comments share one weight, the thread size law's range already reaches at $\kappa = 0.9$ the largest law of any labeling that flags gold's share (+45.72), and at $\kappa = 0.5$ also the smallest.}
\label{tab:bounds}
\footnotesize
\begin{tabular}{@{}llcccccc@{}}
\toprule
 & & & \multicolumn{2}{c}{Gold's flag rate} & \multicolumn{3}{c}{Haiku's counts} \\
\cmidrule(lr){4-5}\cmidrule(l){6-8}
Population & Law & Gold & $\kappa = 0.9$ & $\kappa = 0.5$ & $\kappa$ & Labeler's law & Gold's law, given Haiku's flags \\
\midrule
Civil Comments & Depth & $-$0.12 & $-$1.30 to +1.65 & $-$3.02 to +7.12 & 0.25 & $-$9.64 to +15.39 & $-$3.57 to +8.81 \\
 & Thread size & +0.15 & $-$1.91 to +2.22 & $-$9.94 to +5.76 &  & $-$37.43 to +17.08 & $-$13.58 to +6.10 \\
 & Article size & +0.32 & $-$5.26 to +2.07 & $-$19.03 to +2.72 &  & $-$21.91 to +8.16 & $-$18.77 to +2.92 \\
\addlinespace[2pt]
Civil Comments, 5\% sample & Depth & $-$0.16 & $-$1.37 to +1.83 & $-$3.21 to +7.45 & 0.24 & $-$10.37 to +16.26 & $-$3.85 to +9.27 \\
 & Thread size & +0.08 & $-$1.95 to +2.12 & $-$9.87 to +5.82 &  & $-$38.32 to +17.34 & $-$13.54 to +6.12 \\
 & Article size & +0.10 & $-$5.31 to +2.04 & $-$18.77 to +2.73 &  & $-$22.17 to +8.27 & $-$18.64 to +2.94 \\
\addlinespace[2pt]
CAD & Depth & +0.41 & $-$2.85 to +4.67 & $-$9.85 to +14.46 & 0.47 & $-$12.14 to +16.55 & $-$10.13 to +14.49 \\
 & Thread size & +1.64 & $-$6.87 to +5.25 & $-$22.67 to +6.86 &  & $-$24.63 to +8.22 & $-$22.07 to +6.96 \\
\addlinespace[2pt]
Qian et al.'s Reddit & Depth & $-$2.52 & $-$4.64 to $-$0.05 & $-$8.05 to +8.17 & 0.44 & $-$5.45 to +6.16 & $-$7.76 to +9.87 \\
 & Thread size & $-$10.30 & $-$19.46 to $-$1.13 & $-$40.91 to +13.27 &  & $-$34.15 to +9.70 & $-$40.29 to +14.97 \\
\addlinespace[2pt]
Qian et al.'s Gab & Depth & $-$1.95 & $-$5.54 to +2.29 & $-$10.04 to +11.24 & 0.66 & $-$8.35 to +8.53 & $-$8.50 to +9.53 \\
 & Thread size & $-$11.98 & $-$19.86 to $-$4.08 & $-$47.57 to +27.20 &  & $-$36.79 to +15.43 & $-$33.88 to +18.85 \\
\addlinespace[2pt]
CGA-Wiki & Depth & +3.74 & +2.00 to +5.45 & $-$1.87 to +9.84 & 0.65 & $-$1.26 to +11.74 & $-$1.65 to +6.93 \\
 & Thread size & $-$2.51 & $-$7.40 to +2.50 & $-$23.23 to +16.47 &  & $-$25.75 to +16.78 & $-$16.17 to +13.66 \\
\addlinespace[2pt]
BAD & Depth & +1.02 & $-$0.96 to +2.99 & $-$8.14 to +8.90 & -- & -- & -- \\
 & Thread size & +5.91 & $-$64.78 to +45.72 & $-$68.58 to +45.72 &  & -- & -- \\
\bottomrule
\end{tabular}
\end{table*}

\begin{table*}[tp]
\caption{The labeler that agrees best with gold in each population, among the LLM and classifier settings without the parent: its $\kappa$, its flag rate, the law it distorts most in standard-error units, that distortion (points per level for depth laws, points per doubling for size laws, log units for the risk ratio), its $\lvert z\rvert$, and how many of its laws it distorts by $\lvert z\rvert > 3$.}
\label{tab:bestkappa}
\scriptsize
\setlength{\tabcolsep}{2pt}
\begin{tabular}{@{}llcc>{\raggedright\arraybackslash}p{1.6cm}ccc@{}}
\toprule
Population & Labeler & $\kappa$ & Flags (\%) & Most distorted law & Distortion & $\lvert z\rvert$ & $\lvert z\rvert > 3$ \\
\midrule
Civil Comments & Detoxify-unbiased (gold's rate) & 0.71 & 7.6 & Log reply RR & +0.10 & 10.6 & 3 of 5 \\
Civil Comments, 5\% sample & s-nlp RoBERTa (gold's rate) & 0.72 & 7.5 & Log reply RR & +0.17 & 3.7 & 1 of 5 \\
CAD & Sonnet, abuse question & 0.51 & 20.9 & Log reply RR & $-$0.36 & 7.4 & 3 of 4 \\
Qian et al.'s Reddit & Gemma, hate-speech question & 0.62 & 23.6 & Thread size law & +2.25 & 5.8 & 2 of 4 \\
Qian et al.'s Gab & Dynabench hate (gold's rate) & 0.73 & 43.3 & Thread size law & +4.25 & 17.1 & 3 of 4 \\
CGA-Wiki & Detoxify (gold's rate) & 0.91 & 10.9 & Within-thread depth law & $-$0.54 & 4.9 & 2 of 3 \\
BAD & s-nlp RoBERTa (gold's rate) & 0.42 & 39.2 & Log reply RR & $-$0.36 & 19.5 & 3 of 4 \\
\bottomrule
\end{tabular}
\end{table*}

\begin{table*}[tp]
\caption{Validating the labeler's error rates on 2{,}000 uniformly sampled gold labels and correcting the law (divided by $\mathrm{TPR} - \mathrm{FPR}$ for linear laws; the matrix method for the reply risk ratio), over the combinations of labeler, population and law of the evidence, without labelers shown the parent: the share of combinations whose corrected 95\% interval covers gold's law in fewer than 90\% of 500 simulated validations, the median coverage, the share whose corrected law (at the true error rates) lies further from gold's than the uncorrected one (combinations with a defined correction in parentheses), and the median $\lvert$bias$\rvert$ relative to $\lvert$gold's law$\rvert$ after and before correction.}
\label{tab:correct}
\scriptsize
\setlength{\tabcolsep}{2.2pt}
\begin{tabular}{@{}lcccccc@{}}
\toprule
Law & Combinations & Cover $<$ 0.90 & Median coverage & Further from gold & Bias, corrected & Bias, raw \\
\midrule
Depth law & 113 & 82\% & 0.00 & 47\% (113) & 62\% & 53\% \\
Within-thread depth law & 113 & 84\% & 0.00 & 50\% (113) & 40\% & 61\% \\
Thread size law & 113 & 92\% & 0.00 & 48\% (113) & 40\% & 48\% \\
Article size law & 35 & 100\% & 0.00 & 91\% (35) & 190\% & 53\% \\
Log reply RR & 97 & 96\% & 0.00 & 79\% (67) & 166\% & 52\% \\
\midrule
All laws & 471 & 89\% & 0.00 & 56\% (441) & 69\% & 53\% \\
\bottomrule
\end{tabular}
\end{table*}

\begin{table*}[tp]
\caption{How far each law estimate falls from gold's, over every labeler setting, population and law of the evidence (depth, within-thread depth, thread size, article size, log reply risk ratio), without labelers shown the parent (on the other corpora the classifiers measure toxicity or offensiveness, not the corpus's construct): the number of estimates; the share off by more than a quarter of gold's law and the median $\lvert$distortion$\rvert$ relative to $\lvert$gold's law$\rvert$; estimates of the opposite sign with the labeler's 95\% interval excluding 0 (in parentheses: gold's interval also excludes 0); significance where gold's law is not significant (manufactured) and the reverse (lost); and the median share of the distortion that is unstructured error (Proposition~\ref{prop:decomp}). One rater: one random rater against the majority of the others, 20 draws of four laws (all but the within-thread law), Civil Comments reply threads only.}
\label{tab:census}
\scriptsize
\setlength{\tabcolsep}{2.2pt}
\begin{tabular}{@{}lccccccc@{}}
\toprule
Labelers & $N$ & $>$25\% off & Median off & Opposite sign & Manufactured & Lost & Unstructured \\
\midrule
All labelers & 471 & 74\% & 52\% & 23 (20) & 31 & 36 & 53\% \\
LLMs & 65 & 86\% & 63\% & 7 (4) & 10 & 4 & 56\% \\
Classifiers & 406 & 72\% & 50\% & 16 (16) & 21 & 32 & 53\% \\
$\kappa \ge 0.6$ & 130 & 44\% & 23\% & 0 (0) & 0 & 4 & 61\% \\
\midrule
One rater & 80 & 94\% & 47\% & -- & -- & -- & 65\% \\
\bottomrule
\end{tabular}
\end{table*}

\paragraph{The best-agreeing labeler and the error-rate correction.} Table~\ref{tab:bestkappa} gives, for each population, the labeler that agrees best with gold and the law it distorts most (Section~\ref{sec:agreement}), in the law's units and in standard errors. Table~\ref{tab:correct} gives the simulated correction for validated error rates by law.

\paragraph{Every law estimate.} Table~\ref{tab:census} gives the census of Section~\ref{sec:consequences} for all labelers, the LLMs, the classifiers and the labelers with $\kappa \ge 0.6$, with the single rater for reference: how far the estimates fall from gold's, how often they reverse a sign, manufacture or lose significance, and how much of their distortion is unstructured.

\section{Algorithms}
\label{app:alg}

\paragraph{Laws.} The working model's features are the labeler's score levels, depth, thread and article size, the number of replies and the auxiliary scores. Wave $j$ samples $n_j$ comments of $F_j$ with Algorithm~\ref{alg:item}'s probabilities, and the Wald interval's variance of $\hat t_m$ is estimated by $\sum_j \sum_{S_j} (1 - \pi_{ji}) b_{mi}^2 (Z_i - p_{ji})^2/\pi_{ji}^2$. The tempered prediction is $p_i = \sigma(\eta_i/\sqrt{1 + 0.39\, s_i^2})$, with $\eta_i$ the linear predictor and $s_i^2$ its variance. The priorities $\rho_m$ (1 for laws, 0.25 for placebo statistics) divide each target by its own optimal variance, and a guard doubles, up to six times, the priority of a target whose variance exceeds four times its optimum. The audited placebo size slope weights each gold negative by one over its unit's size, so that each unit counts about once.

\paragraph{Contagion.} The edge design targets the reply risk ratio and the labeler's placebo risk ratio, functions of eight edge totals. After a pilot, three waves draw reply edges with probabilities proportional to the working model's standard deviation of the log risk ratios' linearized weights, mixed with a uniform share of 0.15; one ridge logistic model gives the parent's gold label and another the reply's given the parent's. An edge costs the gold labels of its ends not yet bought. The difference estimator is post-stratified by the labeler's flags on the edge's two ends, and the interval is log-Wald, with MOVER-R~\cite{donner2012mover} on Wilson limits~\cite{wilson1927} for small counts.
\FloatBarrier
\section{Evaluation Details}
\label{app:eval}

\paragraph{The cells.} Table~\ref{tab:cells} lists the 55 evaluation cells, the 48 of them that also audit contagion, and the 17 development cells on which the design's settings were chosen. One cell, Qwen + parent on Civil Comments, applies the cut of Qwen without the parent to its scores with the parent, so it flags 21.9\% of comments; it enters the coverage and width results as one more labeler that steers an audit, but not Table~\ref{tab:neighty}, which tests the labeler's own law.

\begin{table*}[tp]
\caption{The audit cells, one labeler on one population each: the evaluation cells for laws and placebo slopes, those that also audit contagion (``same'': all of them; CGA-Wiki's conversations end at the attack, so it has none), and the 17 development cells on which the design's settings were chosen. Haiku and Sonnet at their own cut and Qwen and Gemma at gold's flag rate unless stated; ``+ parent'': replies labeled with the parent shown and roots without it. The four classifiers that played no part in developing \ts{} are s-nlp RoBERTa, martin-ha DistilBERT, the Dynabench hate model and CardiffNLP's offensive-language model.}
\label{tab:cells}
\scriptsize
\begin{tabular}{@{}>{\raggedright\arraybackslash}p{1.9cm}>{\raggedright\arraybackslash}p{8.1cm}>{\raggedright\arraybackslash}p{1.6cm}>{\raggedright\arraybackslash}p{4.6cm}@{}}
\toprule
Population & Evaluation cells & Contagion & Development cells \\
\midrule
Civil Comments & CardiffNLP offensive (gold's rate); Detoxify (gold's rate); Dynabench hate (gold's rate); Haiku + parent; IMSYPP (argmax); IMSYPP (gold's rate); martin-ha DistilBERT (gold's rate); Qwen; Qwen + parent (cut without the parent); s-nlp RoBERTa (gold's rate) & same & Detoxify (0.6); Detoxify-unbiased (gold's rate); Haiku; Qwen (own cut) \\
Civil Comments, 5\% sample & CardiffNLP offensive (gold's rate); Dynabench hate (gold's rate); Gemma; martin-ha DistilBERT (gold's rate); s-nlp RoBERTa (gold's rate); Sonnet; Sonnet + parent & same & Haiku + parent; IMSYPP (gold's rate); Qwen + parent \\
CAD & CardiffNLP offensive (gold's rate); Detoxify (0.6); Dynabench hate (gold's rate); Haiku; IMSYPP (gold's rate); martin-ha DistilBERT (gold's rate); s-nlp RoBERTa (gold's rate); Sonnet & same & Detoxify (gold's rate); Haiku + parent \\
Qian et al.'s Reddit & CardiffNLP offensive (gold's rate); Detoxify (0.6); Dynabench hate (gold's rate); Gemma; Haiku; IMSYPP (gold's rate); martin-ha DistilBERT (gold's rate); s-nlp RoBERTa (gold's rate) & same & Haiku + parent; IMSYPP (argmax) \\
Qian et al.'s Gab & CardiffNLP offensive (gold's rate); Detoxify (0.6); Dynabench hate (gold's rate); Gemma; Haiku; IMSYPP (gold's rate); martin-ha DistilBERT (gold's rate); s-nlp RoBERTa (gold's rate) & same & Detoxify (gold's rate); Haiku + parent \\
CGA-Wiki & CardiffNLP offensive (gold's rate); Dynabench hate (gold's rate); Haiku; IMSYPP (gold's rate); martin-ha DistilBERT (gold's rate); Qwen; s-nlp RoBERTa (gold's rate) & none & Detoxify (gold's rate); Haiku + parent \\
BAD & CardiffNLP offensive (gold's rate); Detoxify (gold's rate); Dynabench hate (gold's rate); Haiku + parent; IMSYPP (gold's rate); martin-ha DistilBERT (gold's rate); s-nlp RoBERTa (gold's rate) & same & Detoxify (0.6); IMSYPP (argmax) \\
\bottomrule
\end{tabular}
\end{table*}

\begin{table*}[tp]
\caption{Coverage of the 95\% intervals by target and budget: mean over the combinations of cell, budget and target, with the lowest in parentheses, for laws (depth, thread size, article size) and placebo slopes at 250, 500 and 1{,}000--8{,}000 labels (to 2{,}000 on CGA-Wiki), and for the reply and placebo risk ratios at 2{,}000--16{,}000. Calibrated within strata: calibrate and aggregate within strata of score level, depth and size (laws and placebo slopes only).}
\label{tab:covtarget}
\footnotesize
\begin{tabular}{@{}lcccccccc@{}}
\toprule
 & \multicolumn{3}{c}{Laws} & \multicolumn{3}{c}{Placebo slopes} & Reply RR & Placebo RR \\
\cmidrule(lr){2-4}\cmidrule(lr){5-7}\cmidrule(lr){8-8}\cmidrule(l){9-9}
Design & 250 & 500 & 1k--8k & 250 & 500 & 1k--8k & 2k--16k & 2k--16k \\
\midrule
\ts & .948 (.92) & .949 (.92) & .949 (.92) & .950 (.92) & .949 (.92) & .949 (.92) & .949 (.91) & .953 (.93) \\
\midrule
PPI++ & .888 (.56) & .908 (.66) & .939 (.75) & .894 (.55) & .909 (.58) & .937 (.68) & .950 (.92) & .951 (.92) \\
DSL & .927 (.79) & .933 (.88) & .943 (.88) & .936 (.81) & .940 (.87) & .946 (.87) & .950 (.92) & .956 (.92) \\
Active inference & .939 (.85) & .945 (.88) & .948 (.91) & .950 (.92) & .951 (.93) & .949 (.82) & .950 (.92) & .956 (.92) \\
Stratified & .928 (.69) & .941 (.81) & .948 (.89) & .934 (.78) & .943 (.83) & .948 (.89) & .949 (.91) & .953 (.92) \\
Uniform & .880 (.59) & .906 (.68) & .939 (.80) & .930 (.79) & .935 (.84) & .945 (.86) & .950 (.93) & .951 (.92) \\
Calibrate \& aggregate & .219 (.00) & .170 (.00) & .086 (.00) & .209 (.00) & .177 (.00) & .094 (.00) & .955 (.91) & .949 (.77) \\
Calibrated within strata & .650 (.00) & .643 (.00) & .742 (.00) & .754 (.00) & .814 (.00) & .874 (.00) & -- & -- \\
\bottomrule
\end{tabular}
\end{table*}

\begin{table*}[tp]
\caption{Width of \ts's 95\% interval divided by each design's (median over matched cell, budget and target entries; below 1: \ts's is narrower), by target and budget as in Table~\ref{tab:covtarget}. Calibrate and aggregate is not shown: it covers too rarely for its width to be comparable.}
\label{tab:widthtarget}
\footnotesize
\begin{tabular}{@{}lcccccccc@{}}
\toprule
 & \multicolumn{3}{c}{Laws} & \multicolumn{3}{c}{Placebo slopes} & Reply RR & Placebo RR \\
\cmidrule(lr){2-4}\cmidrule(lr){5-7}\cmidrule(lr){8-8}\cmidrule(l){9-9}
Design & 250 & 500 & 1k--8k & 250 & 500 & 1k--8k & 2k--16k & 2k--16k \\
\midrule
PPI++ & 0.87 & 0.76 & 0.63 & 0.66 & 0.60 & 0.46 & 0.85 & 0.40 \\
DSL & 0.94 & 0.89 & 0.75 & 0.61 & 0.58 & 0.50 & 0.88 & 0.58 \\
Active inference & 0.96 & 0.89 & 0.81 & 0.84 & 0.80 & 0.74 & 0.86 & 0.71 \\
Stratified & 0.90 & 0.88 & 0.76 & 0.87 & 0.91 & 0.80 & 0.91 & 0.89 \\
Uniform & 0.67 & 0.58 & 0.51 & 0.37 & 0.35 & 0.30 & 0.80 & 0.40 \\
\bottomrule
\end{tabular}
\end{table*}

\begin{table*}[tp]
\caption{Five labelers built from gold and structure: one minus Detoxify's score, a random score, a constant score, flags on exactly the comments at depth 2 or more, and a gold-aware trap whose score is the gold label at depths 0 and 1 and its opposite deeper. Coverage of \ts, PPI++ and uniform sampling (mean, lowest in parentheses) for laws and placebo slopes at 1{,}000--8{,}000 labels and for the risk ratios at 2{,}000--16{,}000, and the width of \ts's interval divided by uniform sampling's (median). The constant labeler flags nothing, so its placebo statistics are undefined.}
\label{tab:adversaries}
\footnotesize
\setlength{\tabcolsep}{3pt}
\begin{tabular}{@{}lcccccccccc@{}}
\toprule
 & \multicolumn{3}{c}{Laws and placebo slopes} & \multicolumn{3}{c}{Risk ratios} & \multicolumn{4}{c}{\ts's width $\div$ uniform's} \\
\cmidrule(lr){2-4}\cmidrule(lr){5-7}\cmidrule(l){8-11}
Labeler & \ts & PPI++ & Uniform & \ts & PPI++ & Uniform & Laws & Placebo slopes & Reply RR & Placebo RR \\
\midrule
Reversed score & .950 (.92) & .943 (.84) & .943 (.84) & .948 (.93) & .953 (.93) & .948 (.93) & 0.45 & 0.50 & 0.91 & 0.43 \\
Random score & .950 (.93) & .942 (.86) & .944 (.81) & .948 (.92) & .952 (.93) & .950 (.91) & 0.46 & 0.40 & 0.93 & 0.38 \\
Constant score & .974 (.91) & .969 (.83) & .970 (.83) & .947 (.93) & .948 (.92) & .949 (.94) & 0.44 & -- & 0.89 & -- \\
Flags every comment at depth $\ge 2$ & .949 (.93) & .941 (.80) & .942 (.81) & .973 (.93) & .957 (.93) & .959 (.93) & 0.45 & 0.50 & 0.92 & 0.98 \\
Gold-aware trap & .954 (.92) & .942 (.81) & .943 (.84) & .974 (.93) & .957 (.94) & .960 (.94) & 0.48 & 0.27 & 0.79 & 1.11 \\
\midrule
All five & .955 (.91) & .947 (.80) & .948 (.81) & .959 (.92) & .954 (.92) & .954 (.91) & 0.46 & 0.39 & 0.89 & 0.75 \\
\bottomrule
\end{tabular}
\end{table*}

\begin{table*}[tp]
\caption{Labels needed to detect a labeler's distortion of a law: the smallest budget (250, 500, 1{,}000, \dots{} labels for laws; 500 to 16{,}000 for the reply risk ratio; to 2{,}000 on CGA-Wiki) at which at least 80\% of 500 audits give an interval that excludes the labeler's own value of the law, for \ts{} and the stratified design. ${>}$: not reached at the largest budget; --: the law is not audited there. Rows: the audit cells in which some design reaches 80\% for some law. Rows ``+ parent'': replies labeled with their parent and roots without it; the audited law pools both modes.}
\label{tab:neighty}
\scriptsize
\setlength{\tabcolsep}{2.5pt}
\begin{tabular}{@{}llcccccccc@{}}
\toprule
 & & \multicolumn{2}{c}{Depth law} & \multicolumn{2}{c}{Thread size law} & \multicolumn{2}{c}{Article size law} & \multicolumn{2}{c}{Reply RR} \\
\cmidrule(lr){3-4}\cmidrule(lr){5-6}\cmidrule(lr){7-8}\cmidrule(l){9-10}
Population & Labeler & \ts & Stratified & \ts & Stratified & \ts & Stratified & \ts & Stratified \\
\midrule
Civil Comments & Haiku + parent & 250 & 500 & 500 & 500 & 500 & 500 & 4{,}000 & 8{,}000 \\
 & Qwen & ${>}$8{,}000 & ${>}$8{,}000 & ${>}$8{,}000 & ${>}$8{,}000 & ${>}$8{,}000 & ${>}$8{,}000 & 16{,}000 & ${>}$16{,}000 \\
 & IMSYPP (argmax) & 2{,}000 & 2{,}000 & ${>}$8{,}000 & ${>}$8{,}000 & 1{,}000 & 1{,}000 & 8{,}000 & 16{,}000 \\
 & IMSYPP (gold's rate) & 4{,}000 & 8{,}000 & ${>}$8{,}000 & ${>}$8{,}000 & ${>}$8{,}000 & ${>}$8{,}000 & ${>}$16{,}000 & ${>}$16{,}000 \\
 & s-nlp RoBERTa (gold's rate) & ${>}$8{,}000 & ${>}$8{,}000 & ${>}$8{,}000 & ${>}$8{,}000 & ${>}$8{,}000 & ${>}$8{,}000 & 16{,}000 & ${>}$16{,}000 \\
 & martin-ha DistilBERT (gold's rate) & ${>}$8{,}000 & ${>}$8{,}000 & ${>}$8{,}000 & ${>}$8{,}000 & ${>}$8{,}000 & ${>}$8{,}000 & 16{,}000 & ${>}$16{,}000 \\
\addlinespace[2pt]
Civil Comments, 5\% sample & Sonnet & 500 & 500 & 1{,}000 & 2{,}000 & 500 & 1{,}000 & 16{,}000 & 16{,}000 \\
 & Sonnet + parent & 500 & 500 & 1{,}000 & 1{,}000 & 500 & 500 & 8{,}000 & 16{,}000 \\
 & Gemma & ${>}$8{,}000 & ${>}$8{,}000 & ${>}$8{,}000 & ${>}$8{,}000 & 8{,}000 & ${>}$8{,}000 & ${>}$16{,}000 & ${>}$16{,}000 \\
 & s-nlp RoBERTa (gold's rate) & ${>}$8{,}000 & ${>}$8{,}000 & ${>}$8{,}000 & ${>}$8{,}000 & ${>}$8{,}000 & ${>}$8{,}000 & 16{,}000 & 16{,}000 \\
 & martin-ha DistilBERT (gold's rate) & ${>}$8{,}000 & ${>}$8{,}000 & ${>}$8{,}000 & ${>}$8{,}000 & ${>}$8{,}000 & ${>}$8{,}000 & 8{,}000 & 16{,}000 \\
 & Dynabench hate (gold's rate) & ${>}$8{,}000 & ${>}$8{,}000 & ${>}$8{,}000 & ${>}$8{,}000 & 8{,}000 & ${>}$8{,}000 & ${>}$16{,}000 & ${>}$16{,}000 \\
 & CardiffNLP offensive (gold's rate) & ${>}$8{,}000 & ${>}$8{,}000 & ${>}$8{,}000 & ${>}$8{,}000 & ${>}$8{,}000 & ${>}$8{,}000 & 16{,}000 & 16{,}000 \\
\addlinespace[2pt]
CAD & Haiku & ${>}$8{,}000 & ${>}$8{,}000 & 2{,}000 & 2{,}000 & -- & -- & 4{,}000 & 2{,}000 \\
 & Sonnet & ${>}$8{,}000 & ${>}$8{,}000 & 2{,}000 & 2{,}000 & -- & -- & 2{,}000 & 2{,}000 \\
 & Detoxify (0.6) & ${>}$8{,}000 & ${>}$8{,}000 & 8{,}000 & ${>}$8{,}000 & -- & -- & 4{,}000 & 2{,}000 \\
 & IMSYPP (gold's rate) & ${>}$8{,}000 & ${>}$8{,}000 & 8{,}000 & 8{,}000 & -- & -- & 4{,}000 & 4{,}000 \\
 & s-nlp RoBERTa (gold's rate) & ${>}$8{,}000 & ${>}$8{,}000 & 4{,}000 & 4{,}000 & -- & -- & 2{,}000 & 2{,}000 \\
 & martin-ha DistilBERT (gold's rate) & ${>}$8{,}000 & ${>}$8{,}000 & 4{,}000 & 8{,}000 & -- & -- & 4{,}000 & 2{,}000 \\
 & Dynabench hate (gold's rate) & ${>}$8{,}000 & ${>}$8{,}000 & 8{,}000 & 8{,}000 & -- & -- & 4{,}000 & 4{,}000 \\
 & CardiffNLP offensive (gold's rate) & ${>}$8{,}000 & ${>}$8{,}000 & 4{,}000 & 8{,}000 & -- & -- & 4{,}000 & 2{,}000 \\
\addlinespace[2pt]
Qian et al.'s Reddit & Haiku & 1{,}000 & 1{,}000 & 500 & 1{,}000 & -- & -- & 4{,}000 & 4{,}000 \\
 & Gemma & 8{,}000 & 8{,}000 & 2{,}000 & 4{,}000 & -- & -- & 16{,}000 & 16{,}000 \\
 & Detoxify (0.6) & 1{,}000 & 2{,}000 & 1{,}000 & 1{,}000 & -- & -- & 16{,}000 & 16{,}000 \\
 & IMSYPP (gold's rate) & 1{,}000 & 1{,}000 & 500 & 1{,}000 & -- & -- & 16{,}000 & 16{,}000 \\
 & s-nlp RoBERTa (gold's rate) & 1{,}000 & 1{,}000 & 500 & 1{,}000 & -- & -- & 4{,}000 & 4{,}000 \\
 & martin-ha DistilBERT (gold's rate) & 500 & 500 & 250 & 500 & -- & -- & 4{,}000 & 4{,}000 \\
 & Dynabench hate (gold's rate) & 2{,}000 & 4{,}000 & 1{,}000 & 2{,}000 & -- & -- & ${>}$16{,}000 & ${>}$16{,}000 \\
 & CardiffNLP offensive (gold's rate) & 1{,}000 & 1{,}000 & 500 & 1{,}000 & -- & -- & 4{,}000 & 4{,}000 \\
\addlinespace[2pt]
Qian et al.'s Gab & Haiku & 1{,}000 & 2{,}000 & 500 & 500 & -- & -- & 1{,}000 & 1{,}000 \\
 & Gemma & 2{,}000 & 4{,}000 & 1{,}000 & 2{,}000 & -- & -- & 2{,}000 & 4{,}000 \\
 & Detoxify (0.6) & 500 & 500 & 500 & 1{,}000 & -- & -- & 2{,}000 & 2{,}000 \\
 & IMSYPP (gold's rate) & 500 & 500 & 250 & 500 & -- & -- & 1{,}000 & 1{,}000 \\
 & s-nlp RoBERTa (gold's rate) & 500 & 500 & 500 & 1{,}000 & -- & -- & 2{,}000 & 2{,}000 \\
 & martin-ha DistilBERT (gold's rate) & 500 & 500 & 250 & 500 & -- & -- & 500 & 500 \\
 & Dynabench hate (gold's rate) & 500 & 1{,}000 & 500 & 1{,}000 & -- & -- & 8{,}000 & 8{,}000 \\
 & CardiffNLP offensive (gold's rate) & 500 & 500 & 500 & 500 & -- & -- & 2{,}000 & 2{,}000 \\
\addlinespace[2pt]
CGA-Wiki & Haiku & 500 & 1{,}000 & 2{,}000 & ${>}$2{,}000 & -- & -- & -- & -- \\
 & Qwen & 250 & 500 & 2{,}000 & ${>}$2{,}000 & -- & -- & -- & -- \\
 & IMSYPP (gold's rate) & 500 & 500 & ${>}$2{,}000 & ${>}$2{,}000 & -- & -- & -- & -- \\
 & s-nlp RoBERTa (gold's rate) & 1{,}000 & 1{,}000 & ${>}$2{,}000 & ${>}$2{,}000 & -- & -- & -- & -- \\
 & martin-ha DistilBERT (gold's rate) & 1{,}000 & 1{,}000 & 2{,}000 & ${>}$2{,}000 & -- & -- & -- & -- \\
 & Dynabench hate (gold's rate) & 250 & 500 & 2{,}000 & ${>}$2{,}000 & -- & -- & -- & -- \\
 & CardiffNLP offensive (gold's rate) & 1{,}000 & 1{,}000 & ${>}$2{,}000 & ${>}$2{,}000 & -- & -- & -- & -- \\
\addlinespace[2pt]
BAD & Haiku + parent & 4{,}000 & 4{,}000 & ${>}$8{,}000 & ${>}$8{,}000 & -- & -- & 1{,}000 & 500 \\
 & Detoxify (gold's rate) & 4{,}000 & 4{,}000 & ${>}$8{,}000 & ${>}$8{,}000 & -- & -- & 2{,}000 & 2{,}000 \\
 & IMSYPP (gold's rate) & ${>}$8{,}000 & ${>}$8{,}000 & 2{,}000 & 8{,}000 & -- & -- & 4{,}000 & 4{,}000 \\
 & s-nlp RoBERTa (gold's rate) & 4{,}000 & 4{,}000 & 8{,}000 & ${>}$8{,}000 & -- & -- & 1{,}000 & 1{,}000 \\
 & martin-ha DistilBERT (gold's rate) & 4{,}000 & 4{,}000 & 8{,}000 & ${>}$8{,}000 & -- & -- & 16{,}000 & 16{,}000 \\
 & Dynabench hate (gold's rate) & 1{,}000 & 1{,}000 & 4{,}000 & ${>}$8{,}000 & -- & -- & 4{,}000 & 4{,}000 \\
 & CardiffNLP offensive (gold's rate) & ${>}$8{,}000 & ${>}$8{,}000 & 8{,}000 & ${>}$8{,}000 & -- & -- & 4{,}000 & 4{,}000 \\
\bottomrule
\end{tabular}
\end{table*}

\begin{table}[tp]
\caption{Fresh gold: each gold label bought is a fresh judgment drawn from the comment's raters (seven cells of the 5\% sample, 1{,}000 and 4{,}000 labels, 500 audits): coverage of the law of the expected labels with Theorem~\ref{thm:valid}(c)'s rater-variance term added to the interval and without it, coverage of the law of the labels actually drawn, the median width relative to fixed gold (laws; placebo slopes), and the raters bought per label (adaptive: mean over cells, range in parentheses). Coverage: mean, lowest in parentheses.}
\label{tab:freshgold}
\scriptsize
\setlength{\tabcolsep}{2pt}
\begin{tabular}{@{}>{\raggedright\arraybackslash}p{2.2cm}cccccc@{}}
\toprule
 & \multicolumn{2}{c}{Coverage} & Realized & \multicolumn{2}{c}{Width $\div$ fixed} & Raters \\
\cmidrule(lr){2-3}\cmidrule(lr){5-6}
Gold process & with term & without & law & Laws & Placebo & per label \\
\midrule
Fixed gold (all raters, as in the data) & -- & .949 (.93) & .949 (.93) & 1.00 & 1.00 & all \\
\midrule
One rater & .952 (.93) & .933 (.84) & .949 (.92) & 1.53 & 1.43 & 1 \\
Majority of three raters & .958 (.93) & .934 (.78) & .950 (.93) & 1.16 & 1.18 & 3 \\
Adaptive (three, and two more if they disagree) & .957 (.93) & .933 (.81) & .948 (.92) & 1.07 & 1.12 & 3.6 (3.5--3.8) \\
\bottomrule
\end{tabular}
\end{table}

\begin{table}[tp]
\caption{Removing one part of \ts{} at a time: the median width of the variant's 95\% interval divided by \ts's (above 1: wider) for laws and placebo slopes at 1{,}000--8{,}000 labels and the reply and placebo risk ratios at 2{,}000--16{,}000, and the variant's coverage (mean, lowest in parentheses). The edge design has no tempering variant.}
\label{tab:ablation}
\scriptsize
\setlength{\tabcolsep}{2pt}
\begin{tabular}{@{}>{\raggedright\arraybackslash}p{2.6cm}cccccc@{}}
\toprule
 & \multicolumn{4}{c}{Width $\div$ \ts's} & \multicolumn{2}{c}{Coverage} \\
\cmidrule(lr){2-5}\cmidrule(l){6-7}
Part removed (replacement) & Laws & Placebo & Reply RR & Placebo RR & Laws, placebo & Risk ratios \\
\midrule
Weighted sampling (uniform probabilities instead) & 1.45 & 2.11 & 1.17 & 1.73 & .947 (.90) & .948 (.89) \\
Auxiliary classifiers (the labeler's scores only) & 1.11 & 1.08 & 1.06 & 1.06 & .950 (.92) & .949 (.92) \\
Four waves (one wave: the model fitted on the pilot only) & 1.08 & 1.08 & 1.03 & 1.03 & .950 (.86) & .952 (.92) \\
Tempered predictions (plain predictions) & 0.99 & 0.99 & -- & -- & .948 (.86) & -- \\
\bottomrule
\end{tabular}
\end{table}

\paragraph{By target and budget.} Tables~\ref{tab:covtarget} and~\ref{tab:widthtarget} break Table~\ref{tab:confirm} down by target and budget. \ts{} covers at the nominal rate at every budget, from 250 labels on; the comparators that sample uniformly or by the labeler's score fall short at small budgets, mostly on the size laws, and calibrate and aggregate falls short everywhere except on the risk ratios, for which it calibrates the flags of an edge's two ends jointly. The cells of the four classifiers that played no part in developing \ts{} and the other evaluation cells behave alike: on the other 27 evaluation cells \ts{} covers 94.9\% for laws (lowest 91.6\%; Section~\ref{sec:eval} gives the four's), and its intervals are 0.62 times as wide as PPI++'s there and 0.64 times on the four's cells.

\paragraph{Labelers built to break the audit.} Table~\ref{tab:adversaries} gives coverage under each of the five labelers built from gold and structure. Under all of them \ts{} keeps nominal coverage and stays at most half as wide as uniform sampling on the laws. PPI++ and uniform sampling fall below 90\% coverage in some cells under every one of them, because their uniform samples rarely draw the comments that weigh most. Under the trap, the stratified design's strata are pure and it returns the exact value.

\paragraph{Labels needed to detect a distortion.} Table~\ref{tab:neighty} gives, for every audit cell in which some design detects a distortion, the smallest budget at which at least 80\% of 500 audits give an interval that excludes the labeler's own law. For contagion, \ts{} needs fewer labels than the stratified design in eight of 40 cases and more in five, all on CAD or BAD (BAD: Haiku + parent; CAD: Haiku, Detoxify at 0.6, and martin-ha DistilBERT and CardiffNLP offensive at gold's rate); Section~\ref{sec:eval} gives the counts for laws.

\paragraph{How the design was chosen.} The design's settings (the working model's features, tempering, the uniform share, the number of waves and the priorities) were chosen on the 17 development cells of Table~\ref{tab:cells}, separate from the evaluation cells. We also chose them again seven times for laws and six times for contagion, each time leaving out one population's development cells: the design adopted with all cells was adopted again in four of seven runs for laws and in none of six for contagion. Against the oracle of Section~\ref{sec:eval}, \ts{} reaches 0.54 of its precision on the placebo risk ratio.

\paragraph{Fresh gold and ablations.} Table~\ref{tab:freshgold} gives the audits that buy fresh judgments, and Table~\ref{tab:ablation} the variants of \ts{} with one part removed (Section~\ref{sec:eval}).

\FloatBarrier
\section{Proofs}
\label{app:proofs}

\paragraph{Proposition~\ref{prop:decomp}.} Write $Y_i = Z_iY_i + (1 - Z_i)Y_i$ and add and subtract TPR and FPR: $L(Y) = \mathrm{TPR}\sum_i w_iZ_i + \mathrm{FPR}\sum_i w_i(1 - Z_i) + S_{\mathrm{TP}} + S_{\mathrm{FP}}$, and $\sum_i w_i(1 - Z_i) = -L(Z)$ since $\sum_i w_i = 0$. For escalation, $w_i = (x_i - \bar x)/\mathrm{SS}$ with $x_i$ the capped depth, and $\sum_{Z_i = 0}(Y_i - \mathrm{FPR}) = 0$, so $S_{\mathrm{FP}} = \sum_{Z_i = 0}(x_i - \bar x_0)(Y_i - \mathrm{FPR})/\mathrm{SS} = (\mathrm{SS}_0/\mathrm{SS})\beta_0$. \emph{Permutations.} For any law $\ell$, let $Y^\sigma$ be $Y$ permuted uniformly within gold positives and within gold negatives, $Y^{F}$ equal to $Y$ on gold negatives and to $Y^\sigma$ on gold positives, and $Y^{H}$ the reverse. With $u = \E\ell(Y^\sigma) - \ell(Z)$, the identity $\ell(Y) - \ell(Z) = u + \tfrac12[(\E\ell(Y^{F}) - \E\ell(Y^\sigma)) + (\ell(Y) - \E\ell(Y^{H}))] + \tfrac12[(\E\ell(Y^{H}) - \E\ell(Y^\sigma)) + (\ell(Y) - \E\ell(Y^{F}))]$ splits the remainder into the Shapley values~\cite{shapley1953value} of the two players that restore the actual false positives and the actual hits. For a linear law, $\E L(Y^\sigma) = \mathrm{TPR}\sum_{Z_i = 1}w_i + \mathrm{FPR}\sum_{Z_i = 0}w_i = (\mathrm{TPR} - \mathrm{FPR})L(Z)$, both differences in the first bracket equal $S_{\mathrm{FP}}$ and both in the second $S_{\mathrm{TP}}$. \qed

\paragraph{Proposition~\ref{prop:bounds}.} The counts fix only how many flags fall among gold positives and among gold negatives, so $L(Y)$ is largest when they take each class's largest weights and smallest with its smallest; $\kappa$ and the flag rate $q$ fix TPR and FPR, since, with $\bar z$ gold's prevalence, $q = \bar z\,\mathrm{TPR} + (1 - \bar z)\mathrm{FPR}$ and the observed agreement is $p_o = \bar z\,\mathrm{TPR} + (1 - \bar z)(1 - \mathrm{FPR})$. Moving one flag to the next larger weight in its class moves $L$ by one gap, so the attained values form a chain with steps at most the largest gap. The statement with the roles exchanged follows by the same argument with $Y$ and $Z$ swapped, since the counts are symmetric in the two labelings. Agreement reported per score level fixes these counts within each level, and the argument applies level by level. \qed

\paragraph{Designs.} A \emph{design} is a nonanticipating, possibly randomized query policy, which picks each next comment using only the labels revealed so far, with an output $T$ computed from the revealed labels. It is \emph{unbiased} for a target $f(Z) \in \R^m$ if $\E[T \mid Z = z] = f(z)$ for every $z \in \{0,1\}^N$, and its \emph{risk} under a full-support working model $p$ is $\E_p\lVert T - \E_p f\rVert_H^2$; the main text's priorities are $H = \mathrm{diag}(\rho_m)$. Expectations are under $p$ and the design's randomness unless $Z = z$ is fixed. A transcript $O$ reveals a partial assignment $s(O) \in \{*, 0, 1\}^N$, and $z \succeq s$ means that $z$ agrees with $s$ where $s$ is revealed. Let $\mu = \E_p f$, $g = f - \mu$, $\ip{\xi}{g} = \E_p[\xi(Z)^\top g(Z)]$ and $H$ positive definite. \refstepcounter{lemma}\label{lem:duality}\emph{Lemma~\thelemma\ (weak duality).} For every unbiased design and every $\xi$ with $\E\xi = 0$, $\ip{\xi}{g}^2 \le I_O(\xi)\,\E\lVert T - \mu\rVert_H^2$ with $I_O(\xi) = \E\lVert\E[\xi(Z) \mid O]\rVert^2_{H^{-1}}$: by unbiasedness $\E[(T - \mu)^\top\xi(Z)] = \ip{\xi}{g}$, which equals $\E[(T - \mu)^\top\E[\xi \mid O]]$, and Cauchy--Schwarz applies.

\paragraph{Proposition~\ref{prop:ident}.} If no unit has two labeled comments, each $\ind\{z \succeq s\}$, and hence $\E[T \mid Z = z]$, is a multilinear polynomial with at most one label per unit. Multilinear expansions on $\{0,1\}^N$ are unique, so a target with a product of two labels of one unit is not recovered by any estimator, while every target in the span is recovered by labeling one uniformly chosen comment per unit and inverting. Fr\'echet's bounds are attained by nested events and by events as disjoint as the marginals allow. \qed

\paragraph{Proposition~\ref{prop:sqrt}.} Write $a_i = (b_{mi})_m$ for the weights of comment $i$ in a vector of targets and $\alpha_i = p_i(1 - p_i)\lVert a_i\rVert_H^2$ ($\alpha_i = b_i^2p_i(1 - p_i)$ for one target), and $q_i$ for the main text's $\pi_i$. Under a product law an additive target separates into single labels with targets $a_i(Z_i - p_i)$, since the cross terms of the pairing and of the information in Lemma~\ref{lem:duality} vanish by independence. This holds for adaptive designs too. Whether label $i$ is bought may depend on the history $\mathcal H$ of labels bought before it, but under the product law $Z_i$ is independent of that history, so conditioning on $\mathcal H$ leaves the single-label problem below with buying probability $q_i(\mathcal H)$; since $\alpha_i/q$ is convex in $q$, $\E_{\mathcal H}[\alpha_i/q_i(\mathcal H)] \ge \alpha_i/\E_{\mathcal H}[q_i(\mathcal H)]$, and the design that buys each label independently with probability $\E_{\mathcal H}[q_i(\mathcal H)]$ has the same expected cost and no larger risk. A label bought with probability $q$, with output $t_0$ if not bought and $t_z$ if bought, is unbiased if and only if $(1 - q)t_0 + qt_z = a_i(z - p_i)$; its risk $\alpha_i/q + \lVert t_0\rVert_H^2(1 - q)/q$ is smallest at $t_0 = 0$, and minimizing $\sum_i\alpha_i/q_i$ subject to $\sum_ic_iq_i \le B$ and $q_i \le 1$ gives the square-root rule. For a smooth $g$, the delta method at $\E_p t$ gives the first-order risk of $g(\hat t)$ as the risk of $\nabla g^\top \hat t$, which is unbiased for the additive target with weights $a_i = \nabla g^\top \partial t/\partial Z_i$ whenever $\hat t$ is unbiased for $t$; the square-root rule with these weights estimates every total with the same draws, without bias, and attains the minimum. \qed

\paragraph{Theorem~\ref{thm:valid}.} \emph{(a)} Let $\mathcal G_j$ be everything revealed before wave $j$ is drawn. The flags, scores and auxiliary scores depend only on $Z$ and on randomness independent of the design, so given them and $Z$ only the design is random. Wave $j$'s $\pi_{ji}$ and $p_{ji}$ are $\mathcal G_j$-measurable, and systematic sampling in random order includes $i \in F_j$ with probability exactly $\pi_{ji}$, so $\E[\sum_{S_j} b_{mi}(Z_i - p_{ji})/\pi_{ji} \mid \mathcal G_j, Z, Y] = \sum_{F_j} b_{mi}(Z_i - p_{ji})$ and wave $j$'s term has conditional mean $\sum_{F_j} b_{mi}Z_i$ whatever the model. The pilot's total is exact and the waves partition the other comments, so the tower property gives $\E[\hat t_m \mid Z, Y] = t_m(Z)$; nothing restricts how $Y$ depends on $Z$. The centered wave terms are martingale differences, which gives the variance; the Poisson formula approximates each wave's conditional variance, since systematic sampling has no unbiased variance estimator. Since $\pi_{ji} \ge \delta n_j/|F_j|$, the wave's Poisson variance is at most $(|F_j|/(\delta n_j) - 1)/(|F_j|/n_j - 1)$, about $1/\delta$ for a small sampling fraction, times that of uniform sampling with the same predictions, whatever the labeler~\cite{hesterberg1995defensive}. \emph{(b)} $\E[\hat R\hat b] = \E\hat a = Rb$, so $\E\hat R\,b + \Cov(\hat R, \hat b) = Rb$, and Cauchy--Schwarz bounds the covariance by $\mathrm{sd}(\hat R)\,\mathrm{sd}(\hat b)$~\cite{hartley1954ratio}. \emph{(c)} Draw, for every comment, the label $\tilde Z_i$ it would receive if bought, independently across comments and of the design; the audit reveals $\tilde Z_i$ exactly when it buys label $i$, so it runs as in (a) with gold $\tilde Z$. Then (a) gives $\E[\hat t_m \mid \tilde Z, Y] = t_m(\tilde Z)$, and averaging over $\tilde Z$ gives $\E[\hat t_m \mid Y] = \sum_i b_{mi}\E\tilde Z_i = t_m(\bar Z)$. The total variance adds $\Var(t_m(\tilde Z)) = \sum_i b_{mi}^2 \Var(\tilde Z_i)$ to the expected design variance, which the waves' variances estimate. \qed
\clearpage
\end{document}